\documentclass[paperpaper,twocolumn,english,superscriptaddress,aps,prx,floatfix,amsfonts,amssymb]{revtex4-1}

\usepackage{epsfig}
\usepackage{graphicx}
\usepackage{dcolumn}
\usepackage{bm}

\newcommand{\EQ}{\begin{equation}}
\newcommand{\EN}{\end{equation}}
\newcommand{\ea}{\end{eqnarray}}
\newcommand{\ba}{\begin{eqnarray}}

\newcommand{\bear}{\begin{eqnarray}}
\newcommand{\ear}{\end{eqnarray}}

\begin{document}
\title{Diffusive spin transport of the spin-$1/2$ $XXZ$ chain in the Ising regime at zero magnetic field and finite temperature}


\author{J. M. P. Carmelo}
\affiliation{Center of Physics of University of Minho and University of Porto, LaPMET, P-4169-007 Oporto, Portugal}
\affiliation{CeFEMA, Instituto Superior T\'ecnico, Universidade de Lisboa, LaPMET, Av. Rovisco Pais, P-1049-001 Lisboa, Portugal}
\author{P. D. Sacramento}
\affiliation{CeFEMA, Instituto Superior T\'ecnico, Universidade de Lisboa, LaPMET, Av. Rovisco Pais, P-1049-001 Lisboa, Portugal}

\begin{abstract}
The studies of this paper on the spin-$1/2$ $XXZ$ chain at finite temperatures $T>0$
have two complementary goals. The first is to identify the spin carriers of all its $S_q>0$ energy eigenstates and to show that their 
spin elementary currents fully control the spin-transport quantities. Here $S_q$ is the $q$-spin of 
the continuous $SU_q(2)$ symmetry of the model for anisotropy $\Delta >1$. To achieve this goal, 
our studies rely on a suitable exact {\it physical-spin representation} valid in throughout the Hilbert space
and associated with normal spin operators, rather than fractionalized particles such as spinons.
Both the spin stiffness and the zero-field spin-diffusion constant are expressed in 
terms of thermal expectation values of the square of the elementary currents carried by the spin carriers.
Our second goal is to confirm that the zero-field and finite-temperature spin transport is normal diffusive for
$\Delta >1$. We use two complementary methods that rely on an inequality for the $T>0$ 
spin stiffness and the above thermal expectation values, respectively, 
to show that the contributions to ballistic spin transport vanish. Complementarily, for
$T>0$ and $\Delta >1$, the spin-diffusion constant is found to be  finite and enhanced upon lowering $T$, 
reaching its largest yet finite values at low temperatures. Evidence suggests that it diverges in the 
$\Delta \rightarrow 1$ limit for $T>0$, consistent with $T>0$ anomalous superdiffusive spin transport at $\Delta = 1$ 
and zero field.
\end{abstract}
\maketitle

\section{Introduction}
\label{SECI}

There has been significant recent interest in the spin transport properties of the one-dimensional (1D) spin-$1/2$ $XXZ$ 
chain in the limit of high temperature \cite{Znidaric_11,Ljubotina_17,Medenjak_17,Ilievski_18,Gopalakrishnan_19,Weiner_20,Gopalakrishnan_23}.
In this paper we consider that spin chain for anisotropy $\Delta > 1$. It is a correlated quantum system of great 
interest for different physical systems. For instance, this includes systems of ultra-cold atoms where
spin transport in a tunable spin-$1/2$ $XXZ$ chain has been realized \cite{Jepsen_20}.
The theoretical complex Bethe strings of lengths two and three
\cite{Gaudin_71,Takahashi_99,Carmelo_22,Carmelo_23} in the spin-conducting phase of the spin-$1/2$ $XXZ$ chain for
spin densities $m = 2S^z/N \in [0,1]$, corresponding magnetic fields $h\in [h_{c1},h_{c2}]$, and anisotropy $\Delta \approx 2$ 
have been observed in experimental studies of the dynamical properties 
of spin chains in quasi-1D materials \cite{Carmelo_23,Bera_20}. [The critical fields $h_{c1}$ and $h_{c2}$ are defined in Eq. (\ref{criticalfields}) of Appendix \ref{A}.] Indeed, the low-temperature dynamical properties of such materials is well described by this and related 
1D integrable quantum systems \cite{Carmelo_23,Bera_20,Wang_18,Carmelo_06,Carmelo_19,Carmelo_19A}. 

Studies considering the limit of high temperature reveal that at zero magnetic field the spin-$1/2$ $XXZ$ chain has 
ideal ballistic spin transport for $-1<\Delta <1$, anomalous superdiffusive behavior for $\Delta =1$, and normal diffusive spin transport 
for $\Delta >1$ \cite{Znidaric_11,Ljubotina_17,Medenjak_17,Ilievski_18,Gopalakrishnan_19,Weiner_20,Gopalakrishnan_23}. 
Finite contributions to the spin-diffusion constant exist for anisotropy $0<\Delta<1$ and $T>0$ at zero
field \cite{Sirker_11,Nardis_18,Nardis_19}. However, for $0<\Delta<1$ the finite-temperature dominant spin transport is ballistic, 
the spin stiffness being finite \cite{Zotos_99}. Finite contributions to the spin-diffusion constant at finite temperature 
are also known to exist for anisotropy $\Delta >1$ and zero magnetic field \cite{Nardis_19,Steinigeweg_11}. 
The issue, however, is whether normal diffusive spin transport is dominant at finite temperature. This requires that the contributions 
to ballistic spin transport vanish in the thermodynamic limit.

The absence of spin-flip odd charges besides the 
total magnetization for $\Delta >1$, zero field, and all finite temperatures $T>0$ \cite{Nardis_19} proves that the Mazur lower bound \cite{Mazur_69}
of the spin stiffness vanishes. Whether this, combined with arguments involving the use of
the hydrodynamic assumption (local equilibration) provides a rigorous proof that the spin stiffness itself vanishes 
\cite{Nardis_18,Nardis_19} or merely a strong heuristic argument that it vanishes \cite{Tomaz_24} remains under debate.

Complementarily to the studies of Ref. \onlinecite{Nardis_19}, in this paper we handle the problem using
completely different methods. In the last three to four decades, representations in terms of spinons and similar quasi-particles such as psinons 
and antipsinons \cite{Karbach_02} have been widely used to successfully describe the 
static and dynamical properties of both spin-chain models and the physics of the materials they represent. 
Hence such representations became the paradigm of the spin-chains physics. 

However, at zero field and arbitrary finite temperature, the contributions to spin transport involve
a huge number of energy eigenstates. Many of those are generated in the thermodynamic limit 
from ground states by an infinite number of elementary microscopic processes. This renders the usual spinon and alike 
representations unsuitable to handle that very complex finite-temperature quantum problem. 

The results of this paper are for zero magnetic field, $h=0$, unless specified.
Alternatively, we use a suitable exact {\it physical-spin representation} for
the $N$ physical spins $1/2$ of the spin-$1/2$ $XXZ$ chain for $\Delta >1$ in its full Hilbert space.
It is associated with normal spin operators, rather than fractionalized particles such as spinons, psinons,
and antipsinons. It accounts for the $q$-spin continuous $SU_q(2)$ symmetry of that
chain for $\Delta >1$ \cite{Carmelo_22,Carmelo_23,Pasquier_90}.

Such a representation naturally emerges from the relation of the
$q$-spin continuous $SU_q(2)$ symmetry's irreducible representations
to a complete set of $2^N$ energy eigenstates for all field values, $h\geq 0$. 
It is expressed in terms of a number ${\cal{M}} = N-2S_q$ of paired physical spins in the $q$-spin singlet configurations of 
the $S_q>0$ energy eigenstates and a complementary number $M=2S_q$ of unpaired physical spins in the $q$-spin multiplet configuration of
these states. 

A result that plays a major role in our studies is that
only the latter unpaired physical spins are {\it the spin carriers}.
In their number, $M=2S_q$, $S_q$ is the $q$-spin whose values are the same as for spin $S$.
For simplicity, in this paper we consider that $N$ and $2S_q$ are even integer numbers, yet in the thermodynamic limit
the same results are reached for $N$ and $2S_q$ odd integer numbers. 

Our studies have two complementary goals. The first is to show that the spin elementary 
currents carried by the spin carriers of all $S_q>0$
energy eigenstates fully control spin-transport quantities such as the spin stiffness and the
spin-diffusion constant. Our second goal is to confirm that $T>0$ spin transport is normal
diffusive for $\Delta >1$. 

We express both the spin stiffness for fields $h\geq 0$ and the spin-diffusion constant in terms of the 
spin elementary currents $j_{\pm 1/2}$ carried by the spin carriers of projection $\pm 1/2$. 
That gives $D^z (T) = {m^2\over 2T} \Omega_{m} (T)$ for the spin stiffness and 
$D (T) = C (T)\,\Pi (T)$ for the spin-diffusion constant, respectively, both for finite
temperatures $T>0$. Here $m = 2S^z/N$, the coefficient $C (T)$ is finite, and
$\Omega_{m} (T) = N\,\langle\vert j_{\pm 1/2}\vert^2\rangle_{S^z,T}$ and 
$\Pi (T) = N\,\langle\vert j_{\pm 1/2}\vert^2\rangle_{T}$ are thermal
expectation values at fixed $S^z$ and temperature $T$ and at fixed $T$, respectively. 

Most recent results on spin transport in the spin-$1/2$ $XXZ$ chain refer to the limit of high temperature 
\cite{Znidaric_11,Ljubotina_17,Medenjak_17,Ilievski_18,Gopalakrishnan_19,Weiner_20,Gopalakrishnan_23}.
We derive an inequality for the spin stiffness. It provides strong evidence that for 
all finite temperatures $T>0$  and $\Delta >1$ it vanishes both at $h=0$ and for $h\rightarrow 0$. 

We use our general expression $D (T) = C (T)\,\Pi (T)$ for the spin-diffusion constant in terms of the 
spin elementary currents $j_{\pm 1/2}$ to calculate its anisotropy $\Delta$ and temperature $T$ dependence 
in the limit of very low temperatures. We find nearly ballistic spin transport, but with vanishing 
spin stiffness. Similar results also involving nearly ballistic 
transport at low temperature were obtained in the case of charge transport for
the half-filled 1D Hubbard model \cite{Carmelo_24}.
In addition, we combine our general expression for the spin-diffusion constant
with previous results \cite{Gopalakrishnan_19,Steinigeweg_11} to study its behaviors
in the opposite limit of high finite temperatures. 

Our overall results on the spin-diffusion constant reveal it is enhanced upon lowering the temperature and
reaches its largest yet finite values for very low temperatures. For $\Delta >1$, it only diverges in
the $T\rightarrow\infty$ limit. This shows that the spin-diffusion constant 
$D (T) = C (T)\,\Pi (T)$ is finite for $\Delta >1$ and $T>0$. Since the coefficient 
$C (T)$ is finite, this implies as well that $\Pi (T)$ is finite and thus $\langle\vert j_{\pm 1/2}\vert^2\rangle_{T} = \Pi (T)/N$ is of the 
order of $1/N$. We also show that $\Omega_0 (T) < \Pi (T)$
in the spin stiffness limiting expression, $\lim_{m\rightarrow 0}D^z (T) = {m^2\over 2T} \Omega_0 (T)$. 
In agreement with the inequality we found it to obey, this shows and confirms that the spin stiffness 
vanishes at zero field for $\Delta >1$ and $T>0$. 

Combination of all our results thus confirms the occurrence of normal diffusive spin transport 
for $T>0$ and $\Delta >1$. On the other hand, evidence is found that for the spin-$1/2$ $XXZ$ chain
the spin-diffusion constant diverges in the $\Delta \rightarrow 1$ limit for all temperatures, which is
consistent with anomalous superdiffusive spin transport at $\Delta = 1$.

For spin anisotropy $\Delta=\cosh\eta > 1$ and thus $\eta > 0$, spin densities $m \in [0,1]$, exchange integral $J$, and lattice length 
$L\rightarrow\infty$ for $N/L$ finite, the Hamiltonian of the spin-$1/2$ $XXZ$ chain in a longitudinal magnetic field $h$ reads,
\begin{equation}
\hat{H} = J\sum_{j=1}^{N}\left({\hat{S}}_j^x{\hat{S}}_{j+1}^x + {\hat{S}}_j^y{\hat{S}}_{j+1}^y + 
\Delta\,{\hat{S}}_j^z{\hat{S}}_{j+1}^z\right) - g\mu_B\,h\,\hat{S}^z \, .
\label{HD1}
\end{equation}
Here $\hat{\vec{S}}_{j}$ is the spin-$1/2$ operator at site $j=1,...,N$ with components $\hat{S}_j^{x,y,z}$,
$g$ is the Land\'e factor, and $\mu_B$ is the Bohr magneton. This Hamiltonian describes the correlations of
$N=\sum_{\sigma =\uparrow,\downarrow}N_{\sigma}$ physical spins $1/2$. 
We use natural units in which the Planck constant 
and the lattice spacing are equal to one, so that $N=L$.

The paper is organized as follows. The physical-spin representation used in our
studies and the corresponding identification of the spin carriers are the issues
addressed in Sec. \ref{SECII}. This includes providing information on both the relation 
of that representation to the Bethe-ansatz quantum numbers and its description of states beyond
that ansatz. In Sec. \ref{SECIII} we use an inequality for the spin stiffness to
provide strong evidence that the ballistic contributions to spin transport 
vanish in the thermodynamic limit for anisotropy $\Delta > 1$ and all finite temperatures $T> 0$. 
In Sec. \ref{SECIV} we derive the $T=0$ spin stiffness for anisotropies $\Delta > 1$ and spin densities $m\in [0,1]$.
The spin-diffusion constant is expressed for $\Delta >1$ and $T>0$ in terms of the spin elementary currents
carried by the spin carriers in the multiplet configuration of all $S_q>0$ states in Sec. \ref{SECV}. 
We then discuss its behavior in the limit of both low and high temperatures and its enhancement upon lowering the 
temperature for finite temperatures $T>0$. The spin-diffusion constant is found
to remain finite for $T>0$. We use suitable thermal averages of the square
of the spin elementary currents carried by the spin carriers to show that for $\Delta >1$
and $T>0$ the spin stiffness indeed vanishes both at $h=0$ and in the $h\rightarrow 0$ limit,
consistently with the results obtained in Sec. \ref{SECIII} by the use of an inequality.
This combined with the found finiteness of the spin-diffusion constant for $\Delta >1$
and $T>0$ confirms the dominance of normal diffusive spin transport.
Finally, the concluding remarks are presented in 
Sec. \ref{SECVI}. Three Appendices provide useful information needed for our studies.

\section{physical-spin representation and the spin carriers elementary currents}
\label{SECII}

We start by introducing the information needed for the studies of this paper on the physical-spins 
representation of the spin-$1/2$ $XXZ$ chain with anisotropy $\Delta > 1$ \cite{Carmelo_22,Carmelo_23}.
It applies to the whole Hilbert space and directly refers to {\it all} $N$ physical spins $1/2$ described by 
the Hamiltonian, Eq. (\ref{HD1}). 

The usual representations in terms of spinons, psinons, and antipsinons 
\cite{Karbach_02} refer only to limited subspaces \cite{Carmelo_23} and are not
suitable for our goals. Indeed, our study involves contributions to spin-current expectation values 
of $S_q >0$ energy eigenstates that are generated from ground states by an infinite
number of elementary processes and span subspaces where such representations do not apply.

In the case of anisotropy $\Delta >1$, the physical-spin representation is a generalization 
of that for the $\Delta =1$ isotropic case \cite{Carmelo_15,Carmelo_17,Carmelo_20}. 
For $\Delta >1$, the spin projection $S^z$ remains a good quantum number whereas spin $S$ is not. 
It is replaced by the $q$-spin $S_q$ in the eigenvalues of the Casimir generator of the continuous 
$SU_q(2)$ symmetry \cite{Carmelo_22,Pasquier_90}. The values of $q$-spin $S_q$ are exactly the 
same for anisotropy $\Delta >1$ as spin $S$ for $\Delta =1$. This includes their relation to the values 
of $S^z$. Hence {\it singlet} and {\it multiplet} refer in this paper to physical-spins configurations with 
zero and finite $q$-spin $S_q$, respectively.

In the following we use the physical-spin representation to identify the spin carriers and 
to introduce their spin elementary currents.

\subsection{Physical spins in two types of configurations}
\label{SECIIA}

The representation in terms of unpaired and paired physical spins used in this paper for the spin-$1/2$
$XXZ$ chain applies to anisotropies in the range $\Delta \geq 1$. In the case of the isotropic point, Refs. \onlinecite{Carmelo_15,Carmelo_17}
provide useful information on that representation and corresponding notation in the more standard $SU(2)$ case. 

That the irreducible representations of the $\Delta >1$ continuous $SU_q(2)$ symmetry are isomorphic 
to those of the $\Delta =1$ $SU(2)$ symmetry \cite{Carmelo_22}, with the spin $S$ being replaced by the 
$q$-spin $S_q$, refers to a useful symmetry. A related issue that plays a key role in the studies of this paper 
follows from the use of the corresponding $\Delta >1$ continuous $SU_q(2)$ symmetry algebra. 
For that algebra the operator $\hat{S}^z$ remains the same as at $\Delta = 1$. In contrast,
the $SU (2)$ symmetry operators $(\hat{\vec{S}})^2$ and $\hat{S}^{\pm}$ are
replaced by $\eta$-dependent operators $(\hat{\vec{S}}_{\eta})^2$ and $\hat{S}^{\pm}_{\eta}$, respectively.
The $\eta >0$ and $\Delta >1$ continuous $SU_q(2)$ symmetry algebra is then such that \cite{Pasquier_90},
\begin{eqnarray}
&& (\hat{\vec{S}}_{\eta})^2 = \hat{S}^+_{\eta}\hat{S}^-_{\eta} - {\sinh^2 (\eta/2)\over\sinh^2 \eta} 
+ {\sinh^2 (\eta\,(\hat{S}^z+1/2))\over\sinh^2 \eta} 
\nonumber \\
&& [\hat{S}^{+}_{\eta},\hat{S}^{-}_{\eta}] = {\sinh (\eta\,2\hat{S}^z)\over\sinh \eta} \, .
\label{Seta2+-}
\end{eqnarray}
Note that the usual $SU (2)$ symmetry expressions are achieved in the $\eta\rightarrow 0$ limit.

Accounting for the isomorphism of the irreducible representations of the continuous $SU_q(2)$ symmetry
to those of the $SU(2)$ symmetry, we find, as for the spin-$1/2$ $XXX$ chain \cite{Carmelo_15,Carmelo_17,Carmelo_20}, 
that all quantum-problem $2^N$ energy eigenstates whose $q$-spin is in the range $0\leq S_q\leq N/2$ are 
populated by physical spins in two types of configurations \cite{Carmelo_22,Carmelo_23}:\\

(i) A number $M=M_{+1/2}+M_{-1/2} = 2S_q$ of {\it unpaired physical spins} in the multiplet configuration
 of any $S_q > 0$ energy eigenstate. The number $M_{\pm 1/2}$ of such unpaired physical spins of projection $\pm 1/2$
 is solely determined by that state values of $S_q$ and $S^z$, as it reads $M_{\pm 1/2} = S_q \pm S^z$, so that,
\begin{equation}
2S^z = M_{+1/2} - M_{-1/2} \hspace{0.20cm}{\rm and}\hspace{0.20cm}2S_q = M_{+1/2} + M_{-1/2} \, .
\label{2Sz2Sq}
\end{equation}

(ii) A complementary even number ${\cal{M}} = {\cal{M}}_{+1/2} + {\cal{M}}_{-1/2}$
of {\it paired physical spins} in the energy eigenstates's singlet configurations where
${\cal{M}}_{+1/2} = {\cal{M}}_{-1/2} = N/2-S_q$.\\

That this holds for {\it all} $2^N$ energy eigenstates of the Hamiltonian, Eq. (\ref{HD1}), much simplifies 
our study. Our designation {\it $n$-pairs} refers both to {\it $1$-pairs} and {\it $n$-string-pairs} for $n>1$:\\

(i) The internal degrees of freedom of a $1$-pair correspond to one unbound singlet
pair of physical spins. It is described by a $n=1$ single real Bethe rapidity. Its translational 
degrees of freedom refer to the $1$-band momentum  carried by each such a pair whose 
discrete values belong to an interval $q_j \in [q_1^-,q_1^+]$ where $j = 1,...,L_1$.\\

(ii) The internal degrees of freedom of the $n$-string-pairs refer to a number $n>1$ of bound singlet pairs of physical spins. They are
bound within a configuration described by a complex Bethe $n$-string given below in Eq. (\ref{LambdaIm}). 
Their translational degrees of freedom refer to the $n>1$ $n$-band momentum carried by each such $n$-pairs
whose discrete values belong to an interval $q_j \in [q_n^-,q_n^+]$ where $j = 1,...,L_n$ .\\

For both $n=1$ and $n>1$, the number $L_n$ of $j=1,...,L_n$ discrete momentum 
values $q_j$ in each $n$-band is given by,
\begin{equation}
L_ n = N_n + N_n^h \hspace{0.20cm}{\rm where}\hspace{0.20cm}
N_n^h = 2S_q + \sum_{n'=n+1}^{\infty}2(n'-n)N_{n'} \, .
\label{LnNnh}
\end{equation}
Here $N_n$ is the number of occupied $q_j$'s and thus of $n$-pairs and $N^h_{n}$ 
is that of unoccupied $q_j$'s. We call them $n$-holes. 
The numbers $M_{\pm 1/2}$ and ${\cal{M}}_{\pm 1/2}$ of unpaired and paired
physical spins of projection $\pm 1/2$, respectively, and the corresponding
total numbers of such two types of physical spins
of an energy eigenstate can then be exactly expressed as,
\begin{eqnarray} 
M & = & M_{+1/2}+M_{-1/2} = 2S_q  
\nonumber \\
M_{\pm 1/2} & = & S_q \pm S^z = N/2 - \sum_{n=1}^{\infty}n\,N_n \pm S^z
\nonumber \\
{\cal{M}} & = & {\cal{M}}_{+1/2} + {\cal{M}}_{-1/2} = N - 2S_q = \sum_{n=1}^{\infty}2n\,N_n
\nonumber \\
{\cal{M}}_{\pm 1/2} & = & N/2-S_q = \sum_{n=1}^{\infty}n\,N_n \, .
\label{MM}
\end{eqnarray}

Such numbers obey the following sum-rule that gives the total number of physical spins:
\begin{eqnarray} 
N & = & M_{+1/2} + M_{-1/2} + {\cal{M}}_{+1/2} + {\cal{M}}_{-1/2} 
\nonumber \\
& = & M + {\cal{M}} \, .
\label{NMM}
\end{eqnarray}

The Bethe-ansatz quantum numbers $I_j^n$ \cite{Gaudin_71} are actually
the discrete $n$-band momentum values $q_j = {2\pi\over N}I_j^n$
in units of ${2\pi\over N}$. They read $I_j^n = 0,\pm 1,...,\pm {L_n -1\over 2}$ for $L_n$ odd
and $I_j^n =\pm 1/2,\pm 3/2,...,\pm {L_n -1\over 2}$ for $L_n$ even. 

Each $n$-band $j=1,...,L_n$ set of $q_j$'s such that $q_{j+1}-q_j = {2\pi\over N}$ have values 
in the range $q_j \in [q_n^-,q_n^+]$ where,
\begin{equation}
q_n^{\pm} = q_n (\pm\pi) = \pm {\pi\over N}(L_n -1) + q_n^{\Delta} \, ,
\nonumber \\
\label{qqq}
\end{equation}
and $q_n^{\Delta}$ has limiting values given in Eq. (\ref{qDelta}) of Appendix \ref{A}.
Such a set $\{q_j\}$ of discrete $n$-band $q_j$'s have Pauli-like 
occupancies: The corresponding $n$-band momentum distributions read $N_n (q_j) = 1$ and 
$N_n (q_j) = 0$ for occupied and unoccupied $q_j$'s, respectively.
Since $q_{j+1}-q_j = {2\pi\over N}$, the $n$-band discrete momentum values $q_j={2\pi\over N}I_j^n$'s
can in the thermodynamic limit be described by continuous $n$-band momentum variables $q \in [q_n^-,q_n^+]$.

Each of the $2^N$ energy eigenstates that span the full Hilbert space is specified by 
the numbers $M_{\pm 1/2} = S_q \pm S^z$ of unpaired physical spins of projection $\pm 1/2$ and
a set of $n=1,...,\infty$ $n$-band momentum distributions $\{N_n (q_j)\}$. 

Each such state is described by a corresponding set of $n$-band rapidity functions $\{\varphi_{n} (q_{j})\}$.
For $n >1$ they are the real part of a complex rapidity $\varphi_{n,l,j} = \varphi_{n,l} (q_j)$.
The corresponding $n>1$ $n$-string structure depends on the system size. In the thermodynamic limit in which that structure simplifies, 
a rapiditity can both for $n=1$ and $n>1$ be expressed as,
\begin{equation}
\varphi_{n,l} (q_j) = \varphi_{n} (q_j) + i(n + 1 -2l)\,\eta \hspace{0.20cm}{\rm where}\hspace{0.20cm}l=1,...,n \, .
\label{LambdaIm}
\end{equation}
The $n$-band rapidity functions $\varphi_{n} (q_{j})$ where $q_j \in [q_n^-,q_n^+]$ associated with its real part
can be defined in terms of their inverse functions $q_n (\varphi)$ where $\varphi \in [-\pi,\pi]$. The latter are for 
each energy eigenstate solutions of the coupled Bethe-ansatz equations \cite{Gaudin_71,Carmelo_22}.
They are given in functional form in Eq. (\ref{BAqn}) of Appendix \ref{A}.

Concerning the connection of the Bethe-ansatz quantum numbers and quantities to the
physical-spin representation, a singlet $n$-pair of $n$-band momentum $q_j$, which
contains a number $n\geq 1$ of singlet pairs of physical spins, is described by a corresponding
rapidity $\varphi_{n,l} (q_j)$, Eq. (\ref{LambdaIm}). When $n>1$, that rapidity imaginary part, $i(n + 1 -2l)\,\eta$ 
where $l=1,...,n$, is finite and describes the binding of the $l=1,...,n$ singlet pairs of physical spins within 
the corresponding $n$-string-pair. 

The information contained in the $n=1,...,\infty$ $n$-band momentum distributions $\{N_n (q)\}$ of any energy
eigenstate defines the corresponding distribution of the Bethe ansatz $N_n$ occupied discrete 
momentum values $q_j = {2\pi\over N}I_j^n$ in each of the $n=1,...,\infty$ $n$-bands 
for which $N_n>0$ over the available $L_n$ discrete $n$-band momentum values, Eq. (\ref{LnNnh}).
These distributions describe the translational degrees of freedom of the 
$\sum_{n=1}^{\infty} N_n$ $n$-pairs that populate an energy eigenstate and thus of the
corresponding ${\cal{M}} = {\cal{M}}_{+1/2} + {\cal{M}}_{-1/2} =\sum_{n=1}^{\infty} 2n\,N_n$ 
paired physical spins. 

In the thermodynamic limit, the set of $n$-band momentum distributions 
$\{N_n (q)\}$ where $q \in [q_n^-,q_n^+]$ can alternatively be represented by a 
corresponding set of $n$-band rapidity-variable distributions $\{\tilde{N}_n (\varphi)\}$ where $\varphi \in [-\pi,\pi]$.
Both alternative representations describe occupancies of the $n$-pairs, each containing $2n$ paired physical spins. 
The rapidity-variable distributions representation is defined by the relation $\tilde{N}_{n} (\varphi_n (q)) = N_n (q)$, Eq. (\ref{tildeNN}) of Appendix \ref{A}.
It is directly defined by the Bethe-ansatz equations, Eq. (\ref{BAqn}) of Appendix \ref{A},
in terms of the inverse functions, $\{q_n (\varphi)\}$, of the rapidity
functions, $\{\varphi_n (q)\}$. Their solution provides the later set of rapidity functions
of each energy eigenstate in the argument of $\tilde{N}_{n} (\varphi_n (q))$.
 
The real part $\varphi_{n} (q)$ of the rapidity $\varphi_{n,l} (q)$, Eq. (\ref{LambdaIm}), 
provides the $n$-band rapidity-variable value $\varphi$ of a $n$-pair of $n$-band momentum $q$ within
the alternative representation in terms of $n$-band rapidity-variable distributions $\{\tilde{N}_n (\varphi)\}$.
That representation is useful because both the Bethe-ansatz equations, Eq. (\ref{BAqn}) of Appendix \Ref{A}, and
the expressions of several physical quantities provided by the Bethe ansatz have a simpler form in terms
of rapidity variables $\varphi \in [-\pi,\pi]$. Important examples for the studies of this paper refer to the general 
expressions of the energy eigenvalues, second expression in Eq. (\ref{Energy}) given below, and
of the spin-current expectation values, second expression in Eq. (\ref{jznn}) of Appendix \ref{B}.

The above analysis clarifies the relation of the number ${\cal{M}} = {\cal{M}}_{+1/2} + {\cal{M}}_{-1/2} =\sum_{n=1}^{\infty} 2n\,N_n$ 
of paired physical spins to the Bethe ansatz. The question is thus where in that ansatz is the remaining number $M = 2S_q$ 
of unpaired physical spins in the multiplet configuration of any finite-$S_q$ energy eigenstate?

On the one hand, the translational degrees of freedom of the $M= M_{+1/2} + M_{-1/2} = 2S_q$ 
unpaired physical spins are described in each $n$-band of a $S_q >0$ energy eigenstate
for which $N_n>0$ by its $N^h_{n} = 2S_q +\sum_{n'=n+1}^{\infty}2(n'-n)N_{n'}$ $n$-holes.
Below in Sec. \ref{SECIIB} the spin carriers are shown to be such $M= 2S_q$ unpaired physical spins. 
How the quantum system identifies the $M=2S_q$ unpaired physical spins out of the 
$N^h_{n} = 2S_q +\sum_{n'=n+1}^{\infty}2(n'-n)N_{n'}$ $n$-holes is an issue clarified in Appendix \ref{C}.
It combines a squeezed-space construction \cite{Ogata_90,Penc_97,Kruis_04} and
the Hamiltonian, Eq. (\ref{HD1}), in the presence of a vector potential.

On the other hand, the spin internal degrees of freedom of the $M_{\pm 1/2} = S_q \pm S^z$ 
unpaired physical spins of projection $\pm 1/2$ in the multiplet configuration of a $S_q>0$ energy eigenstate
is an issue beyond the Bethe ansatz. Indeed, that ansatz refers only to subspaces spanned either by the 
highest-weight states (HWSs) or the lowest-weight states (LWSs) of the continuous $SU_q(2)$ symmetry 
\cite{Gaudin_71,Carmelo_22}. For such states, all the $M=2S_q$ unpaired physical spins 
have the same projection $+1/2$ or $-1/2$, respectively. This implies that $S^z = S_q$ and $S^z = -S_q$, 
respectively. In this paper we use a HWS Bethe ansatz. 

In contrast, the physical-spin representation applies to the whole Hilbert space.
It thus accounts for the internal degrees of freedom of the $M_{\pm 1/2} = S_q \pm S^z$ 
unpaired physical spins. Let $\left\vert l_{\rm r}^{\eta},S_q,S^z\right\rangle$ be an energy eigenstate 
of the Hamiltonian $\hat{H}$, Eq. (\ref{HD1}), whose set of quantum numbers beyond $S_q$ and $S^z$ 
needed to specify it are here generally denoted by $l_{\rm r}^{\eta}$. Consider a HWS 
$\left\vert l_{\rm r}^{\eta},S_q,S_q\right\rangle$. A number $2S_q$ of continuous $SU_q(2)$ symmetry 
non-HWSs outside the Bethe-ansatz solution referring to different multiplet configurations of the
$M= M_{+1/2} + M_{-1/2} = 2S_q$ unpaired physical spins are generated from that HWS as,
\begin{equation} 
\left\vert l_{\rm r}^{\eta},S_q,S_q-n_z\right\rangle 
= {1\over \sqrt{{\cal{C}}_{\eta}}}({\hat{S}}^{+}_{\eta})^{n_z}\left\vert l_{\rm r}^{\eta},S_q,S_q\right\rangle \, .
\label{state}
\end{equation} 
Here $S_q - n_z = S^z$, $n_z = M_{-1/2}$, and,
\begin{equation}
{\cal{C}}_{\eta} = 
\prod_{l=1}^{n_z}{\sinh^2 (\eta\,(S_q+1/2)) - \sinh^2 (\eta\,(l - S_q - 1/2))\over\sinh^2 \eta} \, ,
\label{nonBAstatesDelta1}
\end{equation}
for $n_z = M_{-1/2} = 1,...,2S_q$. Similarly to the $\Delta =1$ bare ladder spin operators $\hat{S}^{\pm}$, the 
action of the $\Delta >1$ $q$-spin continuous $SU_q(2)$ symmetry ladder operators $\hat{S}^{\pm}_{\eta}$,
Eq. (\ref{Seta2+-}), on $S_q > 0$ energy eigenstates flips one {\it unpaired physical-spin}. 

For the non-HWSs, Eq. (\ref{state}), the two sets of $M_{-1/2} = n_z = S_q - S^z$ 
and $M_{+1/2} = 2S_q-n_z = S_q + S^z$ unpaired physical spins have opposite 
projections $-1/2$ and $+1/2$, respectively. The multiplet configurations associated
with the internal degrees of freedom of the $M_{\pm 1/2} = S_q \pm S^z$ 
unpaired physical spins of projection $\pm 1/2$ are thus generated as given in Eq. (\ref{state}). 
Each $S_q >0$ energy eigenstate has a multiplet configuration uniquely defined by the 
values of $M_{+1/2}$ and $M_{-1/2}$.

Within the present functional representation, the energy eigenvalues of any energy
eigenstate $\left\vert l_{\rm r}^{\eta},S_q,S^z\right\rangle$, Eq. (\ref{state}), of
the Hamiltonian, Eq. (\ref{HD1}), are in the thermodynamic limit given by,
\begin{eqnarray}
E (l_{\rm r}^{\eta},S_q,S^z) & = & - \sum_{n=1}^{\infty}\sum_{l=1}^{n}\sum_{j=1}^{L_n}
{J\sinh^2(\eta)\,N_{n} (q_{j})\over \cosh\eta - \cos\varphi_{n,l} (q_j)} 
\nonumber \\
& + & g\mu_B\,h\,{1\over 2}\left(M_{+1/2} - M_{-1/2}\right) \, .
\label{Energy0}
\end{eqnarray}

The summation $\sum_{l=1}^{n}$ runs here over the $n$-pairs number $n =2,...,\infty$ of 
bound singlet pairs, each containing a number $2n$ of paired physical spins.
In Appendix \ref{A} that summation is performed, with the result,
\begin{eqnarray}
&& E (l_{\rm r}^{\eta},S_q,S^z) = - J\sinh\eta\sum_{n=1}^{\infty}\sum_{j=1}^{L_n}
{\sinh (n\,\eta)\,N_{n} (q_{j})\over \cosh (n\,\eta) - \cos\varphi_{n} (q_{j})} 
\nonumber \\
&& \hspace{1cm} - g\mu_B\,h\,{1\over 2}\left(M_{+1/2} - M_{-1/2}\right)
\nonumber \\
&& = - {JN\sinh\eta\over 2\pi}\sum_{n=1}^{\infty}\int_{-\pi}^{\pi}d\varphi\,2\pi\sigma_n (\varphi) 
{\sinh (n\,\eta)\,\tilde{N}_n (\varphi)\over \cosh (n\,\eta) - \cos\varphi} 
\nonumber \\
&& \hspace{1cm} - g\mu_B\,h\,{1\over 2}\left(M_{+1/2} - M_{-1/2}\right) \, .
\label{Energy}
\end{eqnarray}
The distributions $2\pi\sigma_n (\varphi) = d q_n (\varphi)/d\varphi$ appearing here are in the
thermodynamic limit the Jacobians of the transformations from $n$-band momentum values $q$ to $n$-band 
rapidity variables $\varphi$. They are defined by Eq. (\ref{sigmanderivative}) of Appendix \ref{A}.

The continuous $SU_q (2)$ symmetry is behind the energy of the $n_z\equiv S_q - S^z = 1,...,2S_q$ 
non-HWSs, Eq. (\ref{state}), outside the Bethe ansatz differing from that of the corresponding HWS only in the presence 
of a magnetic field $h$. As given in Eq. (\ref{Energy}), this difference refers to the values of the numbers of unpaired
physical spins of projection $\pm 1/2$, such that $M_{+1/2} - M_{-1/2} = 2S^z$, Eq. (\ref{2Sz2Sq}).
That symmetry indeed imposes that at zero field the number $2S_q +1$ of states of the same $q$-spin tower
have exactly the same energy.

Consistently, states of the same $q$-spin tower have exactly the same $n$-pairs occupancy 
configurations and thus the same values for the set of $n=1,...,\infty$ $n$-band momentum distributions $\{N_n (q_j)\}$ and 
corresponding rapidity functions $\{\varphi_{n} (q_{j})\}$. 

\subsection{Spin carriers and their spin elementary currents}
\label{SECIIB}

The Hamiltonian, Eq. (\ref{HD1}), in the presence of a uniform vector potential (twisted boundary conditions),
$\hat{H}=\hat{H} (\Phi/L)$ with $\Phi = \Phi_{\uparrow} = -\Phi_{\downarrow}$, remains solvable by the Bethe ansatz \cite{Shastry_90}.
After some straightforward algebra using the corresponding $\Phi\neq 0$ Bethe-ansatz equations \cite{Carmelo_18}, 
which refer to Eq. (\ref{BAqn}) of Appendix \ref{A} for $n=1,...,\infty$ with $q_n (\varphi)$ replaced by $q_n (\varphi) -2n{\Phi\over N}$, 
one finds that the momentum eigenvalues of HWSs are in the thermodynamic limit given by,
\begin{equation}
P = \pi\sum_{j=1}^{L_n}N_{n} + \sum_{n=1}^{\infty}\sum_{j=1}^{L_n}N_{n} (q_j)\,q_j + {\Phi\over N}\,(N - \sum_{n=1}^{\infty}2n\,N_{n}) \, .
\label{PPhi}
\end{equation}
The number of physical spins that couple to the vector potential is given by the factor that multiplies ${\Phi\over N}$ 
in Eq. (\ref{PPhi}). The use of the exact sum rules, Eq. (\ref{MM}), shows that such a 
number is actually given by $M= 2S_q = N - \sum_{n=1}^{\infty}2n\,N_{n}$.

The term ${\Phi\over N}\,N$ in ${\Phi\over N}\,M = {\Phi\over N}\,(N - \sum_{n=1}^{\infty}2n\,N_{n})$,
Eq. (\ref{PPhi}), refers to {\it all} $N$ physical spins coupling to the vector potential in the absence of physical-spin
singlet pairing. Indeed, the negative coupling counter terms $-\sum_{n=1}^{\infty}2n\,N_n$ refer to the 
number $2n$ of paired physical spins in each of the $n$-pairs that populate a state.
This applies to both $1$-pairs and $n$-string-pairs for which $n>1$. 

These counter terms {\it exactly cancel} 
the positive coupling of the corresponding $2n$ paired physical spins in 
each such $n$-pairs. Physically, this results from the $n$-pairs singlet nature.
As a result of such counter terms, only the $M = 2S_q = N - \sum_{n=1}^{\infty}2n\,N_n$ 
unpaired physical spins couple to the vector potential and thus carry spin current.

Consistent with $M_{\pm 1/2} =  N/2 - \sum_{n=1}^{\infty}n\,N_{n} \pm S^z$, Eq. (\ref{MM}),
a similar analysis for non-HWSs, Eq. (\ref{state}), gives Eq. (\ref{PPhi}) with 
${\Phi\over N}\,(N - \sum_{n=1}^{\infty}2n\,N_{n})={\Phi\over N}\,M = {\Phi\over N}\,2S_q$ replaced by 
${\Phi\over N}(M_{+1/2} - M_{-1/2})={\Phi\over N}2S^z$, as given in Eq. (\ref{PPhinonHWS})
of Appendix \ref{C}. This confirms that only the $M_{\pm 1/2} = S_q \pm S^z$ 
unpaired physical spins of projection $+1/2$ and $-1/2$ couple to a uniform vector potential,
their coupling having opposite sign, respectively. Hence they are 
indeed the spin carriers. 

Each of the $M_{\pm 1/2}$ spin carriers of projection $\pm 1/2$ of a $S_q>0$ energy eigenstate
carries a spin elementary current $j_{\pm 1/2}$ that we evaluate in the following. The 
$z$ component of the spin-current operator reads,
\begin{equation}
\hat{J}^z = -i\,J\sum_{j=1}^{N}(\hat{S}_j^+\hat{S}_{j+1}^- - \hat{S}_{j+1}^+\hat{S}_j^-)  \, .
\label{c-s-currents}
\end{equation}
We use the following notations for its zero-temperature expectation values,
\begin{eqnarray}
\langle \hat{J}^z (l_{\rm r}^{\eta},S_q,S^z) \rangle & = &
\langle  l_{\rm r}^{\eta},S_q,S^z\vert\hat{J}^z\vert  l_{\rm r}^{\eta},S_q,S^z\rangle
\nonumber \\
\langle \hat{J}^z_{HWS} (l_{\rm r}^{\eta},S_q) \rangle & = &
\langle l_{\rm r}^{\eta},S_q,S_q\vert\hat{J}^z\vert  l_{\rm r}^{\eta},S_q,S_q\rangle \, .
\label{not-currents}
\end{eqnarray}

There is a unitary transformation associated with unitary operators $\hat{U}_{\eta}^{\pm}$
that maps any $\Delta =1$ energy eigenstate $\vert l_{\rm r}^{0},S,S^z\rangle$ 
onto a $\Delta >1$ energy eigenstate $\vert l_{\rm r}^{\eta},S_q,S^z\rangle$ such that $S_q = S$ and vice versa \cite{Carmelo_22},
\begin{eqnarray}
\vert l_{\rm r}^{\eta},S_q,S^z\rangle\vert_{S_q=S} & = & \hat{U}_{\eta}^+\vert l_{\rm r}^{0},S,S^z\rangle
\nonumber \\
\vert l_{\rm r}^{0},S,S^z\rangle\vert_{S=S_q} & = & \hat{U}_{\eta}^-\vert l_{\rm r}^{\eta},S_q,S^z\rangle \, .
\label{trans-states}
\end{eqnarray}

The generation of a number $n_z= 1,...,2S_q$ of non-HWSs from one HWS also holds for
the $\Delta = 1$ $SU(2)$ spin algebra, Eq. (\ref{state}) for $\eta\rightarrow 0$ and $S_q  = S$.
By combining the systematic use of the commutators given in Eq. (21) of Ref. \onlinecite{Carmelo_15} with known state 
transformation laws, we find that the spin-current expectation values of the number $n_z= 1,...,2S$ 
of $SU(2)$ spin non-HWSs $\vert l_{\rm r}^{0},S,S^z\rangle$ of the spin-$1/2$ $XXX$ chain are 
defined in terms of that of the corresponding HWS $\vert l_{\rm r}^{0},S,S\rangle$ by the following 
{\it exact} relation,
\begin{equation}
\langle \hat{J}^z (l_{\rm r}^{0},S,S^z) \rangle =
{S^z\over S}\langle \hat{J}^z_{HWS} (l_{\rm r}^{0},S) \rangle  \, .
\label{rel-currents-XXX}
\end{equation}
A similar relation was obtained  in Ref. \onlinecite{Carmelo_15} within a LWS representation.

As reported in Sec. \ref{SECIIA}, the irreducible representations of the $\eta >0$ 
continuous $SU_q(2)$ symmetry \cite{Pasquier_90} are isomorphic to those of the $\eta=0$ $SU(2)$ symmetry. 
The spin projection $S^z$ remains a good quantum number under the unitary transformation, Eq. (\ref{trans-states}),
whereas spin $S$ is mapped onto $q$-spin $S_q$ such that $S_q=S$. It follows that under it the 
factor ${S^z\over S}$ in the exact relation, Eq. (\ref{rel-currents-XXX}), is mapped onto a corresponding factor 
${S^z\over S_ q}$ such that $S_q=S$. 

The $\Delta =1$ relation, Eq. (\ref{rel-currents-XXX}), is thus mapped onto a corresponding exact relation
for $\Delta >1$,
\begin{eqnarray}
&& \langle \hat{J}^z (l_{\rm r}^{\eta},S_q,S^z) \rangle =
{S^z\over S_ q}\langle \hat{J}^z_{HWS} (l_{\rm r}^{\eta},S_q) \rangle 
\nonumber \\
&& = {M_{+1/2} - M_{-1/2}\over M_{+1/2} + M_{-1/2}}
\langle \hat{J}^z_{HWS} (l_{\rm r}^{\eta},S_q) \rangle \, .
\label{rel-currents-XXZ}
\end{eqnarray}
To reach the second expression, we have used Eq. (\ref{2Sz2Sq}).

Combination of the exact relation, Eq. (\ref{rel-currents-XXZ}), with the generation of 
unpaired physical spin flips described by Eq. (\ref{state}), gives,
\begin{equation}
\langle \hat{J}^z (l_{\rm r}^{\eta},S_q,S^z) \rangle = \sum_{\sigma =\pm 1/2}M_{\sigma}\,j_{\sigma} (l_{\rm r}^{\eta},S_q) \, ,
\label{rel-currents}
\end{equation}
where $M_{\sigma} = M_{\pm 1/2} = S_q \pm S^z$ and $j_{\sigma} = j_{\pm 1/2}$ is given by,
\begin{eqnarray}
j_{\pm 1/2} & = & j_{\pm 1/2} (l_{\rm r}^{\eta},S_q) = \pm {\langle \hat{J}^z_{HWS} (l_{\rm r}^{\eta},S_q) \rangle \over 2S_q} 
\nonumber \\
& = & \pm {\langle \hat{J}^z_{HWS} (l_{\rm r}^{\eta},S_q) \rangle \over (M_{+1/2} + M_{-1/2})} \, .
\label{elem-currents}
\end{eqnarray}

It is straightforward to confirm that $j_{\pm 1/2} = j_{\pm 1/2} (l_{\rm r}^{\eta},S_q)$, Eq. (\ref{elem-currents}), 
is the spin elementary current carried by each of the $M_{\pm 1/2} = S_q \pm S^z$ unpaired physical spins of projection $\pm 1/2$
in the multiplet configuration of a $S_q>0$ energy eigenstate of spin-current expectation value $\langle \hat{J}^z (l_{\rm r},S_q,S^z) \rangle =
{S^z\over S_q}\langle \hat{J}^z_{HWS} (l_{\rm r},S_q) \rangle$. Here $S^z$ can have $2S_q +1$ values, 
$S^z = S_q - n_z$ where $n_z = 0,1,...,2S_q$. 

Indeed, under each unpaired physical spin flip generated by the operator 
${\hat{S}}^{+}_{\eta}$ in Eq. (\ref{state}) the spin-current expectation value $\langle \hat{J}^z (l_{\rm r},S_q,S^z) \rangle$, 
Eqs. (\ref{rel-currents-XXZ}) and (\ref{rel-currents}), exactly changes by $-2j_{+1/2} = 2j_{-1/2}$.

This reveals the deep physical meaning of the relation, Eqs. (\ref{rel-currents-XXZ}) and
(\ref{rel-currents}): It expresses the spin-current expectation value 
$\langle \hat{J}^z\rangle = \langle \hat{J}^z (l_{\rm r}^{\eta},S_q,S^z)\rangle$
of any $S_q >0$ energy eigenstate $\vert l_{\rm r}^{\eta},S_q,S^z\rangle$ in terms of the spin elementary currents
$j_{\pm 1/2}$ carried by each of the $M_{\pm 1/2} = S_q \pm S^z$ spin carriers of projection $\pm 1/2$ in
the multiplet configuration of that state. 

It also reveals that when $M = M_{+1/2} + M_{-1/2} = 0$ or $M_{+1/2} = M_{-1/2}$ the spin-current expectation value 
$\langle \hat{J}^z \rangle = \sum_{\sigma =\pm 1/2}M_{\sigma}\,j_{\sigma}$, Eq. (\ref{rel-currents}), 
vanishes, i.e. $S_q=0$ states and $S_q >0$ states for which $S^z =0$ have vanishing spin-current expectation value.

An important symmetry is that the spin elementary currents $j_{\pm 1/2}$, Eq. (\ref{elem-currents}), 
have the same value for the number $2S_{q} + 1$ of states of the same $q$-spin tower, Eq. (\ref{state}). 
Hence they do not depend on $S^z$ and $m = 2S^z/N$, so that,
\begin{equation}
{\partial  j_{\pm 1/2} \over \partial\,m} = 0 \, .
\label{deriv-elem-currents}
\end{equation}

The spin elementary currents, Eq. (\ref{elem-currents}), play a major role in our study.
Only the corresponding $M = M_{+1/2} + M_{-1/2} = 2S_q$ 
unpaired physical spins in the multiplet configuration of $S_q >0$ energy
eigenstates contribute to spin transport. 

As reported in Appendix \ref{C}, this occurs through 
effective fluxes piercing the rings associated with the $n$-squeezed effective lattices
for which $N_n >0$. Each such a real-space $n$-squeezed effective lattice with a 
number $L_n$ of sites and length $L = N$ emerges from the above-mentioned
squeezed-space construction \cite{Ogata_90,Penc_97,Kruis_04}. A
$n$-squeezed effective lattice is associated 
with a corresponding momentum-space $n$-band with a number $L_n$ of discrete
momentum values $q_j$. 

\section{Vanishing of the ballistic contributions to spin transport for $\Delta > 1$ and $T> 0$}
\label{SECIII}

The vanishing at zero field of the spin stiffness for $\Delta > 1$ and $T> 0$ is in this paper confirmed
by two complementary methods. In this section we use an inequality for that stiffness. 
It provides strong evidence that the spin stiffness vanishes.

Below in Sec. \ref{SECV} we use the relation between two suitable chosen 
thermal averages of the square of the spin elementary currents carried
by the spin carriers to confirm that the spin stiffness indeed vanishes 
at zero field for $\Delta > 1$ and $T> 0$.

\subsection{Spin stiffness in terms of spin elementary currents}
\label{SECIIIA}

The real part of the spin conductivity is for finite temperatures given by,
\begin{eqnarray}
\sigma (\omega,T) & = & 2\pi D^z (T)\,\delta (\omega) + \sigma_{{\rm reg}} (\omega,T) 
\hspace{0.20cm}{\rm where}
\nonumber \\
D^z (T) & = & {1\over 2 N T}\sum_{S_q=\vert S^z\vert}^{N/2}\sum_{l_{\rm r}^{\eta}} p_{l_{\rm r}^{\eta},S_q,S^z}
\nonumber \\
& \times & \vert \langle \hat{J}^z (l_{\rm r}^{\eta},S_q,S^z) \rangle  \vert^2 \, ,
\label{D-all-T-simpA}
\end{eqnarray}
is the $T>0$ spin stiffness. In its expression given here \cite{Mukerjee_08}
the summations run over {\it all} quantum-problem energy eigenstates $\vert l_{\rm r}^{\eta},S_q,S^z\rangle$
with fixed $S^z$ and $T$ values and $p_{l_{\rm r}^{\eta},S_q,S^z}$ are the corresponding Boltzmann weights.
When the spin stiffness in the conductivity singular part is finite, the dominant spin transport is ballistic. 

The use of Eq. (\ref{rel-currents}) in Eq. (\ref{D-all-T-simpA}) accounting for the equality 
$\vert j_{-1/2}\vert = \vert j_{+1/2}\vert$, leads to the following expression of the spin stiffness $D^z (T)$ directly 
in terms of the spin elementary currents carried by the spin carriers,
\begin{equation}
D^z (T) = {m^2\over 2T} \Omega_m (T) \, ,
\label{D-all-T-simp-m}
\end{equation}
where $m = 2S^z/N$ and,
\begin{eqnarray}
\Omega_m (T) & = & N\,\langle\vert j_{\pm 1/2} \vert^2\rangle_{S^z,T} 
\nonumber \\
& = & N\,\sum_{S_q=\vert S^z\vert}^{N/2}\sum_{l_{\rm r}^{\eta}} 
p_{l_{\rm r}^{\eta},S_q,S^z}\vert j_{\pm 1/2} (l_{\rm r}^{\eta},S_q) \vert^2 \, .
\label{jz2T}
\end{eqnarray}
Here $\langle ...\rangle_{S^z,T}$ denotes canonical equilibrium averages at fixed $S^z$ and $T$.
Hence $\langle\vert j_{\pm 1/2} (l_{\rm r}^{\eta},S_q) \vert^2\rangle_{S^z,T} = \Omega_m (T)/N$ is the
thermal expectation value in the subspace spanned by states with fixed $S^z$
of the square of the absolute value $\vert j_{\pm 1/2}\vert = \vert j_{\pm 1/2} (l_{\rm r}^{\eta},S_q)\vert$ 
of the spin elementary current carried by one spin carrier of projection $\pm 1/2$. 

Due to the factor $m^2/(2T) = m\,S^z/(2T N)$ on the right-hand side of
Eq. (\ref{D-all-T-simp-m}), the spin stiffness $D^z (T)$ vanishes both at $S^z =0$
and in the $m\rightarrow 0$ limit provided that $\Omega_m (T)/N$ is finite in that limit.
That the spin stiffness $D^z (T)$ vanishes in the thermodynamic limit within the 
canonical ensemble requires indeed that $\lim_{m\rightarrow 0}D^z (T) =0$.
Note that the corresponding $S_q > 0$ energy eigenstates for which $S^z/N\rightarrow 0$ as 
$N\rightarrow\infty$ have $S_q$ values in the whole interval 
$S_q\in [\vert S^z\vert,N/2]$. 

In contrast to the isotropic case \cite{Carmelo_15,Carmelo_17}, the absence of a critical 
point at zero field for $\Delta >1$ ensures that in the thermodynamic limit the canonical ensemble and 
grand canonical ensemble lead to the same value for the spin stiffness $D^z (T)$. Hence
if $D^z (T)$ vanishes at $m=0$ and in the $m\rightarrow 0$ limit for $\Delta >1$ and all finite temperatures $T>0$,
it also vanishes at zero field and in the $h\rightarrow 0$ limit within the grand canonical ensemble. 

\subsection{Useful statement for the finiteness of $\vert j_{\pm 1/2}\vert$}
\label{SECIIIB}

Following Eqs. (\ref{D-all-T-simp-m}) and (\ref{jz2T}) and accounting for the subspace normalization 
$\sum_{S_q=\vert S^z\vert}^{N/2}\sum_{l_{\rm r}} p_{l_{\rm r}^{\eta},S_q,S^z} = 1$, one has
that $\lim_{m\rightarrow 0}D^z (T) =0$ for $T>0$ provided that $\langle\vert j_{\pm 1/2}\vert^2\rangle_T$ 
is finite for all energy eigenstates for which $S^z/N\rightarrow 0$ for $N\rightarrow\infty$. 
We find, below in Sec. \ref{SECVA}, that for $T>0$ such a thermal average is actually smaller than 
a related quantity of the order of $1/N$.
\begin{figure*}
\includegraphics[width=0.495\textwidth]{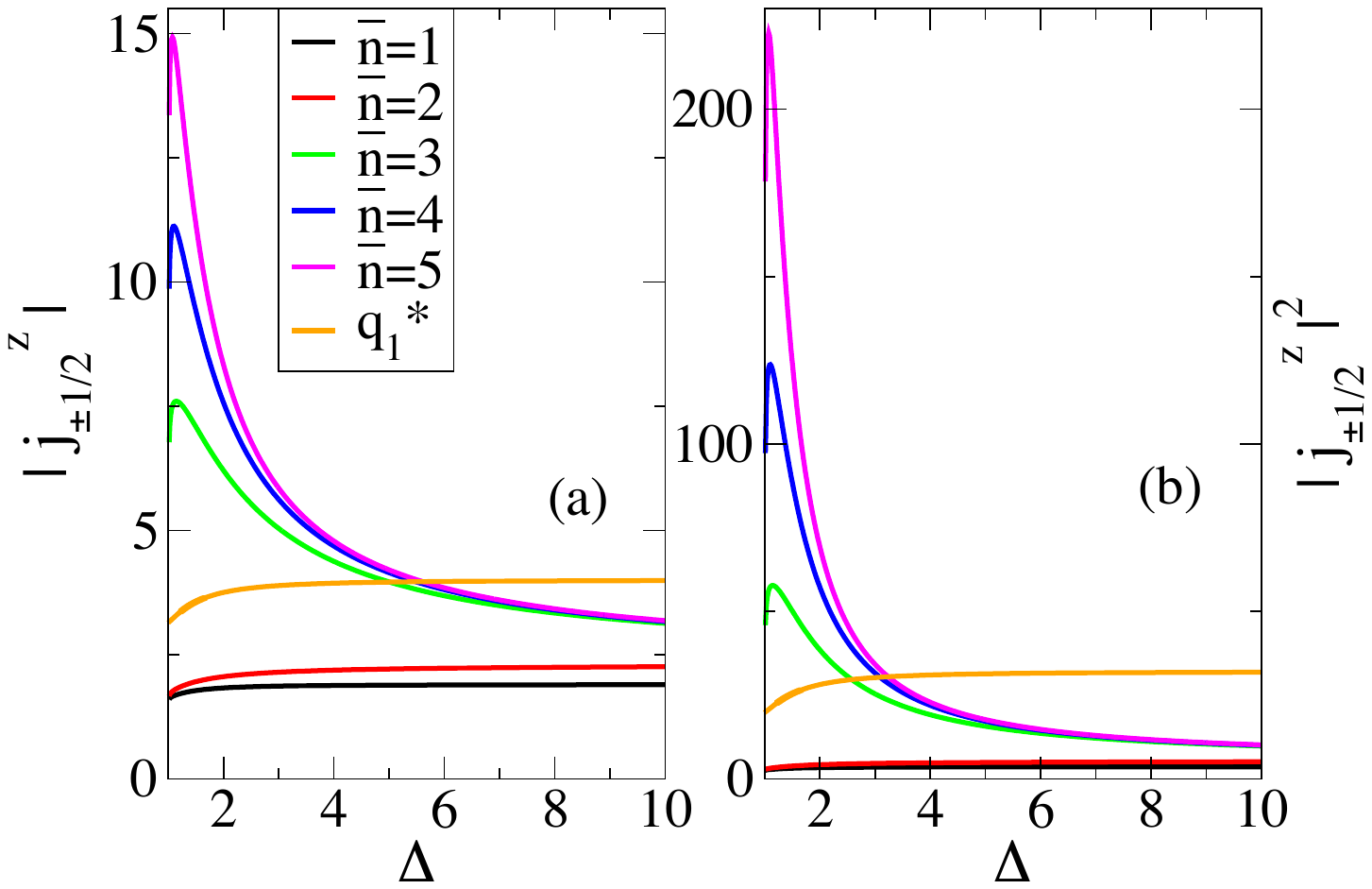}
\includegraphics[width=0.495\textwidth]{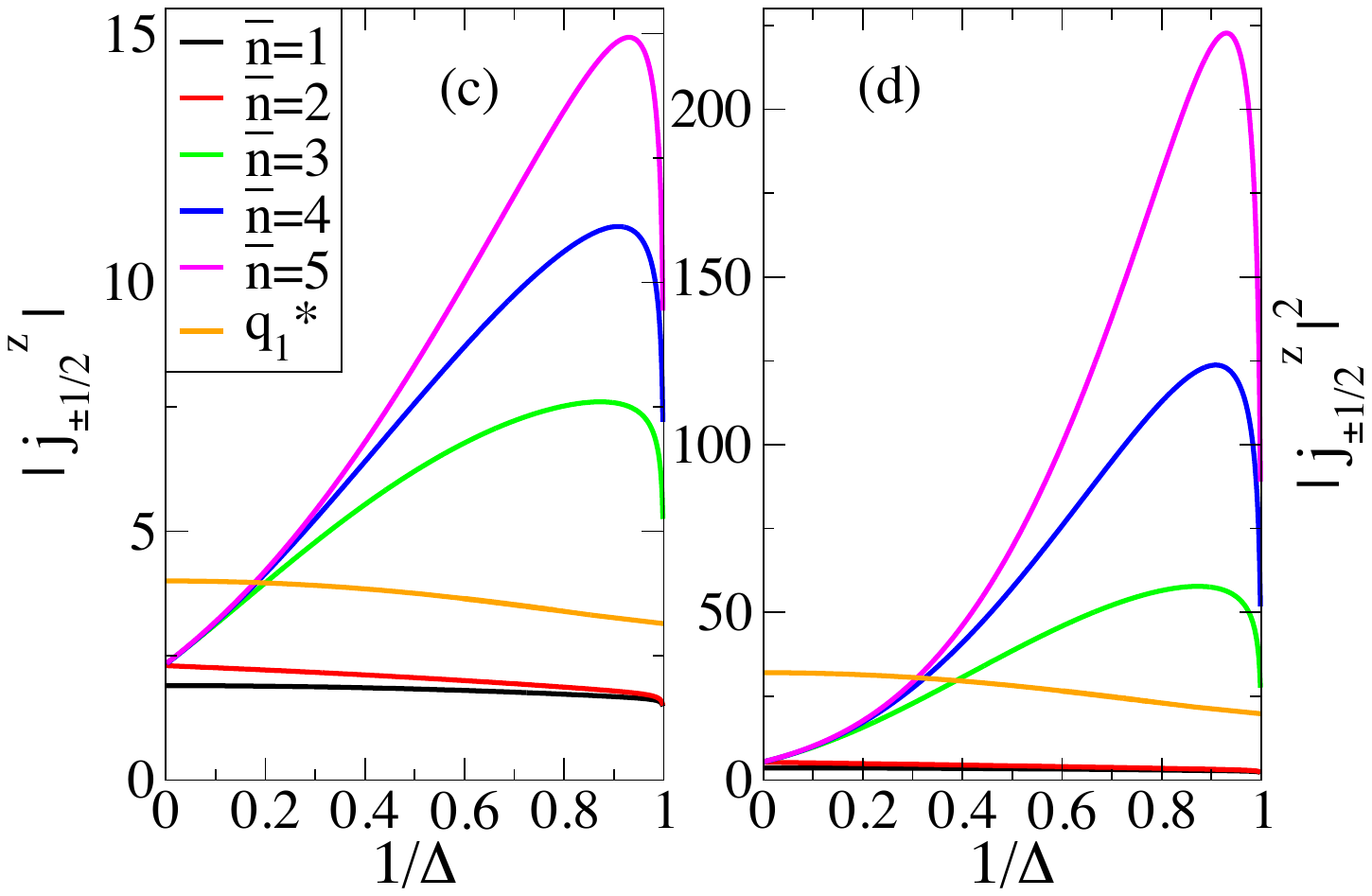}
\caption{The absolute values $\vert j_{\pm 1/2}(\bar{n})\vert$ of the spin elementary current carried by one 
spin carrier for $\bar{n}$-states, Eq. (\ref{jzlarger}) of Appendix \ref{B}, with $\bar{n}=1,2,3,4,5$
and the absolute value $\vert j_{\pm 1/2}(q_1^*)\vert$ for that of the $q_1^*$-states, Eq. (\ref{jzq1*}) of that Appendix,
(a) and its square (b) as a function of anisotropy $\Delta$ and (c) and (d) as a function of $1/\Delta$, respectively.}
\label{figure1PR}
\end{figure*}

All finite-$S_q$ energy eigenstates, including those considered in Appendix \ref{B}, are 
in the thermodynamic limit, $N\rightarrow\infty$, of two classes (i) and (ii) whose 
concentration $M/N=2S_q/N$ of spin carriers vanishes and is finite, respectively.
For these two classes of states, also the 
ratio $\vert\langle \hat{J}^z\rangle\vert/N$ that involves the absolute value of the 
spin-current expectation value, Eqs. (\ref{rel-currents-XXZ}) and (\ref{rel-currents}), 
vanishes and is finite, respectively, for $\Delta >1$ in the $N\rightarrow\infty$ limit. 

A useful statement involving the spin elementary currents carried by the spin carriers identified within 
the physical-spin representation is then the following:\\

{\it That for finite-$S_q$ states of class (i) and (ii) both the ratio $\vert\langle \hat{J}^z_{HWS}\rangle\vert/N$ 
and the concentration of spin carriers $M/N=2S_q/N$ (i) vanish and (ii) are finite, respectively, 
for $\Delta >1$ in the thermodynamic limit, $N\rightarrow\infty$, implies that in both cases 
the largest values of the ratio $\vert\langle\hat{J}^z_{HWS}\rangle\vert/(2S_q)$, which is the spin elementary current, 
Eq. (\ref{elem-currents}), absolute value $\vert j_{\pm 1/2}\vert$, are finite in that limit.}\\

Such a statement though does not apply in the $\Delta\rightarrow 1$ limit for a few states, 
as confirmed by the results of this paper. However, our main interest is for anisotropy $\Delta >1$.

\subsection{States with large spin elementary currents}
\label{SECIIIC}

Here, we discuss issues concerning the large absolute values $\vert j_{\pm 1/2}\vert$ of selected 
finite-$S_q$ energy eigenstates considered in Appendix \ref{B}. The spin elementary currents carried by
spin carriers in the multiplet configuration of the states named $\bar{n}_*$-states in Appendix \ref{B} have for
an arbitrarily small, yet finite, anisotropy range $\Delta \in ]1,\Delta_*]$ the largest 
absolute value $\vert j_{\pm 1/2}\vert$ of all finite-$S_q$ energy eigenstates identified in that Appendix.
This is used to introduce the inequality, Eq. (\ref{ineAll}) of Appendix \ref{B}. 

In that Appendix we use clear criteria to identify finite-$S_q$ states of both classes (i) and (ii) 
considered in the following, some of which with huge spin-current expectation values.
This includes the $\bar{n}_*$-states. They are a particular type of more general states that in Appendix \ref{B} are 
called ${\bar{n}}$-states. The latter states have finite occupancies, $N_n>0$, for a set of 
successive $n=1,...,\bar{n}$ $n$-bands and zero occupancy, $N_n = 0$, for $n>\bar{n}$. 
They are generated from the ground states by an infinite number of elementary processes and except 
when both ${\bar{n}}\rightarrow\infty$ and $\Delta\rightarrow\infty$, they are class (ii) states. 
Their $n$-band occupancies are simpler to express in terms of $n$-band rapidity-variable distributions $\tilde{N}_{n} (\varphi)$
defined through the relation $\tilde{N}_{n} (\varphi_n (q)) = N_n (q)$, Eq. (\ref{tildeNN}) of Appendix \ref{A}.

For ground states we have that $N_n = 0$ for $n>1$ and the $1$-band rapidity-variable distribution function 
$\tilde{N}_{1} (\varphi)$ is symmetric around $\varphi=0$, reading $\tilde{N}_{1} (\varphi)=1$
for $\varphi \in [-B,B]$ and $\tilde{N}_{1} (\varphi)=0$ for $\vert\varphi\vert \in [B,\pi]$.
Here $\pm B = \varphi_1 (\pm k_{F\downarrow})$
and $k_{F\downarrow} ={\pi\over 2}(1-m)$, in the thermodynamic limit. 

On the other hand, ${\bar{n}}$-states 
have for all successive $n=1,...,\bar{n}$ $n$-bands with finite occupancy fully asymmetric 
half-filled occupancies $\tilde{N}_{n} (\varphi) = 0$ (or $=1$) for $\varphi\in [-\pi,0]$ and 
$\tilde{N}_{n} (\varphi) = 1$ (or $=0$) for $\varphi\in [0,\pi]$ in terms of rapidity variables. 
For such states, $\tilde{N}_{n} (0) = N_n (q_n^0+q_n^{\Delta})$, where the finite $n$-band {\it separation} momentum 
$q_n^0 = q_n^0 (\eta)$ refers to zero $n$-band rapidity variable, $\varphi =0$. It separates the $q_n^{\Delta}$-shifted $n$-band 
momentum unoccupied $(q - q_n^{\Delta}) \in [-{\pi\over N}(L_n -1), q_n^0]$ from occupied $(q - q_n^{\Delta}) \in [q_n^0,{\pi\over N}(L_n -1)]$ 
intervals, respectively.

As was reported in Sec. \ref{SECIIA}, an energy eigenstate is defined by the occupancies of the Bethe-ansatz quantum numbers $I_j^n$ associated 
with the corresponding discrete $n$-band momentum values $q_j = {2\pi\over N}I_j^n$ in units of ${2\pi\over N}$. 
Hence states with the same $n$-band momentum occupancies for different anisotropy values refer to the same 
energy eigenstates. If $q_n^0=0$ and thus $\tilde{N}_{n} (0) = N_n (q_n^{\Delta})$, one has that $N_n = N_n^h$. 
In contrast, for $\bar{n}$ states the value of the $n$-band separation momentum $q_n^0 = q_n^0 (\eta)$ that corresponds 
to $\varphi=0$ is finite and varies upon changing the anisotropy. Therefore, the numbers $N_n$ and $N_n^h$, Eq. (\ref{2Sq}) 
of Appendix \ref{B}, vary upon changing the anisotropy. This shows that ${\bar{n}}$-states are {\it different 
energy eigenstates} for different values of anisotropy $\Delta = \cosh\eta$. 

As justified in Appendix \ref{B}, upon decreasing the anisotropy from (i) $\Delta\rightarrow\infty$ to (ii) 
$\Delta\rightarrow 1$, the separation momentum values $q_1^0$ and $q_n^0$ for $n>1$ and the 
concentration of spin carriers $M=2S_q/N$ increase from (i) $q_1^0=q_n^0=0$ and 
$M=2S_q/N = c_{\bar{n}}^{-1}(2 + \sqrt{3})^{-\bar{n}}$ to (ii) $q_1^0 \rightarrow \pi/4$, $q_n^0 \rightarrow \pi/2$, and 
$M=2S_q/N \rightarrow 1/2$, respectively. Here $c_{\bar{n}}$ is the coefficient $c_n$ defined in Eq. (\ref{Anxnnj})
of Appendix \ref{B} for $n = \bar{n}$. (Its values for $n=0,1,...,11$ are given in Table \ref{table1} of that
Appendix.)
\begin{figure*}
\includegraphics[width=0.495\textwidth]{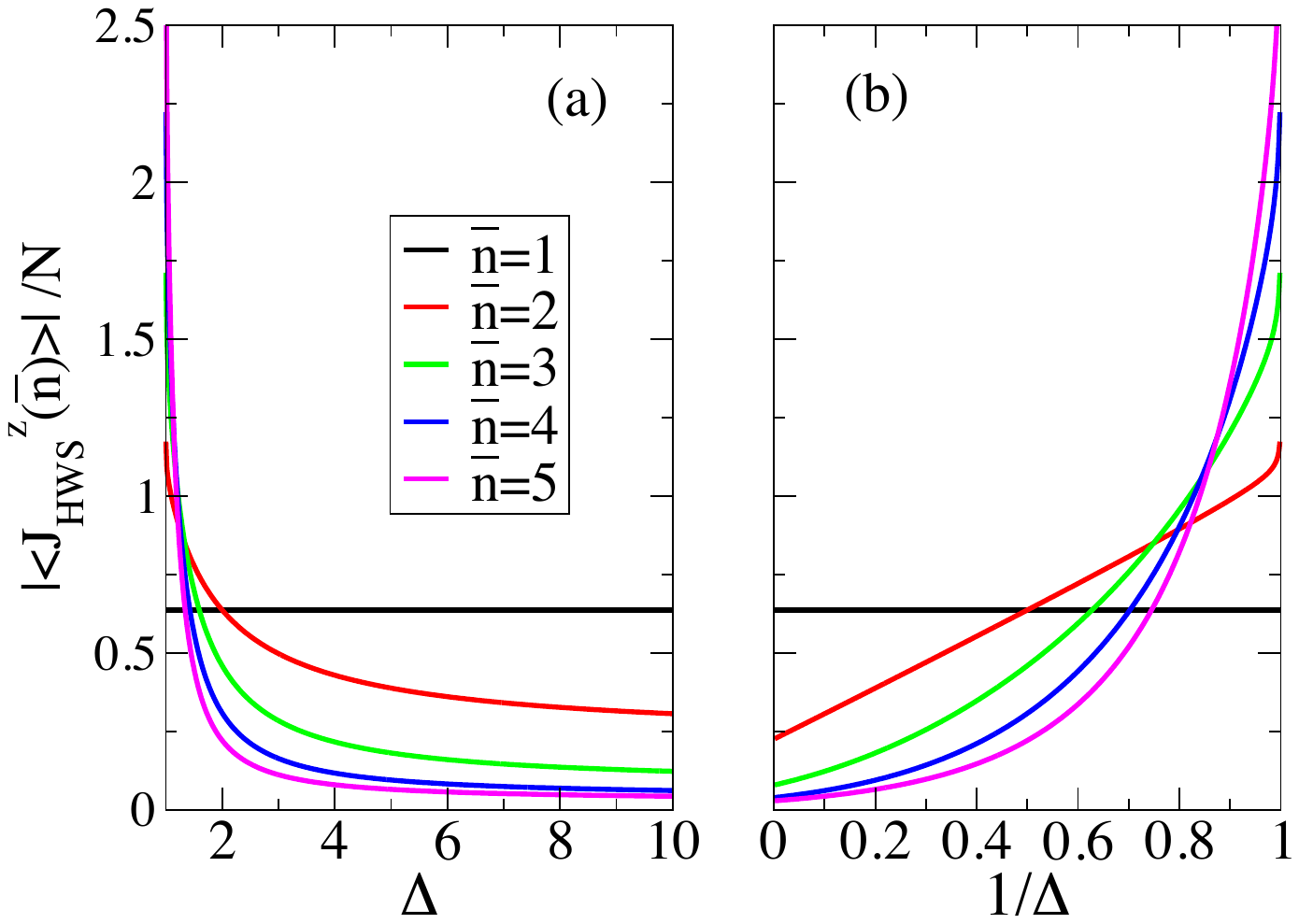}
\includegraphics[width=0.495\textwidth]{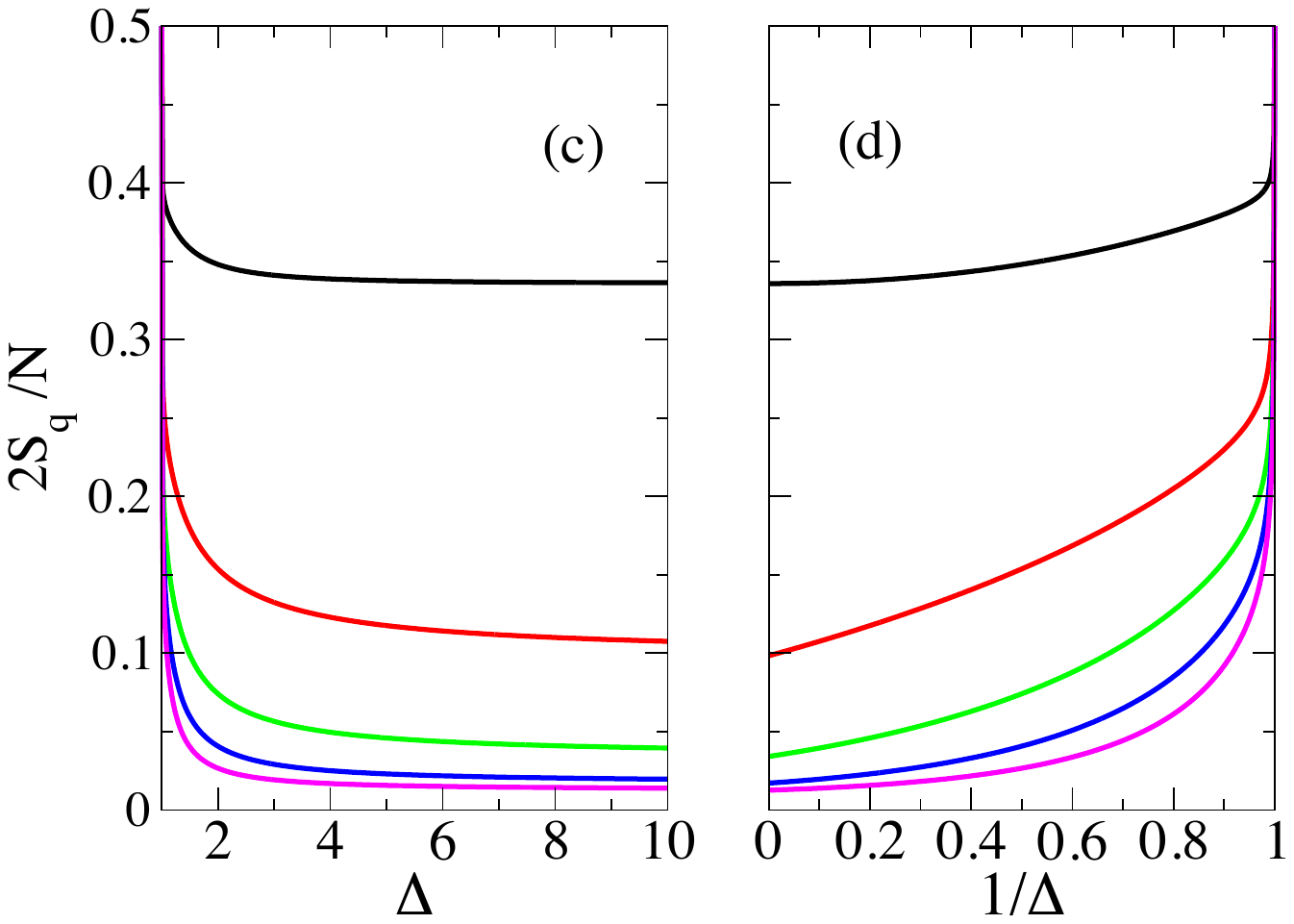}
\caption{The ratio $\vert\langle \hat{J}^z_{HWS}(\bar{n})\rangle\vert/N$ of the spin-current expectation 
value of $\bar{n}$-states over $N$ for $\bar{n}=1,2,3,4,5$ as a 
function of $\Delta$ (a) and $1/\Delta$ (b) and the corresponding concentration of spin carriers 
$M/N=2S_q/N$, Eq. (\ref{2Sq}) of Appendix \ref{B}, for these states as a function of $\Delta$ (c) and $1/\Delta$ (d).
While for $\Delta >1$ the latter decreases for increasing $\bar{n}$, for $\Delta \rightarrow 1$ it has the universal 
value $M/N=2S_q/N=1/2$ for all $\bar{n}$. The minimum values reached in the $\Delta \rightarrow\infty$ limit
read $M/N=2S_q/N={1\over c_{\bar{n}}}(2+\sqrt{3})^{-\bar{n}}$ where here $\bar{n}=1,2,3,4,5$.}
\label{figure2PR}
\end{figure*}

The $\bar{n}_*$-states are ${\bar{n}}$-states whose integer number ${\bar{n}}=\bar{n}_*$ has the maximum physically allowed
value. For ${\bar{n}}$-states it reads $\bar{n} = {\ln\left(N/(c_{\bar{n}}\,2S_q^{\infty})\right)\over \ln (2+\sqrt{3})}$,
Eq. (\ref{2Sinfgaban}) of Appendix \ref{B}. Here $2S_q^{\infty}/N = \lim_{\Delta \rightarrow\infty}2S_q/N = 2/N$
for $\bar{n}_*$-states, so that $\bar{n}_*$ is the integer number closest to
\begin{equation}
\bar{n}_* = {\ln\left({N\over 2c_{\bar{n}_*}}\right)\over \ln (2+\sqrt{3})} \rightarrow \infty
\hspace{0.20cm}{\rm for}\hspace{0.20cm} N\rightarrow \infty 
\hspace{0.20cm}{\rm and}\hspace{0.20cm}\Delta > 1 \, ,
\label{n*}
\end{equation}
where $c_{\bar{n}_*} = 0.78867513459481...$. The expression of the spin carrier concentration in
Eq. (\ref{setEq}) of Appendix \ref{B} shows that the $\bar{n_*}$-band separation momentum is
fully controlled by that concentration as it reads $q_{\bar{n_*}}^0(\eta)/\pi = 2S_q (\eta)/N$. 
This is because $2S_q^{\infty}/N = 0$ in the $N\rightarrow\infty$ limit.

The $\bar{n}_*$-states are class (ii) states for finite values $\Delta >1$.
In the $\Delta\rightarrow\infty$ limit they become class (i) states populated by only $M=2S_q^{\infty}=2$ spin carriers. 
We thus have also selected in Appendix \ref{B} a much simpler type of 
class (i) states populated by only $M=2S_q=2$ spin carriers for $\Delta >1$. 
We call them $q_1^*$-states. Their choice is justified because for large $\Delta$ values the absolute value 
$\vert j_{\pm 1/2}(q_1^*)\vert$ of their spin elementary current, Eq. (\ref{jzq1*}) 
of Appendix \ref{B}, is larger than that of the $\bar{n}_*$-states
given below in Eq. (\ref{limitsjz12n*}).

Such $q_1^*$-states have only two holes in the $1$-band. Their two
momentum values are suitably chosen in Appendix \ref{B}, so that 
$\vert j_{\pm 1/2}(q_1^*)\vert$ is the largest absolute value of 
a class of states generated from the full $1$-band of ground states by 
creation of two $1$-holes.

In Fig. \ref{figure1PR} the absolute values $\vert j_{\pm 1/2}(q_1^*)\vert$
of the spin elementary current, Eq. (\ref{jzq1*}) of Appendix \ref{B},
and $\vert j_{\pm 1/2}(\bar{n})\vert$ for $\bar{n}=1,2,3,4,5$, Eq. (\ref{jzlarger}) of that Appendix,
for $q_1^*$-states and $\bar{n}$-states, respectively,
of the spin elementary current carried by one spin carrier (a)
and its square (b) are plotted as a function of anisotropy $\Delta$ and
(c) and (d) as a function of $1/\Delta$, respectively. 

For $q_1^*$-states, the absolute value $\vert j_{\pm 1/2}(q_1^*)\vert$ of the
spin elementary current, Eq. (\ref{jzq1*}) of Appendix \ref{B},
smoothly increases upon increasing $\Delta$ from 
$\vert j_{\pm 1/2}(q_1^*)\vert = \pi J$ for $\Delta\rightarrow 1$ to $\vert j_{\pm 1/2}(q_1^*)\vert = 4J$ 
for $\Delta\rightarrow\infty$. For $\bar{n}$-states, $\vert j_{\pm 1/2}(\bar{n})\vert$, Eq. (\ref{jzlarger}) of Appendix \ref{B},
reads ${2c_{\bar{n}}(2+\sqrt{3})J\over \pi\,c_{\bar{n}-1}}$ for $\Delta\rightarrow\infty$.
For $\bar{n} > 2$, it increases upon decreasing $\Delta$ until reaching a maximum
at small $(\Delta-1)$, as shown in Fig. \ref{figure1PR}. It then decreases
to $\vert j_{\pm 1/2}(\bar{n})\vert = {4(\bar{n}-1) J\over\pi}$ as $\Delta\rightarrow 1$.
For $\bar{n} =1,2$ the absolute value $\vert j_{\pm 1/2}(\bar{n})\vert$ rather 
smoothly and continuously increases from ${4J\over\pi}$ for $\Delta\rightarrow 1$ to 
${2c_{\bar{n}}(2+\sqrt{3})J\over \pi\,c_{\bar{n}-1}}$ for $\Delta\rightarrow\infty$.

The absolute value $\vert j_{\pm 1/2}(\bar{n})\vert$ increases 
upon increasing $\bar{n}$. Its maximum value reached by $\bar{n}_*$-states is
given by Eq. (\ref{jzlarger}) of Appendix \ref{B} for $\bar{n} = \bar{n}_*$. This gives,
\begin{eqnarray}
\vert j_{\pm 1/2}(\bar{n}_*)\vert & = & \sum_{n=1}^{\infty}
{{2nJ\over\pi}\sinh\eta\over
({2S_q (\eta)\over N} + {c_{\bar{n}_* -n}\over c_{\bar{n}_*}}{(1 - {4S_q (\eta)\over N})\over (2+\sqrt{3})^{n}}) \sinh (n\eta)}
\nonumber \\
& = & {2J(2+\sqrt{3})\over\pi}
\hspace{0.20cm}{\rm for}\hspace{0.20cm}\Delta\rightarrow\infty
\nonumber \\
& = & {4(\bar{n}_* - 1) J\over\pi} = \infty 
\hspace{0.20cm}{\rm for}\hspace{0.20cm}\Delta\rightarrow 1 \, .
\label{limitsjz12n*}
\end{eqnarray}

This absolute value is not plotted in Fig. \ref{figure1PR}. Indeed, in the case of finite occupancy in an infinite 
number $\bar{n}_*\rightarrow\infty$ of $n$-bands the evaluation of the spin carrier concentration
$2S_q (\eta)/N = q_{\bar{n}_*}^0 (\eta)/\pi \in [0,1/2]$ that appears in the expression, Eq. (\ref{limitsjz12n*}),
is for anisotropy $\Delta = \cosh\eta \in ]1,\infty]$ and thus for $\eta \in ]0,\infty]$ 
a technically complex problem. 

However, from manipulations of the infinite number of coupled integral 
equations that define it, Eq. (\ref{BAqn}) of Appendix \ref{B}, 
we find that $2S_q (\eta)/N$ is for $\eta >0$ a continuous
function that smoothly decreases from $1/2$ to $0$ upon increasing the anisotropy in the related
intervals $\Delta \in ]1,\infty]$ and $\eta\in ]0,\infty]$. 
Its anisotropy dependence is qualitatively similar to that plotted in Fig. \ref{figure2PR}(c) 
and \ref{figure2PR}(d) for $2S_q (\eta)/N$ of $\bar{n}$-states with $\bar{n} = 1,2,3,4,5$. Except for $\Delta\rightarrow 1$ when 
$2S_q (\eta)/N = 1/2$ for all $\bar{n}$, it runs below {\it all} curves for $2S_q (\eta)/N$ 
of finite-$\bar{n}$ $\bar{n}$-states, including thus those plotted in that figure.

The expression, Eq. (\ref{limitsjz12n*}), actually provides all information needed
for our studies: Consistently with anomalous superdiffusive spin transport at the isotropic point,
the $\vert j_{\pm 1/2}(\bar{n})\vert$'s peak that for $\bar{n} = 3,4,5$ is in Fig. \ref{figure1PR}
located at a small $(\Delta-1)$ value is for $\bar{n} = \bar{n}_* \rightarrow\infty$
shifted to $\Delta\rightarrow 1$. And the absolute value $\vert j_{\pm 1/2}(\bar{n}_*)\vert$ diverges in that limit.
For $\Delta >1$ it is finite and continuously decreases from 
infinity for $\Delta\rightarrow 1$ upon increasing $\Delta$ in the interval $\Delta \in ]1,\infty]$.
It reaches its minimum value, ${2(2+\sqrt{3})J\over\pi}$, as $\Delta\rightarrow\infty$.
The latter is indeed smaller than $\vert j_{\pm 1/2}(q_1^*)\vert = 4J$ for the $q_1^*$-states
in that limit.  

In Figs. \ref{figure2PR}(a) and \ref{figure2PR}(b), the ratio $\vert\langle \hat{J}^z_{HWS}(\bar{n})\rangle\vert/N$
involving the spin-current expectation value and in Figs. \ref{figure2PR}(c) and \ref{figure2PR}(d), the spin 
carrier concentration $M/N=2S_q/N$, Eq. (\ref{2Sq}) of Appendix \ref{B},  
of the $\bar{n}=1,2,3,4,5$ $\bar{n}$-states, are plotted as a function of $\Delta$ and $1/\Delta$, respectively.
Confirming that they are class (ii) states, {\it both} such quantities are finite in the thermodynamic limit.

The ratio $\vert\langle \hat{J}^z_{HWS}(\bar{n})\rangle\vert/N$ 
plotted in Fig. \ref{figure2PR} is for $\bar{n} = 1$ 
the largest of all such states for a large anisotropy interval.
In contrast, the curves plotted in Fig. \ref{figure1PR} show that $\vert j_{\pm 1/2}(\bar{n})\vert$ 
is for $\bar{n} = 1$ the smallest spin elementary current of both all $\bar{n}$-states and $q_1^*$-states.
This reveals that there is no obvious relation between the strengths of the absolute values of
spin-current expectation values and those of corresponding spin elementary currents carried by the
spin carriers.

For $\Delta >1$, the concentration of spin carriers of $\bar{n}$-states
plotted in Figs. \ref{figure2PR}(c) and \ref{figure2PR}(d) for $\bar{n}=1,2,3,4,5$ decreases for increasing $\bar{n}$.
However, its maximum value reached for $\Delta \rightarrow 1$ reads $1/2$ for all $\bar{n}$-states. At fixed $\Delta >1$
its minimum value reads $M = 2S_q/N = q^0_{\bar{n}_*} (\eta)/\pi \in [0,1/2]$, being reached by
the $\bar{n}_*$-states. That minimum value vanishes in the $\Delta\rightarrow\infty$ limit in which $2S_q = 2$ and the
$\bar{n}_*$-states become class (i) states.
At fixed $\bar{n}$, the concentration of spin carriers increases from $M = 2S_q/N ={1\over c_{\bar{n}}}(2+\sqrt{3})^{-\bar{n}}$ to 
$M = 2S_q/N =1/2$ upon decreasing $\Delta$ in the interval $\Delta \in ]1,\infty]$.

For the class (i) $q_1^*$-states, both the ratio $\langle \hat{J}^z_{HWS} (q_1^*)\rangle/N$
and the spin carrier concentration $M/N=2S_q/N=2/N$ vanish 
in the thermodynamic limit for the whole anisotropy range $\Delta \in ]1,\infty]$. 
The ratio $\vert\langle \hat{J}^z_{HWS}(\bar{n}_*)\rangle\vert/N$ and the spin carrier concentration $M/N=2S_q/N$
also vanish for $\bar{n}_*$-states in the $\Delta \rightarrow\infty$ limit. 

\subsection{Inequalities associated with the vanishing of the ballistic contributions to spin transport}
\label{SECIIID}

The absolute value $\vert j_{\pm 1/2}(\bar{n}_*)\vert$, Eq. (\ref{limitsjz12n*}), continuously 
decreases upon increasing $\Delta$ in the interval $\Delta \in ]1,\infty]$
and reads $\vert j_{\pm 1/2}(\bar{n}_*)\vert\rightarrow\infty$ in the $\Delta\rightarrow 1$ limit. 
We can thus choose the width of an interval $\Delta \in ]1,\Delta_*]$ to be arbitrarily small yet finite and thus 
$\vert j_{\pm 1/2}(\bar{n}_*^{*})\vert = \vert j_{\pm 1/2}(\bar{n}_*)\vert_{\Delta = \Delta_*}$ to be 
arbitrarily large yet no infinity. Combining that choice with the related inequality, Eq. (\ref{ineAll}) of 
Appendix \ref{B}, we introduce the upper bound,
\begin{eqnarray}
\vert j_{\pm 1/2}(UB)\vert & = & \vert j_{\pm 1/2}(\bar{n}_*)\vert 
\hspace{0.20cm}{\rm for}\hspace{0.20cm} \Delta \in ]1,\Delta_*]
\nonumber \\
& = & \vert j_{\pm 1/2}(\bar{n}_*^{*})\vert
\hspace{0.20cm}{\rm for}\hspace{0.20cm} \Delta > \Delta_* \, .
\label{UB}
\end{eqnarray}
For all finite-$S_q$ energy eigenstates identified in Appendix \ref{B} the following inequality then holds:
\begin{equation}
\vert j_{\pm 1/2}\vert \leq \vert j_{\pm 1/2}(UB)\vert 
\hspace{0.20cm}{\rm for}\hspace{0.20cm} \Delta \in ]1,\infty] \, .
\label{hidineq}
\end{equation}
On the left-hand side of this inequality $\vert j_{\pm 1/2}\vert$ stands for the absolute value
of the spin elementary current carried by the spin carriers in the multiplet configuration
of any such a finite-$S_q$ energy eigenstate. This includes
energy eigenstates for which $S^z/N\rightarrow 0$ for $N\rightarrow\infty$.

Replacement of the absolute values $\vert j_{\pm 1/2}\vert$ in 
the thermal expectation value $\langle\vert j_{\pm 1/2}\vert^2\rangle_{S^z,T}$ in
Eqs. (\ref{D-all-T-simp-m}) and (\ref{jz2T}) by their upper bound $\vert j_{\pm 1/2} (UB)\vert$, 
Eq. (\ref{UB}), then gives the following inequality:
\begin{eqnarray}
\langle\vert j_{\pm 1/2} (l_{\rm r}^{\eta},S_q) \vert^2\rangle_{S^z,T} & < &
\vert j_{\pm 1/2} (UB)\vert^2\sum_{S_q=\vert S^z\vert}^{N/2}\sum_{l_{\rm r}} 
p_{l_{\rm r}^{\eta},S_q,S^z} 
\nonumber \\
& = & \vert j_{\pm 1/2} (UB)\vert^2 \, .
\label{jz2TUB}
\end{eqnarray}
Here, we accounted for the fixed-$S^z$ subspace normalization, $\sum_{S_q=\vert S^z\vert}^{N/2}\sum_{l_{\rm r}} 
p_{l_{\rm r}^{\eta},S_q,S^z} = 1$.

That Eq. (\ref{UB}) actually defines a huge upper bound is confirmed by the result reported below in Sec. \ref{SECVA} that 
for $\Delta >1$  the thermal expectation value $\langle\vert j_{\pm 1/2}\vert^2\rangle_{S^z,T}$ in Eqs. (\ref{D-all-T-simp-m}),
(\ref{jz2T}), and (\ref{jz2TUB}) is for $T>0$ smaller than a related quantity of the order of $1/N$. 

The use of the inequality, Eq. (\ref{jz2TUB}), in the spin-stiffness expression, Eq. (\ref{D-all-T-simp-m}), 
leads to the inequality
\begin{equation}
D^z (T) < m\,{2S^z\over 2T}\vert j_{\pm 1/2} (UB)\vert^2
= {(2S^z)^2\over 2T N}\vert j_{\pm 1/2} (UB)\vert^2 \, .
\label{D-all-T-simp-mUB}
\end{equation}
That $\vert j_{\pm 1/2} (UB)\vert$ is finite for $\Delta > 1$ then implies that
\begin{equation}
\lim_{m\rightarrow 0}D^z (T) = 0 
\hspace{0.20cm}{\rm for}\hspace{0.20cm}\Delta > 1 
\hspace{0.20cm}{\rm and}\hspace{0.20cm}T>0 \, ,
\label{Dz0m0}
\end{equation}
in the thermodynamic limit. This also confirms it vanishes {\it at} $m=0$.

As reported in Sec. \ref{SECIIIA}, accounting for the absence of phase transitions and critical points at zero field in the
case of anisotropy $\Delta >1$, this implies that in the thermodynamic the following result
also holds within the grand-canonical ensemble:
\begin{equation}
\lim_{h\rightarrow 0}D^z (T) = 0 
\hspace{0.20cm}{\rm for}\hspace{0.20cm}\Delta > 1 
\hspace{0.20cm}{\rm and}\hspace{0.20cm}T>0 \, .
\label{Dz0h0}
\end{equation}
In addition, that $D^z (T) = 0$ at $m=0$ implies that $D^z (T)$ also vanishes {\it at} zero field. 

We thus conclude that the use of the huge 
upper bound, Eq. (\ref{UB}), provides strong evidence that for the spin-$1/2$ $XXZ$ chain 
at zero magnetic field the ballistic contributions to spin transport vanish in the thermodynamic limit 
for $\Delta >1$ and {\it all} finite temperatures $T>0$.

Such a vanishing of the spin stiffness implies that the leading spin transport behavior is normal diffusive for 
anisotropy $\Delta >1$ and all temperatures, provided that the spin-diffusion constant
is finite and does not diverge. Its divergence would rather be associated with anomalous superdiffusive 
spin transport. 

Below, in Sec. \ref{SECV}, we find that the spin-diffusion constant is finite for 
$\Delta > 1$ and $T>0$. We also confirm that the spin stiffness vanishes at zero field.
This ensures that the dominant spin transport in the spin-$1/2$ $XXZ$ chain 
is normal diffusive for $\Delta > 1$ and $T>0$.

The spin-diffusion constant is found below in Sec. \ref{SECV} to be related to the $T=0$
spin stiffness in the $T\rightarrow 0$ limit. Hence in the ensuing section we study that spin stiffness.

\section{Zero-temperature spin stiffness for $\Delta \geq 1$, $m\in [0,1]$, and $h\in [0,h_{c2}]$}
\label{SECIV}

The zero-temperature spin stiffness of the 
spin-$1/2$ $XXZ$ chain has been derived for anisotropy 
$-1 \leq \Delta \leq 1$ by use of the Hamiltonian in the presence of a 
vector potential, $\hat{H}=\hat{H} (\Phi/L)$ (twisted boundary conditions) \cite{Shastry_90}. 
However, in the present case of anisotropy $\Delta >1$,
it was only pointed out that it vanishes at $h=0$ \cite{Shastry_90}.
It has not been explicitly calculated for spin densities $m = 2S^z/N \in [0,1]$
and thus magnetic fields $h \in [h_{c1},h_{c2}]$ where the critical fields 
$h_{c1}$ and $h_{c2}$ are defined in Eq. (\ref{criticalfields}) of Appendix \ref{A}.

We have used the same method as Ref. \onlinecite{Shastry_90} involving
twisted boundary conditions to evaluate the $T=0$ spin stiffness for anisotropy
$\Delta>1$. This straightforwardly shows that at $T=0$ it has for $\Delta > 1$ 
and spin densities $m \in [0,1]$ the same general form as for $\Delta =1$,
\begin{equation}
D^z = {\xi^2\over 2\pi}\,v_1 (k_{F\downarrow}) \, .
\label{Dexpression}
\end{equation}
However, and as confirmed below, the quantities in this expression are qualitatively different at $\Delta =1$ and for $\Delta >1$, respectively.
In it, $v_1 (k_{F\downarrow})$ is the $1$-band group velocity $v_1 (q)$, Eq. (\ref{v1q}) 
of Appendix \ref{A}, at $q=k_{F\downarrow}$ and $\xi$ is the parameter defined in Eq. (\ref{x-aaPM}) of that Appendix.

The usual Tomonaga-Luttinger liquid (TLL) parameter $K$ \cite{Horvatic_20} is related to the
parameter $\xi$ as $K = \xi^2$. As given in Eq. (\ref{x-aaPM}) of Appendix \ref{A},
within the physical-spin representation their
expressions involve the phase shift $2\pi\Phi_{1\,1} (q,q')$ defined by Eqs. (\ref{Phi-barPhi}) 
and (\ref{Phis1n}) of that Appendix. $2\pi\Phi_{1\,1} (q,q')$ [and $-2\pi\Phi_{1\,1} (q,q')$] is the phase shift acquired by one $1$-pair scatterer 
of $1$-band momentum $q$ due to creation of one $1$-pair (and $1$-hole) scattering center at $1$-band 
momentum $q'$ under a transition from the ground state to an excited state.

\begin{figure}
\begin{center}
\centerline{\includegraphics[width=9.00cm]{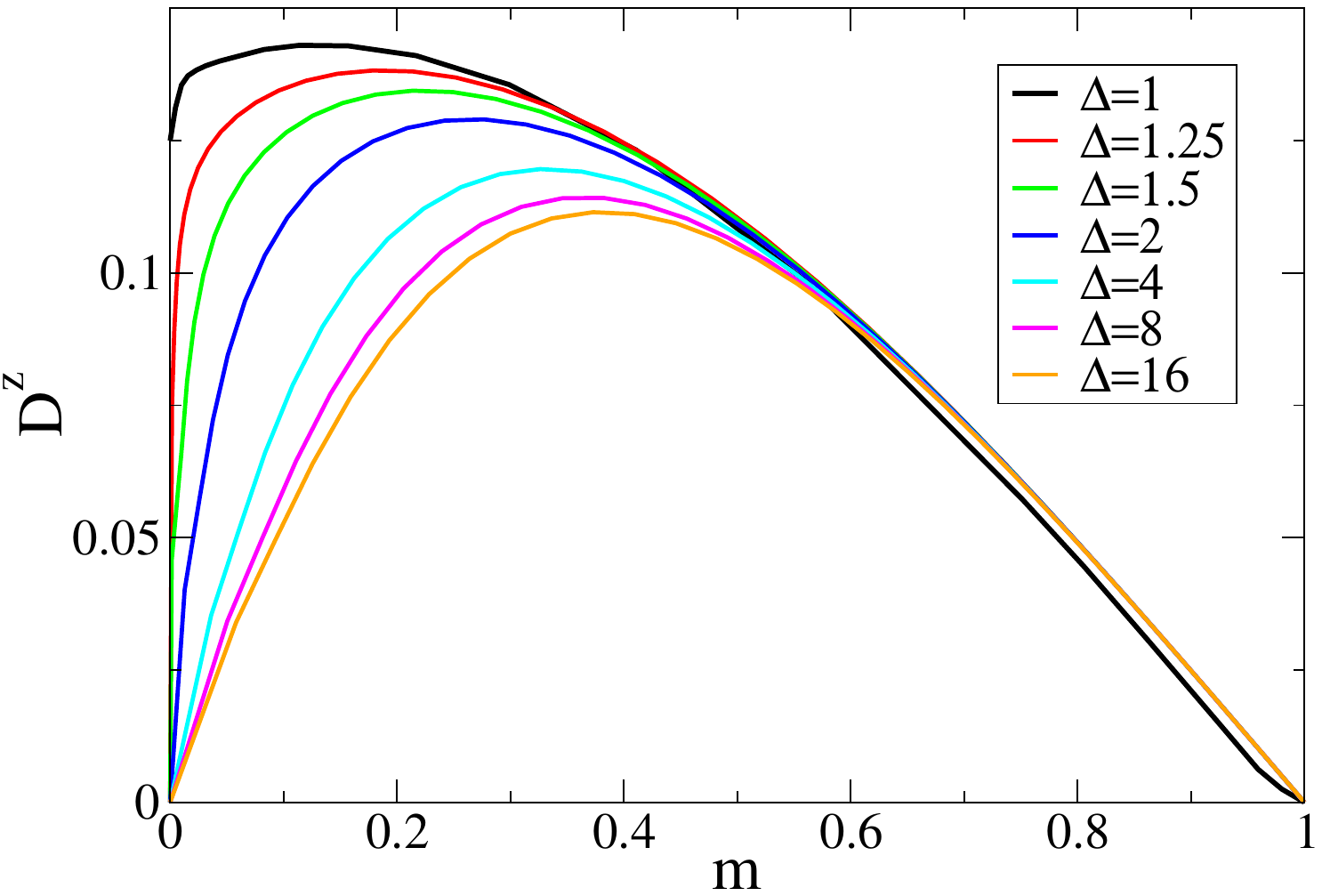}}
\caption{The $T=0$ spin stiffness $D^z (T)$ of the spin-$1/2$ $XXZ$ chain,
Eq. (\ref{Dexpression}), as a function of the spin density in the interval $m = 2S^z/N \in [0,1]$ 
that refers to magnetic fields in the range $h\in [h_{c1},h_{c2}]$
for several values of anisotropy $\Delta \geq 1$. 
The qualitative difference between the $\Delta = 1$ and $\Delta >1$ cases is that 
$D^z (0)$ remains finite and vanishes, respectively, for $m\rightarrow 0$.}
\label{figure3PR}
\end{center}
\end{figure}

As shown in Fig. \ref{figure3PR}, for $\Delta >1$ the $T=0$ spin stiffness vanishes both for 
spin densities $m\rightarrow 0$ and $m\rightarrow 1$. At $\Delta =1$, it only vanishes in the latter limit. This
follows from the behavior of the group velocity $v_{1} (q)$ at $q=k_{F\downarrow}$.
For $\Delta \geq 1$ and (i) $m=0$ and $0\leq h\leq h_{c1}$ and 
(ii) $(1-m)\ll 1$ and $(h_{c2}-h)/(h_{c2}-h_{c1})\ll 1$ the velocity
$v_{1} (q)$ has analytical expressions provided in (i) Eq. (\ref{v1qm0}) and
(ii) Eq. (\ref{v1qm1}) of Appendix \ref{A}.

From the use of such expressions one finds that for $m=0$ and $0\leq h\leq h_{c1}$,
where $h_{c1} = 0$ at $\Delta =1$, the group velocity at $q=k_{F\downarrow}=\pi/2$ reads $v_1(k_{F\downarrow}) = 
v_1(\pi/2) = 0$ for $\Delta >1$ and $v_1(k_{F\downarrow}) = 
v_1(\pi/2) = J{\pi\over 2}$ at $\Delta =1$. This is why the $T=0$ spin stiffness vanishes 
in Fig. \ref{figure3PR} as $m\rightarrow 0$ for $\Delta >1$, yet it is finite and reads
$D^z = J/8$ at $\Delta = 1$. 

On the other hand, one finds 
that $v_1(k_{F\downarrow}) = v_1(0) = 0$ at  $m=1$ and $h= h_{c2}$, 
so that $D^z =0$ for $\Delta \geq 1$, as shown in Fig. \ref{figure3PR}.
(The results for $\Delta = 1$ were already known both for zero field \cite{Zotos_99,Shastry_90}
and $m\in [0,1]$ \cite{Carmelo_15A}.)

The physics behind the finite $T=0$ spin-stiffness expression, Eq. (\ref{Dexpression}), is that
spin transport is ballistic for the spin-conducting quantum phase associated with
the magnetic-field interval $h\in [h_{c1},h_{c2}]$. 

On the other hand, as shown in Fig. \ref{figure3PR}, the $T=0$ spin stiffness vanishes for $\Delta >1$ 
at $m=0$ and thus for $h\in [0,h_{c1}]$. Combination of this $T=0$ result with Eq. (\ref{Dz0h0}) for $T>0$, then gives
\begin{equation}
\lim_{h\rightarrow 0}D^z (T) = 0 
\hspace{0.20cm}{\rm for}\hspace{0.20cm}\Delta > 1 
\hspace{0.20cm}{\rm and}\hspace{0.20cm}T \geq 0 \, .
\label{Dz0h0plusT0}
\end{equation}
This exact $T=0$ result combined with those of Sec. \ref{SECIIID} provides strong
evidence that at zero magnetic field the ballistic contributions to spin transport vanish in the thermodynamic limit for 
$\Delta >1$ and {\it all} temperatures $T \geq 0$.

As shown in Fig. \ref{figure3PR}, for small $m = 2S^z/N \ll 1$ 
the $T=0$ spin stiffness, Eq. (\ref{Dexpression}), grows linearly
with $m$. In that limit it can be written as
\begin{eqnarray}
D^z & = & \xi^2\,I^z \hspace{0.20cm}{\rm where}
\hspace{0.20cm}I^z = {S^z\over 2N\,M_1} = {m\over 4M_1} \hspace{0.20cm}{\rm and}
\nonumber \\
{1\over M_1} & = & {J\over\pi}\sinh \eta\,K (u_{\eta})\,{u_{\eta}^2\over \sqrt{1 - u_{\eta}^2}} \, .
\label{Dexpremll}
\end{eqnarray}
Here $M_1$ is the $1$-hole static mass such that $1/M_1 = \vert d^2\varepsilon_1 (q)/d q^2\vert_{q=\pi/2}$
where $\varepsilon_1 (q)$ is the $1$-band energy dispersion, Eq. (\ref{equA4n}) of Appendix \ref{A} for $h=0$,
$K (u_{\eta})$ is the complete elliptic integral, and the $\eta$ dependence of the parameter 
$u_{\eta}$ is defined by the relation $\eta = \pi K' (u_{\eta}) /K (u_{\eta})$
where $K' (u_{\eta}) = K \left(\sqrt{1 - u_{\eta}^2}\right)$.

\section{Spin-diffusion constant}
\label{SECV}

We start by assuming that the $T>0$ spin stiffness vanishes for $\Delta >1$ both at $h=0$ and for $h\rightarrow 0$,
as found in Sec. \ref{SECIIID} by use of the inequality, Eq. (\ref{D-all-T-simp-mUB}). Our goal is to confirm that 
the dominant spin transport is normal diffusive for $\Delta >1$. 

First, we confirm that the spin-diffusion constant $D (T)$ associated with the regular part of the 
spin conductivity $\sigma_{{\rm reg}} (\omega,T)$ in Eq. (\ref{D-all-T-simpA}) is finite for $\Delta >1$ and $T>0$.
Then we use the relation between two suitable chosen 
thermal averages of the square of the spin elementary currents carried
by the spin carriers to confirm that the spin stiffness indeed vanishes for $\Delta > 1$ and $T> 0$.

We start by showing that the spin-diffusion constant is for $\Delta >1$ and $T>0$ controlled 
by the spin elementary currents carried by the spin carriers, Eq. (\ref{elem-currents}).
For $\Delta >1$, it is found to be enhanced upon lowering the temperature.
It reaches its largest finite values in the limit of very low temperatures, only diverging in
the $T\rightarrow 0$ limit. Most recent studies refer to the opposite limit of high temperature \cite{Znidaric_11,Ljubotina_17,Medenjak_17,Ilievski_18,Gopalakrishnan_19,Weiner_20,Gopalakrishnan_23}.
In Sec. \ref{SECVB}, we derive its expression for $\Delta >1$ and
very low temperatures. It diverges only in the $\Delta\rightarrow 1$ limit.

Complementarily, we combine our general expression for the spin-diffusion constant
with previous results for it at high temperatures \cite{Gopalakrishnan_19,Steinigeweg_11} to access 
the enhancement of that constant upon lowering the temperature
in the opposite limit of high finite temperatures. The corresponding results refer to $\beta J \in [0,1]$ and $\Delta \in ]1,2]$
where $\beta = 1/k_B T$. The finite-temperature spin-diffusion constant is again found to diverge
only in the $\Delta\rightarrow 1$ limit.

\subsection{General finite temperatures}
\label{SECVA}  

From manipulations of the Kubo formula and Einstein relation under the use of the physical-spin 
representation and the spin elementary currents $j_{\pm 1/2}$ emerging from it, we
find that the spin-diffusion constant $D (T)$ associated with the regular part of the 
spin conductivity $\sigma_{{\rm reg}} (\omega,T)$ in Eq. (\ref{D-all-T-simpA}) can be written as,
\begin{equation}
D (T) = C (T)\,\Pi (T) = C (T)\,N\,\langle\vert j_{\pm 1/2}\vert^2\rangle_{T} \, ,
\label{DproptoOmega}
\end{equation}
where the thermal expectation value reads,
\begin{eqnarray}
\langle\vert j_{\pm 1/2}\vert^2\rangle_{T} & = & {\Pi (T)\over N} = \sum_{S_{q}=1}^{N/2}\sum_{S^z=-S_{q}}^{S_{q}}\sum_{l_{\rm r}^{\eta}} 
p_{l_{\rm r}^{\eta},S_{q},S^z}\vert j_{\pm 1/2}\vert^2
\nonumber \\
& = & \sum_{S_{q}=1}^{N/2}(2S_{q}+1)\sum_{l_{\rm r}^{\eta}}p_{l_{\rm r}^{\eta},S_{q},0}\vert j_{\pm 1/2}\vert^2 \, ,
\label{jz2TD}
\end{eqnarray}
and the coefficient $C  (T)$ is finite for $T>0$. Its following expression is either 
exact or a good approximation:
\begin{equation}
C (T) = {1\over 8 v_{\rm LR}\,\chi (T)\,f_{1} (T)} \, .
\label{CalphaT}
\end{equation}
Here $v_{\rm LR}$ and $\chi (T)$ are the Lieb-Robinson velocity and static spin susceptibility, respectively, 
and $f_{1} (T)$ is the second derivative of the free energy density with respect to 
$m$ at $m = 0$. The Lieb-Robinson velocity is the maximal velocity with which the information can travel.

In contrast to the thermal expectation values at fixed $S^z$, Eq. (\ref{jz2T}), 
the thermal expectation values $\langle\vert j_{\pm 1/2}\vert^2\rangle_{T} = \Pi  (T)/N$, Eq. (\ref{jz2TD}),
involve summations over all $S_q>0$ energy eigenstates.
This thus includes those with different $S^z\in [-S_{q},S_{q}]$ values. 

At zero field, the energy eigenvalues, Eq.  (\ref{Energy0}), are the same for a number
$2S_{q} + 1$ of energy eigenstates states of the same $q$-spin tower, Eq. (\ref{state}). 
In addition, there is the symmetry, Eq. (\ref{deriv-elem-currents}), such that 
the spin elementary currents $j_{\pm 1/2}$, Eq. (\ref{elem-currents}), also have the same value for all 
such $2S_{q} + 1$ states. They are thus independent of $S^z$ and $m$. 

Such two symmetries have allowed to perform the summation $\sum_{S^z=-S_{q}}^{S_{q}}$ in Eq. (\ref{jz2TD}). 
The set of $2S_{q}+1$ states of each $q$-spin tower were replaced by a single-tower state. We have
chosen that for which $S^z = 0$. We could though have chosen any other tower state, for instance that for which
$S^z = S_{q}$. However, since all choices give the same value of $\Pi (T)$ in Eq. (\ref{jz2TD}),
for the sake of comparison of the expression of $\Pi (T)$ in that equation
with the expression of $\Omega_0 (T)$, Eq. (\ref{jz2T}) for $m = 2S^z/N =0$, 
we have chosen the tower state for which $S^z = 0$.

The spin-diffusion constant $D (T)$, Eq. (\ref{DproptoOmega}), is known to
obey for $T>0$ the following exact inequality \cite{Medenjak_17}:
\begin{equation}
D (T) \geq {1\over 8\beta v_{\rm LR}\,\chi (T)\,f_{1} (T)}
{\partial^2 D^z (T)\over\partial m^2}\Bigg|_{m = 0} \, .
\label{Dalphaineq}
\end{equation}
From the use of Eqs. (\ref{D-all-T-simp-m}) and (\ref{jz2T}) we find that
\begin{equation}
{\partial^2 D^z (T)\over\partial m^2}\Bigg|_{m = 0} = 
{N\over k_B T} \langle\vert j_{\pm 1/2}\vert^2\rangle_{0,T} = {\Omega_0 (T)\over k_B T} \, ,
\label{2derivD-all-T-simp-m}
\end{equation}
where $\Omega_0 (T)$, Eq. (\ref{jz2T}) at $m=0$, is such that
\begin{equation}
\langle\vert j_{\pm 1/2}\vert^2\rangle_{0,T} = {\Omega_0 (T)\over N} = 
\sum_{S_{q}=1}^{N/2}\sum_{l_{\rm r}^{\eta}}\,p_{l_{\rm r}^{\eta},S_{q},0}\vert j_{\pm 1/2}\vert^2 \, .
\label{jz2T0}
\end{equation}
Here we accounted for $j_{\pm 1/2}= 0$ at $S_{q} = 0$.

The use of the expression, Eq. (\ref{2derivD-all-T-simp-m}), on the right-hand side of Eq. (\ref{Dalphaineq}) then leads to
\begin{equation}
D (T) \geq {N\,\langle\vert j_{\pm 1/2}\vert^2\rangle_{0,T}\over 8 v_{\rm LR}\,\chi (T)\,f_{1} (T)}
= {\Omega_0  (T)\over 8 v_{\rm LR}\,\chi (T)\,f_{1} (T)} \, .
\label{DalphaineqF}
\end{equation}
Comparison of Eq. (\ref{jz2TD}) for $\Pi (T)/N$ with Eq. (\ref{jz2T0}) for $\Omega_0 (T)/N$
shows that $\Pi (T) >  \Omega_0 (T)$. This confirms that the spin-diffusion constant, Eq. (\ref{DproptoOmega}), 
obeys the exact inequality, Eq. (\ref{Dalphaineq}),
\begin{equation}
D (T) = {\Pi (T)\over 8 v_{\rm LR}\,\chi (T)\,f_{1} (T)}
> {\Omega_0 (T)\over 8 v_{\rm LR}\,\chi (T)\,f_{1} (T)} \, .
\label{Dalpha}
\end{equation}

The general expression of the spin-diffusion constant, Eq. (\ref{DproptoOmega}),
involves summations that run over all $S_q>0$ energy eigenstates, Eq. (\ref{jz2TD}).
They are thus difficult to be performed for all finite temperatures.
In the following two sections we study that spin-diffusion constant
for low and high temperatures, respectively. That it reaches its largest
values at low temperatures allows us to conclude it is finite for $\Delta >1$ and
$T>0$.

\subsection{Spin-diffusion constant for low temperatures}
\label{SECVB}

The physical-spin representation includes complementary and alternative representations of the
energy eigenstates in terms of $n$-band momentum values occupancy configurations and 
$n$-squeezed effective lattice sites occupancy configurations, respectively.
The concept of a squeezed effective lattice is well known in 1D correlated systems 
\cite{Ogata_90,Penc_97,Kruis_04}. In Appendix \ref{C}, we provide the information needed for the
studies of this paper on the $n$-squeezed effective lattices of the spin-$1/2$ $XXZ$ chain for anisotropy $\Delta >1$. 

The spin currents are generated by processes where upon moving in each of the $n$-squeezed effective lattices 
for which $N_n >0$ the $M=2S_q$ spin carriers interchange position with a number $N_n$ of  $n$-pairs. The set of active $n$-squeezed effective 
lattices is associated with that of $n$-bands whose number $N_n$ of $n$-pairs is finite and fixed for
a given subspace.

In the case of the $\Delta >1$ spin-$1/2$ $XXZ$ chain, the zero-field ground states are populated by a number $N_1 = N/2$ 
of unbound singlet $1$-pairs containing a number $N=2N_1$ of paired physical spins. There are no unpaired physical spins 
and thus no spin carriers, so that the ground-states spin-current expectation value exactly vanishes. 

We find in the following that in the spin-$1/2$ $XXZ$ chain for $\Delta >1$ nearly ballistic transport occurs in 
the limit of very low temperatures, but with zero spin stiffness.
To derive the spin-diffusion constant for very low temperatures, we start by 
describing the configurations that generate the states that contribute to it in that limit. They are not populated 
by $n>1$ $n$-string pairs. 

Specifically, such states are populated by $S^z=\pm 1$ triplet pairs that involve two unpaired physical spins with the same 
projection $\pm 1/2$. The excitation energy and momentum of one such a $S^z=\pm 1$ 
triplet pair is relative to the ground states given by
\begin{equation}
\delta E (k) = - \varepsilon_{1} (q) - \varepsilon_{1} (q') \hspace{0.20cm}{\rm and}\hspace{0.20cm}
k = \pi - q - q' \, ,
\label{deltaEk20}
\end{equation}
respectively. Here, $-\varepsilon_{1} (q)>0$ and $-\varepsilon_{1} (q')>0$ are the excitation energies associated with 
the creation of two $1$-holes with excitation momentum values $-q$ and $-q'$, respectively. They describe the translational 
degrees of freedom of two emerging unpaired physical spins. The $1$-band energy dispersion $\varepsilon_{1} (q)$ is 
given in Eq. (\ref{equA4n}) of Appendix \ref{A} for $h=0$.

At very low temperatures, the excitation energy of such triplet 
pairs must be very near its minimum allowed value, which refers to the spin gap.
That gap corresponds to the excitation energy of the two $1$-holes having excitation momentum values
$-q = \mp\pi/2$ and $- q' = \mp\pi/2 \pm 2\pi/N$, respectively. Combining the expressions given in Eq. (\ref{deltaEk20})
and Eq. (\ref{equA4n}) of Appendix \ref{A} for $h=0$, we recover the well known expression 
of the spin gap \cite{Takahashi_99},
\begin{equation}
E_{\rm gap} = - 2\varepsilon_{1} (\pi/2)\vert_{h=0} = {2J\over\pi}\sinh \eta\,K (u_{\eta})\,\sqrt{1 - u_{\eta}^2} \, .
\label{Egap}
\end{equation}

$S^z=0$ triplet pairs have the same minimum energy yet involve two unpaired physical spins with {\it opposite} projection.
They do not contribute to spin transport because their spin elementary currents cancel each other. 
The same applies to singlet $2$-string pairs, which are the only $n$-string pairs whose minimum energy is 
also given by $E_{\rm gap}$, Eq. (\ref{Egap}).

Excited states populated by $S^z=\pm 1$ triplet pairs that contribute to spin transport at very low temperatures have pairs 
of $1$-holes with $1$-band excitation momentum values $-q \approx \mp\pi/2$ and $- q' = - q \pm 2\pi/N$ and thus low-excitation momentum 
$k = (\pm\pi - 2q \pm 2\pi/N)$, which reads $k = (\pm\pi - 2q)$ in the thermodynamic limit. (The term $\pm 2\pi/N$ 
in $- q' = - q \pm 2\pi/N$ follows from the $1$-band momentum values having Pauli-like 
occupancies.) 

In the thermodynamic limit, the two corresponding unpaired physical spins with the same projection $+1/2$
or $-1/2$ then move with nearly the same $1$-band momentum. Upon moving in the $1$-squeezed effective lattice 
by interchanging position with the $1$-pairs, the two unpaired physical spins of the same triplet pair are adjacent.

In the $k_B T/E_{\rm gap} \ll 1$ regime, the spin carriers are actually $S^z=\pm 1$ triplet pairs. They refer in the $1$-squeezed 
effective lattice to two adjacent unpaired physical spins with the same projection. We call them {\it triplet spin carriers}
to distinguish from their two unpaired physical spins. 

Such triplet spin carriers play the role of 
low-temperature diffusing spins. They have small excitation momentum $k = (\pm\pi - 2q)$ associated with 
$-q \approx \mp\pi/2$ and $- q' = - q \pm 2\pi/N$ in Eq. (\ref{deltaEk20}) and excitation energy given by
\begin{equation}
\epsilon_p (k) = E_{\rm gap} + 2\,{\left(\pm{\pi\over 2}-q\right)^2\over 2M_1} = E_{\rm gap} + {k^2\over 2M_p} \, .
\label{deltaEk2}
\end{equation}
Their static mass $M_p=2M_1$ is such that $1/M_p = \vert d^2 \epsilon_p (k)/d k^2\vert_{k=0}$. Here
$M_1$ is the $1$-hole static mass defined in Eq. (\ref{Dexpremll}). Their group velocity reads $v_p (k) = {k\over M_p}$.

The lowest excited states have a single $1$-squeezed effective lattice domain wall. It refers to
a single triplet pair. The domain wall or diffusing spin associated with the triplet pair moves ballistically. 
Also for excited states with a finite number and thus vanishing 
concentration of triplet pairs, such pairs move ballistically over large $1$-squeezed effective lattice
distances, which at very low temperature read $x_p = (\beta /M_p)^{1/2}\,e^{E_{\rm gap}\beta}$, 
before interacting with other triplet pairs. 

Since the triplet pairs are the diffusing spins, this nearly ballistic spin transport, but with zero-spin stiffness, 
is associated with a spin-diffusion constant that in the $k_B T/E_{\rm gap} \ll 1$ regime is proportional 
to the inverse of the transport mass $M_t$ of the triplet spin carriers, $D\propto 1/M_t$. Such a proportionality 
is confirmed below by use of a relation between the diffusion constant 
for $T\rightarrow 0$ and the derivative $\partial D^z/\partial m$ of the $T=0$ spin stiffness for 
$m = 2S^z/N \ll 1$, Eq. (\ref{Dexpremll}). 
At general finite temperatures, there is also a relation between 
the diffusion constant and spin stiffness, in that case involving a second derivative associated
with the curvature of the stiffness with respect to the filling parameter \cite{Medenjak_17},
as given in Eq. (\ref{Dalphaineq}).

The inverse of the spin transport mass of the triplet spin carriers
is defined as $1/M_t = \vert\partial j_{\pm 1} (k)/\partial k\vert_{k=0}$.
Here, $j_{\pm 1} (k) = 2j_{\pm 1/2} (q)\vert_{q=(\pm\pi - k)/2}$ is the spin elementary current carried by 
them and $j_{\pm 1/2} (q)$, Eq. (\ref{elem-currents}), is that
carried by the two corresponding adjacent unpaired physical spins of projection $\pm 1/2$.

The triplet pairs that in the $k_B T/E_{\rm gap} \ll 1$ regime contribute
to spin transport are such that $q\approx \pm \pi/2$ in Eq. (\ref{deltaEk2}).
The spin elementary currents of the two corresponding adjacent unpaired physical spins 
thus have absolute values $\vert j_{\pm 1/2} (q,q')\vert$ 
of the form given in Eq. (\ref{jznnqqqq}) of Appendix \ref{B}
with $q\approx \pm \pi/2$ and $q' = q \mp 2\pi/N$, respectively.

The spin elementary currents $j_{\pm 1/2} (q)$ and $j_{\pm 1} (k)$ carried
by the two unpaired physical spins and corresponding triplet pair, respectively,
are in the thermodynamic limit then given by
\begin{eqnarray}
j_{\pm 1/2} (q) & = & \mp {J\over\pi}\sinh \eta\,K (u_{\eta})\,u_{\eta}^2{\sin 2q\over\sqrt{1 - u_{\eta}^2\sin^2 q}}
\hspace{0.20cm}{\rm and}
\nonumber \\
j_{\pm 1} (k) & = & \mp {2J\over\pi}\sinh \eta\,K (u_{\eta})\,u_{\eta}^2{\sin k\over\sqrt{1 - u_{\eta}^2\cos^2 (k/2)}} \, ,
\nonumber \\
\label{jznntoqeq}
\end{eqnarray}
respectively. 

The corresponding inverse of the transport mass of the triplet spin carriers thus reads
\begin{equation}
{1\over M_t} = \vert{\partial j_{\pm 1} (k)\over \partial k}\vert_{k=0}
= {2J\over\pi}\sinh \eta\,K (u_{\eta})\,{u_{\eta}^2\over \sqrt{1 - u_{\eta}^2}} \, ,
\label{DlargXXX}
\end{equation}
so that $j_{\pm 1} (k) \approx \mp {k\over M_t}$ for small momentum $k$.
We then find the relations $1/(4M_t)=1/(2M_1)=1/M_p$ involving that transport mass $M_t$, 
the $1$-hole static mass $M_1$, and the triplet-pair static mass $M_p$.

On the one hand, in the gapless $T=0$ quantum spin conducting phase the spin stiffness,
Eq. (\ref{Dexpression}), does not involve the spin gap $E_{\rm gap}$.
This applies also in the $m = 2S^z/N \ll 1$ limit, Eq. (\ref{Dexpremll}).
On the other hand, there is no TLL physics in the $T\rightarrow 0$ gapped spin 
insulating phase for $h\in [0,h_{c1}]$ and $m=0$.
It follows that the spin-diffusion constant does not involve the
TLL parameter \cite{Horvatic_20}, $K = \xi^2$.

Consistent with both these differences between the $T=0$ and $T\rightarrow 0$ quantum problems
and the nearly ballistic spin-transport of the $k_B T/E_{\rm gap} \ll 1$ regime, but with zero-spin stiffness, 
the leading term of the spin-diffusion constant $D (T)$, Eqs. (\ref{DproptoOmega}) and (\ref{Dalpha}),
can in that regime be expressed in terms of the first derivative with respect to $m$ 
of the $T=0$ spin stiffness $D^z$ for $m\ll 1$, Eq. (\ref{Dexpremll}), as follows:
\begin{equation}
D (T) = {e^{E_{\rm gap}\over K_B T}\over\xi^2}{d D^z\over d m}\Bigg|_{m=0} =
{e^{E_{\rm gap}\over K_B T}\over 8M_t}
\hspace{0.20cm}{\rm for}\hspace{0.20cm}T \ll {E_{\rm gap}\over k_B} \, .
\label{Dlargebeta0}
\end{equation}
Here, $E_{\rm gap}$ and $1/M_t$ are given in Eqs. (\ref{Egap}) and (\ref{DlargXXX}), respectively,
and we used the relation $1/(2M_t)=1/M_1$.

The spin-diffusion constant, Eq. (\ref{Dlargebeta0}), reaches very large values 
in the $k_B T/E_{\rm gap} \ll 1$ regime. It though only diverges in the $T\rightarrow 0$ limit, 
being finite and decreasing upon increasing $T$ in that regime, consistent with the
occurrence of normal diffusive spin transport.

As mentioned above, in the $k_B T/E_{\rm gap} \ll 1$ regime the $1$-squeezed effective lattice 
distance $x_p = (\beta /M_p)^{1/2}\,e^{E_{\rm gap}\beta}$ traveled by one triplet pair before it interacts with other triplet pairs and ceases to be 
ballistic diverges exponentially. Complementarily, the corresponding concentration of triplet pairs is exponentially small,
$S_q/N = 1/x_p = (M_p/\beta)^{1/2}\,e^{-E_{\rm gap}\beta}$. 

Consistently, we can derive the expression, Eq. (\ref{Dlargebeta0}), by 
treating the quantum problem as a dilute gas of excited spin triplet pairs, each with energy $\epsilon_p (k)$, Eq. (\ref{deltaEk2}). 
Their motion and collisions dominate the spin transport properties, their spacing $x_p = (\beta /M_p)^{1/2}\,e^{E_{\rm gap}\beta}$
being for $k_B T/\Delta_{\eta} \ll 1$ much larger than their thermal de Broglie wavelength 
$\lambda_p = (\beta /M_p)^{1/2}$. 

The corresponding quantum problem then involves the collisions of the spin-triplet pairs 
and corresponding $S$ matrix. However, the root mean square thermal velocity of such
pairs, $v_{p,\beta} = 1/\sqrt{M_p\beta}$, tends to zero as $k_B T/E_{\rm gap} \rightarrow 0$,
so that the $S$ matrix that describes the present quantum problem refers to vanishing exchange incoming or outgoing 
momenta. 

This much simplifies the problem, as we can apply the methods of Ref. \onlinecite{Tsvelik_87} to the spin-$1/2$ $XXZ$ 
chain with anisotropy $\Delta >1$ to derive the leading behaviors of the uniform spin
susceptibility $\chi$ and real part of the dc spin conductivity $\sigma$ in the $k_B T/E_{\rm gap} \ll 1$ regime. 

We find that in our units such two quantities are expressed solely in terms of the two above length 
scales as $\chi  = (2/\pi)^{1/2}{\beta\over x_p}$ and 
$\sigma = (2/\pi)^{1/2}{\lambda_p\over 2}$, respectively. The use of the Einstein relation, $\sigma = \chi D$, then 
gives the leading behavior $D = \sigma/\chi = {x_p \lambda_p\over 2\beta}$.
Accounting for the relation $1/(4M_t)=1/M_p$, we find that
$D = {x_p \lambda_p\over 2\beta} = e^{E_{\rm gap}\beta}/(8M_t)$ for $k_B T/E_{\rm gap}\ll 1$, as given in Eq. (\ref{Dlargebeta0}).

Next, we use the expression, Eq. (\ref{CalphaT}), of the coefficient $C (T)$ in Eq. (\ref{DproptoOmega}) 
and the limiting behavior of the spin susceptibility for low temperatures,
$\chi (T) = 2\sqrt{2/\pi}\sqrt{M_t/k_B T}\,e^{-E_{\rm gap}/k_B T}$. We then find that 
$\Pi (T) = N\,\langle\vert j_{\pm 1/2}\vert^2\rangle_{T}$, Eq. (\ref{jz2TD}), 
has for $k_B T/E_{\rm gap}\ll 1$ the following general expression:
\begin{equation}
\Pi  (T) = N\,\langle\vert j_{\pm 1/2}\vert^2\rangle_{T} = 
\sqrt{2\over\pi}\,{2v_{\rm LR}\,f_{1} (T)\over\sqrt{M_t\,k_B T}} 
\hspace{0.20cm}{\rm for}\hspace{0.20cm}\Delta > 1 \, .
\label{PietaTloq}
\end{equation}
Its limiting behaviors for $\Delta \gg 1$ and $(\Delta - 1) \ll 1$ are
\begin{eqnarray}
&& \Pi  (T) \approx 8v_{\rm LR}\,f_{1} (T)\sqrt{J\over \pi k_B T}
\hspace{0.20cm}{\rm for}\hspace{0.20cm}\Delta \gg 1\hspace{0.20cm}{\rm and}
\nonumber \\
&& \approx 2v_{\rm LR}\,f_{1} (T)\sqrt{J\over k_B T}\,e^{{\pi^2\over 4\sqrt{2(\Delta -1)}}}
\hspace{0.20cm}{\rm for}\hspace{0.20cm}(\Delta - 1) \ll 1
\label{PietaTloqLimits}
\end{eqnarray}
respectively. That $\Pi  (T)$ is finite for $\Delta >1$ and diverges in the $\Delta\rightarrow 1$ limit
is consistent with low-temperature anomalous superdiffusive spin transport at $\Delta =1$.

The corresponding limiting behaviors of the spin-diffusion constant, Eq. (\ref{Dlargebeta0}),
are for $\Delta \gg 1$ and $(\Delta - 1) \ll 1$ obtained by the use of those 
of the elliptic integral $K (u_{\eta})$ and $u_{\eta}$ provided below Eq. (\ref{criticalfields})
of Appendix \ref{A}. For $k_B T/E_{\rm gap}\ll 1$ 
it reads in these two limits,
\begin{eqnarray}
D (T) & \approx & J e^{\Delta \beta J}\hspace{0.20cm}{\rm for}\hspace{0.20cm}\Delta \gg 1 
\hspace{0.20cm}{\rm and}
\nonumber \\
 & \approx & {J \pi\over 16}\,e^{{\pi^2\over 2\sqrt{2(\Delta -1)}}}e^{E_{\rm gap}\beta}
\hspace{0.20cm}{\rm for}\hspace{0.20cm}(\Delta -1)\ll 1 \, ,
\label{DDeltasmalllarge}
\end{eqnarray}
respectively. Here, the spin gap, Eq. (\ref{Egap}), is given by
$E_{\rm gap} \approx 2\pi Je^{-\pi^2/(2\sqrt{2(\Delta -1)})}$ for $(\Delta -1)\ll 1$,
vanishing in the $\Delta\rightarrow 1$ limit. As justified below in Sec. \ref{SECVD}, the type of spin transport is 
though controlled by $\Pi  (T) = N\,\langle\vert j_{\pm 1/2}\vert^2\rangle_{T}$, Eq. (\ref{PietaTloq}),
which has no exponential factor $e^{E_{\rm gap}/k_B T}$.

\subsection{Spin-diffusion constant for $\beta J\in [0,1]$}
\label{SECVC}
\begin{figure*}
\includegraphics[width=0.495\textwidth]{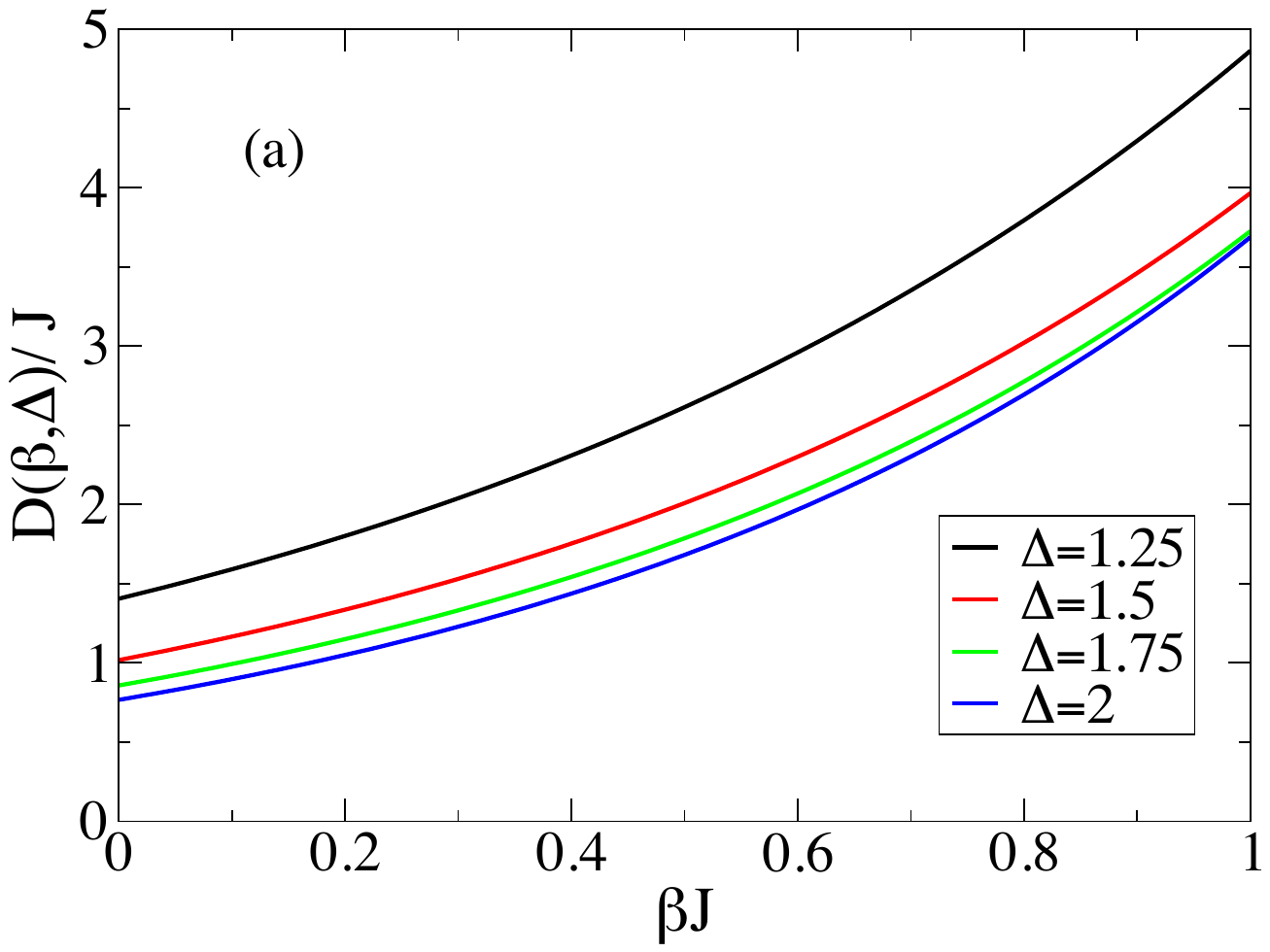}
\includegraphics[width=0.495\textwidth]{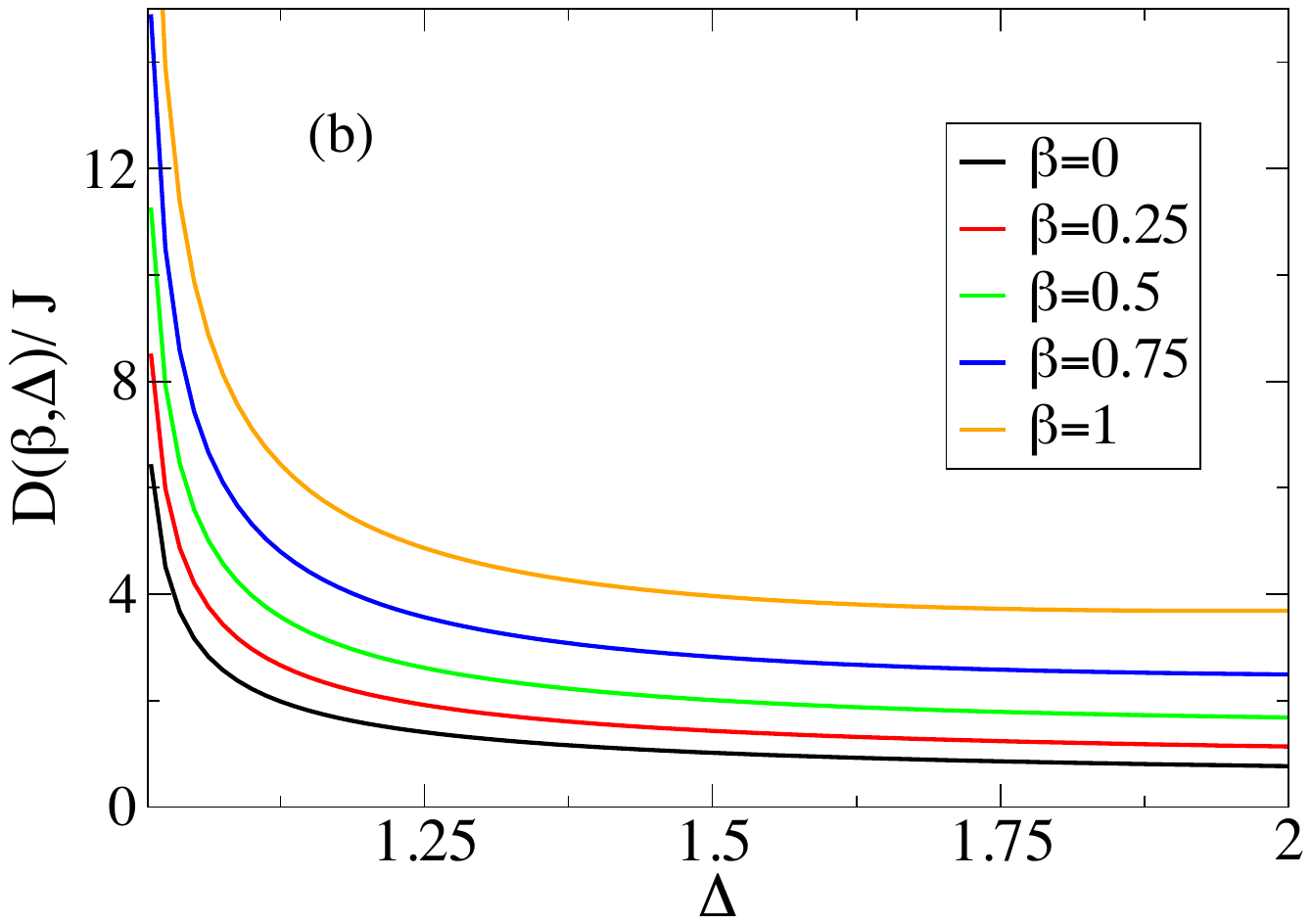}
\caption{The spin-diffusion constant $D (\beta,\Delta) = D (T)$ where $\beta = 1/k_B T$
as given by Eq. (\ref{DiffC}) (a) vs $\beta J \in [0,1]$ 
for anisotropy values $\Delta = 1.25$, $\Delta = 1.5$, $\Delta = 1.75$, and $\Delta = 2.0$
and (b) versus $\Delta \in ]1,2]$ for the inverse temperatures $\beta J = 0$, $\beta J = 0.25$, $\beta J = 0.5$, $\beta J = 0.75$, 
and $\beta J = 1.0$. The curve plotted in (b) for $\beta =0$ exactly coincides with that plotted in Fig. 1
of Ref. \onlinecite{Gopalakrishnan_19}.}
\label{figure4PR}
\end{figure*}

Here we address the enhancement of the spin-diffusion constant, Eqs. (\ref{DproptoOmega}) and (\ref{Dalpha}), 
for $\Delta >1$ upon lowering the temperature in the case of high finite temperatures, starting from infinite 
temperature. To reach that goal, we combine our general expression for that constant given in these equations
with the numerical results of Ref. \onlinecite{Steinigeweg_11} for $\beta J \in [0,1]$ and those of 
Ref. \onlinecite{Gopalakrishnan_19} for $\beta = 0$
derived by a method that combines generalized hydrodynamics and Gaussian fluctuations. 

The following results refer to the behavior of the spin-diffusion constant for high temperatures
associated with the range $\beta J\in [0,1]$ and the anisotropy interval $\Delta \in ]1,2]$ of more interest for spin-chain
materials \cite{Carmelo_23,Bera_20,Wang_18}. 

Figure 4(d) of Ref. \onlinecite{Steinigeweg_11} displays the spin-diffusion constant versus 
$\beta \in [0,2]$ in a semilogarithmic plot for anisotropies $\Delta = 1.5$ and $\Delta =2.0$.  
That constant shows exponential increase in the interval $\beta \in [0,1]$. The
method used in that reference is expected to give a good approximation for 
${1\over D (\infty)}{\partial D (T)\over \partial \beta}$ in that interval.
After careful analysis of the numerical results for the spin-diffusion constant provided in Fig. 4(d)
of Ref. \onlinecite{Steinigeweg_11}, we find that
\begin{equation}
{1\over D (\infty)}{\partial D (T)\over \partial \beta} = c_{\Delta} J\,e^{c_{\Delta}\,\beta J} 
\hspace{0.20cm}{\rm for}\hspace{0.20cm}\beta J \in [0,1] \, ,
\label{Dexpnew}
\end{equation}
where $c_{\Delta} \approx {\pi\over 2}\sqrt{\Delta/2}$ for $\Delta \in ]1,2]$. This then gives
\begin{eqnarray}
&& D (T) = D (\infty)\,e^{c_{\Delta}\,\beta J} 
\nonumber \\
&& = {4J\,e^{c_{\Delta}\,\beta J} \over 9\pi}\sqrt{\Delta^2 - 1}\,\sum_{n=1}^{\infty}(n+1)
\nonumber \\
&& \times \Bigl({n+2\over (\Delta + \sqrt{\Delta^2 -1})^n - (\Delta - \sqrt{\Delta^2 -1})^n}
\nonumber \\
&& - {n\over (\Delta + \sqrt{\Delta^2 -1})^{n+2} - (\Delta - \sqrt{\Delta^2 -1})^{n+2}}\Bigr) 
\label{DiffC}
\end{eqnarray}
for the spin-diffusion constant, Eqs. (\ref{DproptoOmega}) and (\ref{Dalpha}), for $\beta J \in [0,1]$, 
where $D (\infty)$ is its infinite-temperature expression obtained for $\Delta >1$ in Ref. \onlinecite{Gopalakrishnan_19}. 

The main result in the expression given in Eq. (\ref{DiffC}) is indeed the $\beta$ dependence for $\beta >0$
and up to approximately $1/J$: The use of the relation $\Delta = \cosh \eta$ confirms that at $\beta =0$ such an expression 
becomes that already provided in Eq. (11) of Ref. \onlinecite{Gopalakrishnan_19}.

The high-temperature behavior of the spin-diffusion constant $D (T)$, Eq. (\ref{DiffC}), 
is plotted in Fig. \ref{figure4PR}(a) as a function of $\beta J \in [0,1]$ 
for $\Delta = 1.25$, $\Delta = 1.5$, $\Delta = 1.75$, and $\Delta = 2.0$.
In Fig. \ref{figure4PR}(b) it is plotted as a function of $\Delta \in ]1,2]$ for 
$\beta J = 0$, $\beta J = 0.25$, $\beta J = 0.5$, $\beta J = 0.75$, and $\beta J = 1.0$.
$D (T)$ increases exponentially as temperature decreases for $\beta J \in [0,1]$
and decreases at fixed finite temperature in that range upon increasing anisotropy, 
as for infinite temperature \cite{Gopalakrishnan_19}.

In the $T\rightarrow\infty$ limit the expression, Eq. (\ref{CalphaT}), for the coefficient $C (T)$ simplifies to $C (T) = 1/v_{\rm LR}$
and that of $D (\infty)$, Eq. (\ref{DiffC}) for $\beta =0$, is valid for $\Delta \in ]1,\infty]$ \cite{Gopalakrishnan_19}.
We then find that $\lim_{T\rightarrow\infty}\langle\vert j_{\pm 1/2}\vert^2\rangle_{T} = \Pi (\infty)/N = v_{\rm LR}D (\infty)/N$
for $\Delta \in ]1,\infty]$. Here, $\Pi (\infty) = v_{\rm LR}D (\infty)$ is thus finite for $\Delta >1$ in the $T\rightarrow\infty$ limit. 
As for low temperatures, Eqs. (\ref{PietaTloq}) and (\ref{PietaTloqLimits}), $\Pi (\infty)$ diverges in the $\Delta\rightarrow 1$ limit, 
consistent with the occurrence of anomalous superdiffusive transport for all finite temperatures at $\Delta = 1$.

\subsection{Normal diffusive $T>0$ spin transport for $\Delta >1$}
\label{SECVD}  

The coefficient $C (T)$, Eq. (\ref{CalphaT}), is finite for $T>0$. 
Provided that as predicted in Sec. \ref{SECIII} the spin stiffness vanishes
at zero field, the spin-diffusion constant expression, Eq. (\ref{DproptoOmega}), shows 
that the thermal expectation value $\langle\vert j_{\pm 1/2}\vert^2\rangle_{T} = \Pi (T)/L$, Eq. (\ref{jz2TD}), 
controls the type of spin transport. Specifically, according to the following criteria one has
\begin{eqnarray}
&& \Pi (T) = N\,\langle\vert j_{\pm 1/2}\vert^2\rangle_{T}\rightarrow
\infty \Rightarrow {\rm spin}\hspace{0.20cm}{\rm superdiffusion}
\nonumber \\
&& \Pi (T) = N\,\langle\vert j_{\pm 1/2}\vert^2\rangle_{T}
\hspace{0.20cm}{\rm finite} \Rightarrow {\rm spin}\hspace{0.20cm}{\rm diffusion} \, .
\label{noramomD}
\end{eqnarray}

We have found that at low finite temperatures $\Pi (T)$ is finite for $\Delta >1$, as given in Eqs. (\ref{PietaTloq}) and (\ref{PietaTloqLimits}).
It though diverges in the $\Delta\rightarrow 1$ limit. In the opposite limit of high $T\rightarrow\infty$ temperature,
it was found to read $\Pi (\infty) = v_{\rm LR}D (\infty)$ where the $D (\infty)$'s expression is given in Eq.
(\ref{DiffC}). $\Pi (\infty)$ is thus finite for $\Delta >1$ and again diverges in the $\Delta\rightarrow 1$ limit.

On the other hand, the related spin-diffusion constant, Eq. (\ref{DproptoOmega}), was found to decrease upon 
increasing the temperature. Its exact low-temperature expression, Eq. (\ref{Dlargebeta0}),
decreases upon increasing $T$ and is such that $D (T)\rightarrow\infty$ in the $T\rightarrow 0$ limit. 
We can thus choose the width of an interval $T \in ]0,T_0]$ to be arbitrarily small yet finite and thus 
$D (T_0)$ to be arbitrarily large yet no infinity. That such arbitrarily large yet finite value is reached 
at low temperatures combined with spin-diffusion constant decreasing upon 
increasing the temperature then implies it is finite for all finite temperatures $T>0$. 

Since the coefficient $C (T)$, Eq. (\ref{CalphaT}), in $D (T) = C (T)\,\Pi (T)$, Eq. (\ref{DproptoOmega}),
is finite for $T>0$, that also implies that $\Pi (T) = N\,\langle\vert j_{\pm 1/2}\vert^2\rangle_{T}$, Eq. (\ref{jz2TD}), 
is finite for $T >0$ and $\Delta >1$, so that $\langle\vert j_{\pm 1/2}\vert^2\rangle_{T}$ is of the order of $1/N$
for the spin-$1/2$ $XXZ$ chain.

On the other hand, it follows from the inequality, $\Pi (T) > \Omega_0 (T)$, where $\Pi (T)$ is finite for $T>0$ and $\Delta >1$, 
whose validity is justified by Eqs. (\ref{jz2TD}) and (\ref{jz2T0}), that $\Omega_0 (T)$ does not diverge for $\Delta >1$. That 
quantity appears in the spin stiffness expression $D^z (T) = {m^2\over 2T} \Omega_0 (T)$, Eq. (\ref{D-all-T-simp-m}) 
for $m\rightarrow 0$. This then shows that $D^z (T) = 0$ both at $m=0$ and for $m\rightarrow 0$ and thus at $h=0$ and in the $h\rightarrow 0$ limit.
That result confirms the validity of the predictions of Sec. \ref{SECIIID} that for $\Delta >1$ 
and $T>0$ the spin stiffness vanishes at zero field.

That both the ballistic contributions to spin transport vanish at finite temperature $T>0$ for $\Delta >1$ 
and the spin-diffusion constant, Eq. (\ref{DproptoOmega}), and the related quantity $\Pi (T)$, Eq. (\ref{jz2TD}),
are finite, implies according to the criterion, Eq. (\ref{noramomD}), normal diffusive transport.
On the other hand, $\Pi (T)$, Eq. (\ref{jz2TD}), diverges in the $\Delta\rightarrow 1$ limit, consistent with anomalous superdiffusive spin
transport at $\Delta =1$ for $T>0$. 

\subsection{Relation to previous results on the diffusion constant}
\label{SECVE}

Our general expression of the $T>0$ spin-diffusion constant in terms of the spin elementary currents of
the spin carriers, Eqs. (\ref{DproptoOmega}), (\ref{jz2TD}), and (\ref{Dalpha}), is consistent with the general expression,
Eq. (5) of Ref. \onlinecite{Medenjak_17}. In spite of the different representation and notations, the summations 
in such two expressions run over the same states. Concerning the general expression provided in Eq. (6.38) of 
Ref. \onlinecite{Nardis_19}, its agreement with the general expression of Ref. \onlinecite{Medenjak_17}
implies its consistency with our general expression, Eqs. (\ref{DproptoOmega}), (\ref{jz2TD}), and (\ref{Dalpha}).

The leading exponential behavior of the $T>0$ spin-diffusion constant
for the $k_B T/E_{\rm gap}\ll 1$ regime, Eq. (\ref{Dlargebeta0}),
is consistent with results for general gapped 1D models in the quantum sine-Gordon 
universality class studied in Ref. \onlinecite{Damle_05}. 

After we reached our low-temperature results, we learned that the general method used
in that reference applies to the present case of the gapped spin-$1/2$ $XXZ$ chain
with anisotropy $\Delta >1$. Using our notations, the expression of the 
low-temperature spin-diffusion constant obtained by the method of Ref. \onlinecite{Damle_05} 
is for that spin chain given by
\begin{equation}
D (T) = {c^2\over 2E_{\rm gap}}e^{E_{\rm gap}\over k_B T} \, ,
\label{DTc}
\end{equation}
where $c$ is a constant. We then checked that this expression indeed 
quantitatively agrees with that provided in Eq. (\ref{Dlargebeta0})
for the $k_B T/E_{\rm gap}\ll 1$ regime, since the constant $c$ of Ref. \onlinecite{Damle_05}
is for the particular case of the gapped spin-$1/2$ $XXZ$ chain given by
\begin{equation}
c = {J\over\pi}\sinh \eta\,K (u_{\eta})\,u_{\eta} \, .
\label{c}
\end{equation}

The same leading exponential behavior of the $T>0$ spin-diffusion constant
for the $k_B T/E_{\rm gap}\ll 1$ regime, Eqs. (\ref{Dlargebeta0}) and (\ref{DTc}), obtained
in this paper in the thermodynamic limit by a method for nearly ballistic low-temperature transport, but with zero stiffness,
which accounts for the relation to the first derivative with respect to $m$ 
of the $T=0$ spin stiffness $D^z$ for $m\ll 1$, Eq. (\ref{Dexpremll}), and
in Ref. \onlinecite{Damle_05} by an also exact method that accounts for 
the universal relaxational dynamics of gapped 1D models
in the quantum sine-Gordon universality class, could in principle also be
reached by using Lanczos exact diagonalization \cite{Sandvik_10},
under extrapolation to an infinite system.

Finally, the expression of the spin-diffusion constant for $\beta J \in [0,1]$ provided in Eq. (\ref{DiffC}) 
is obviously consistent with that given in Ref. \onlinecite{Gopalakrishnan_19} for $\beta J =0$, as
it was inherently constructed to obey such a $\beta=0$ boundary condition. 
Its slope ${1\over D (\infty)}{\partial D (T)\over \partial \beta}$
is for $\Delta \in ]1,2]$ as well consistent with the spin-diffusion constant plotted in Fig. 4 (d) of
Ref. \onlinecite{Steinigeweg_11} for $\Delta = 1.5$ and $\Delta = 2$. 

\section{Concluding remarks}
\label{SECVI}

In this paper we have found that the spin carriers, which are the unpaired physical
spins in the multiplet configuration of all $S_q>0$ energy eigenstates, fully control the spin-transport 
quantities of the spin-$1/2$ $XXZ$ chain for $\Delta >1$ and $T>0$.
We have then used two complementary methods to show that the spin stiffness 
in the spin conductivity singular part, Eq. (\ref{D-all-T-simpA}), vanishes in the case of
the spin-$1/2$ $XXZ$ chain at zero field for $\Delta >1$ and $T>0$.

On the other hand, the spin-diffusion constant associated with the regular part $\sigma_{{\rm reg}} (\omega,T)$
of the spin conductivity in Eq. (\ref{D-all-T-simpA}) has been expressed for $\Delta >1$ and $T>0$ in 
Eqs. (\ref{DproptoOmega})-(\ref{CalphaT}) 
in terms of the spin elementary currents carried by the spin carriers.
For $\Delta >1$ it was found to be enhanced by lowering the temperature.
It thus reaches its largest yet finite values for very low temperatures. 

We thus derived its leading behavior in that limit, Eq. (\ref{Dlargebeta0}), which only diverges in the $T\rightarrow 0$ limit.
Consistent with nearly ballistic spin transport in the low-temperature regime, but with zero-spin stiffness, 
it was found to be proportional to the inverse spin transport mass of the triplet spin carriers, 
Eq. (\ref{DlargXXX}).

We have also addressed the issue of the spin-diffusion constant enhancement upon lowering the
temperature in the limit of high finite temperatures. 
Its enhancement upon lowering the temperature in the range
$\beta J\in [0,1]$ and its lessening under decreasing anisotropy 
are illustrated in Fig. \ref{figure4PR}(a) and \ref{figure4PR}(b), respectively.

That for $\Delta >1$ the spin-diffusion constant is enhanced 
by lowering the temperature and for very low temperatures reaches it largest yet finite values, 
only diverging in the $T\rightarrow 0$ limit, as given in Eq. (\ref{Dlargebeta0}),
shows that it is finite for all temperatures $T>0$. 

This combined with the vanishing of the spin stiffness at zero field
confirms that the dominant spin transport in the spin-$1/2$ $XXZ$ chain
is for anisotropy $\Delta >1$ normal diffusive for finite temperature $T  > 0$, as for $T\rightarrow\infty$\cite{Znidaric_11,Ljubotina_17,Medenjak_17,Ilievski_18,Gopalakrishnan_19,Weiner_20,Gopalakrishnan_23}. 

The type of spin transport is controlled by the 
thermal expectation value $\langle\vert j_{\pm 1/2}\vert^2\rangle_{T} = \Pi (T)/L$,
as given in Eq. (\ref{noramomD}). Here, $\Pi (T)$ is finite for $\Delta >1$ yet diverges as $\Delta\rightarrow 1$ 
both for low and high temperatures, so that anomalous superdiffusive spin transport at the isotropic point 
$\Delta =1$ is expected to occur at zero field for all finite temperatures.

Provided there is spin $SU(2)$ symmetry or related symmetries with isomorphic irreducible
representations to it, such as the continuous $SU_q(2)$ symmetry,
the physical-spin representation used in our study also applies to other spin models
and to the spin and charge degrees of freedom of electronic models, 
such as the 1D Hubbard model \cite{Carmelo_24,Carmelo_18A}.

In the case of the spin-$1/2$ $XXZ$ chain, its use shows that ballistic spin transport  
is dominant for $0<\Delta <1$ and finite temperature $T>0$, a result that is though
well known \cite{Zotos_99}. On the other hand, it can be used in studies of spin or charge transport
for $T>0$ in other quantum problems whose properties are less understood. 

For instance, the use of our method reveals
that nearly ballistic low-temperature transport, but with zero stiffness,
also occurs in the half-filled 1D Hubbard model in the case of charge transport \cite{Carmelo_24}.

In that model charge transport is actually a more complex quantum problem, as it is normal diffusive for low temperature and
anomalous superdiffusive for $T\rightarrow\infty$ \cite{Carmelo_24}. Our exact low-temperature results
for that model contradict and correct predictions of hydrodynamic theory and KPZ scaling
that charge transport is anomalous superdiffusive for all finite temperatures $T>0$ \cite{Ilievski_18,Fava_20,Moca_23}.

Our results have opened the door to a key advance in the understanding of the spin transport properties
of the spin-$1/2$ $XXZ$ chain with anisotropy $\Delta >1$: They revealed (i) the microscopic processes
that control spin transport in terms of the spin elementary currents carried by the spin carriers in the 
multiplet configuration of all finite-$S_q$ energy eigenstates and (ii) the dominance in the thermodynamic 
limit of normal diffusive spin transport at zero magnetic field for anisotropy $\Delta >1$ and {\it all} finite temperatures $T> 0$.

We have also provided strong evidence that at $h=0$ anomalous superdiffusive spin transport emerges
in the $\Delta\rightarrow 1$ limit for all finite temperatures.

\acknowledgements
We thank Toma\v{z} Prosen and Subir Sachdev for illuminating discussions and the support from Funda\c{c}\~ao para a Ci\^encia e Tecnologia 
through the Grant No. UID/CTM/04540/2019. J. M. P. C. acknowledges support from that Foundation through the Grant 
No. UIDB/04650/2020.\\ \\ 
\appendix

\section{Basic quantities needed for our study}
\label{A}

In the thermodynamic limit, the set of $n=1,...,\infty$ coupled {\it Bethe-ansatz equations}
can be written within a functional representation \cite{Carmelo_22,Carmelo_23}. 
Expressing them in terms of the set of $n$-band momentum functions $q_n (\varphi)$ for $\varphi\in [-\pi,\pi]$
and corresponding $n$-band rapidity-variable distributions $\tilde{N}_n (\varphi)$
that describe each HWS, we find
\begin{eqnarray}
&& q_n (\varphi) = 2\arctan \left(\coth \left({n\eta\over 2}\right)\tan\left({\varphi\over 2}\right)\right)
\nonumber \\
&& - {1\over 2\pi}\int_{-\pi}^{\pi}d\varphi'\,\tilde{N}_n (\varphi')\,2\pi\sigma_n (\varphi')
\nonumber \\
&& \times \Bigl[2\arctan \left(\coth (n\eta)\tan\left({\varphi-\varphi'\over 2}\right)\right)
\nonumber \\
&& + \sum_{l=1}^{n-1}4\arctan \left(\coth (l\eta)\tan\left({\varphi-\varphi'\over 2}\right)\right)\Bigr]
\nonumber \\
&& - {1\over 2\pi}\sum_{n' \neq\,n}\int_{-\pi}^{\pi}d\varphi'\,\tilde{N}_{n'} (\varphi')\,2\pi\sigma_{n'} (\varphi')
\nonumber \\
&& \times \Bigl[2\arctan \left(\coth \left({(n+n')\eta\over 2}\right)\tan\left({\varphi-\varphi'\over 2}\right)\right)
\nonumber \\
&& + 2\arctan \left(\coth \left({\vert n-n'\vert\eta\over 2}\right)\tan\left({\varphi-\varphi'\over 2}\right)\right) 
\nonumber \\
&& + \sum_{l=1}^{{n+n' - \vert n-n'\vert\over 2} -1}
\nonumber \\
&& \times
4\arctan\left(\coth\left({(\vert n-n'\vert + 2l)\eta\over 2}\right)\tan\left({\varphi-\varphi'\over 2}\right)\right)\Bigr] \, .
\nonumber \\
\label{BAqn}
\end{eqnarray}
The $n$-band rapidity-variable distributions $\tilde{N}_n (\varphi)$ appearing in this equation
are uniquely defined by the corresponding $n$-band momentum distributions as
\begin{equation}
\tilde{N}_n (\varphi)\vert_{\varphi = \varphi_n (q)} = N_n (q) \, .
\label{tildeNN}
\end{equation}

The functions $2\pi\sigma_{n} (\varphi) = d q_n (\varphi)/d\varphi$ also appearing in Eq. (\ref{BAqn}) are 
the Jacobians of the transformations from $n$-band momentum values $q$ to $n$-band 
rapidity variables $\varphi$. They are solutions of the following integral equations
obtained from the derivative $d q_n (\varphi)/d\varphi$ of Eq. (\ref{BAqn}):
\begin{eqnarray}
&& 2\pi\sigma_n (\varphi) = {d q_n (\varphi)\over d\varphi} = {\sinh (n\,\eta)\over \cosh (n\eta) - \cos\varphi} 
\nonumber \\
&& + \int_{-\pi}^{\pi}d\varphi^{\prime}\,\sum_{n'=1}^{\infty}\tilde{N}_{n'} (\varphi^{\prime})\,
G_{n\,n'} (\varphi - \varphi^{\prime})\,2\pi\sigma_{n'} (\varphi^{\prime}) 
\label{sigmanderivative}
\end{eqnarray}
whose kernels read
\begin{eqnarray}
G_{n\,n} (\varphi) & = & - {1\over{2\pi}}\Bigl({\sinh (2n\eta)\over \cosh (2n\eta) - \cos\varphi} 
\nonumber \\
& + & 2\sum_{l=1}^{n-1}{\sinh (2l\eta)\over \cosh (2l\eta) - \cos\varphi}\Bigr) \, ,
\label{Gnn}
\end{eqnarray} 
for $n=n'$ and
\begin{eqnarray}
&& G_{n\,n'} (\varphi) = - {1\over{2\pi}}\Bigl({\sinh ((n+n')\eta)\over \cosh (n+n')\eta) - \cos\varphi} 
\nonumber \\
&& + {\sinh (\vert n-n'\vert\eta)\over \cosh (\vert n-n'\vert\eta) - \cos\varphi} 
\nonumber \\
&& + 2\sum_{l=1}^{{n+n' - \vert n-n'\vert\over 2} -1}{\sinh ((\vert n-n'\vert + 2l)\eta)\over \cosh ((\vert n-n'\vert + 2l)\eta) - \cos\varphi}\Bigr) \, ,
\nonumber \\
\label{Gnnp}
\end{eqnarray} 
for $n\neq n'$. 

The functions $2\pi\sigma_{n} (\varphi)$ obey the sum-rules,
\begin{equation}
{1\over 2\pi}\int_{-\pi}^{\pi}d\varphi 2\pi\sigma_n (\varphi) = {N_n\over N} \, .
\label{sumNn}
\end{equation}

In order to study the behavior of $q_n (\varphi)$, Eq. (\ref{BAqn}), 
in the $\Delta\rightarrow 1$ and $\eta\rightarrow 0$ limit,
we find that the following quantities appearing in the expressions provided in 
Eqs. (\ref{Gnn}) and (\ref{Gnnp}) are in that limit given by
\begin{eqnarray}
&& \lim_{b\eta\rightarrow 0}{\sinh (b\,\eta)\over \cosh (b\,\eta) - \cos\varphi} =
\lim_{b\eta\rightarrow 0}{\sinh (b\,\eta)\over \cosh (b\,\eta) - \cos (\Lambda\eta)}
\nonumber \\
&& = \lim_{b\eta\rightarrow 0}{2b\,\eta\over (b^2 + \Lambda^2)\,\eta^2} = 
\lim_{b\eta\rightarrow 0}{2b\,\eta\over (b\,\eta)^2 + \varphi^2} = 2\pi\delta (\varphi) \, ,
\nonumber \\
\label{deltaG}
\end{eqnarray}
where $b$ has in Eqs. (\ref{Gnn}) and (\ref{Gnnp}) the values
$b = n,2n,2l,(n+n'),\vert n - n'\vert$ and we used as intermediate variable that 
usually used for the $\Delta = 1$ spin-$1/2$ $XXX$ chain,
$\Lambda = \varphi/\eta\in [-\infty,\infty]$. 

The use of Eq.  (\ref{deltaG}) in Eqs. (\ref{BAqn}), (\ref{sigmanderivative}), (\ref{Gnn}), and (\ref{Gnnp})
accounting for the $\Delta\rightarrow 1$ boundary
condition, $q_n (\pi) = - q_n (-\pi)$, gives the following limiting behavior
of $q_n (\varphi)$ for $\Delta\rightarrow 1$,
\begin{eqnarray}
q_n (\varphi) & = & 2\pi\Theta (\varphi) - \sum_{n'=1}^{\bar{n}}c_{n,n'} \int_{-\pi}^{\pi}d\varphi'\,\tilde{N}_{n'} (\varphi')
\nonumber \\
& \times & \Theta (\varphi-\varphi')\,2\pi\sigma_{n'} (\varphi')\hspace{0.20cm}{\rm where}
\nonumber \\
c_{n,n'} & = & (n + n' - \vert n - n'\vert - \delta_{n',n})\hspace{0.20cm}{\rm and}
\nonumber \\
\Theta (\varphi) & = & \theta (\varphi) - {1\over 2}
\hspace{0.20cm}{\rm for}\hspace{0.20cm} \Delta \rightarrow 1 \, .
\label{qnDelta1}
\end{eqnarray}
Here $\theta (\varphi)\in [0,1]$ is the Heaviside step function.

In the opposite $\Delta\rightarrow\infty$ limit we find
\begin{equation}
q_n (\varphi) = \varphi {L_n\over N} + q_n^{\Delta}
\hspace{0.20cm}{\rm for}\hspace{0.20cm} \Delta \rightarrow \infty \, ,
\label{qnDeltainfty}
\end{equation}
where $L_ n$ is given in Eq. (\ref{LnNnh}) and
\begin{eqnarray}
q_n^{\Delta} & = & 0\hspace{0.20cm}{\rm for}\hspace{0.20cm}\Delta\rightarrow 1
\nonumber \\
& = & \sum_{n' =1}^{\infty} {n+n' - \vert n-n'\vert\over 2\pi}
\int_{-\pi}^{\pi}\,d\varphi\tilde{N}_{n'} (\varphi)\,2\pi\sigma_{n'} (\varphi)\,\varphi
\nonumber \\
& - & {1\over 2\pi}\int_{-\pi}^{\pi}\,d\varphi\tilde{N}_n (\varphi)\,2\pi\sigma_n (\varphi)\,\varphi 
\hspace{0.20cm}{\rm for}\hspace{0.20cm}\Delta\rightarrow\infty \, .
\label{qDelta}
\end{eqnarray}
Hence $q_n^{\Delta}$ is in Eq. (\ref{qnDeltainfty}) given by the latter expression.

Next, to perform the $n$-string $l$-summation in Eq. (\ref{Energy0}), we
start by replacing it by a $\iota$-summation such that
\begin{eqnarray}
&& \sum_{l=1}^{n}{1\over \cosh\eta - \cos\varphi_{n,l} (q_j)} = 2\sum_{\iota=1}^{(n-\iota_n)/2}
\nonumber \\
&& \times (\cosh\eta - \cosh ((n+1-2\iota)\eta)\cos\varphi_{n} (q_j))
\nonumber \\
&& \times \{(\cosh\eta - \cosh ((n+1-2\iota)\eta)\cos\varphi_{n} (q_j))^2 
\nonumber \\
&& + (\sinh ((n+1-2\iota)\eta)\sin\varphi_{n} (q_j))^{2}\}^{-1}
\nonumber \\
&& + {\iota_n\over \cosh\eta - \cos\varphi_{n} (q_j)} \, ,
\label{equalspectra}
\end{eqnarray}
where $\iota_n = 0$ for $n$ even, $\iota_n = 1$ for $n$ odd, and
the rapidity function $\varphi_{n} (q_j)$ is the real part of the 
complex rapidity, Eq. (\ref{LambdaIm}). The $\iota$-summation then gives
\begin{equation}
\sum_{l=1}^{n}{1\over \cosh\eta - \cos\varphi_{n,l} (q_j)} =
{\sinh^{-1}(\eta)\,\sinh (n\,\eta)\over \cosh (n\,\eta) - \cos\varphi_{n} (q_j)} \, ,
\label{equalspectrafinal}
\end{equation}
for both $n$ even and $n$ odd, where $\sinh^{-1}(\eta) = 1/\sinh\eta$.
The use of this result in Eq. (\ref{Energy0}) leads 
to the simpler expression for the energy eigenvalues given in Eq. (\ref{Energy}).

The calculation of the quantities in the expression of the $T=0$ spin stiffness for anisotropy
$\Delta>1$, Eq. (\ref{Dexpression}), involve those of ground states for spin densities $m\in [0,1]$ and 
corresponding low-energy subspaces. 

For the corresponding energy eigenstates that span such subspaces 
we have that $q_1^{\Delta}=0$ in Eq. (\ref{qqq}) for $n=1$
and the $1$-band momentum values belong to the interval $q \in [-k_{F\uparrow},k_{F\uparrow}]$ 
associated with that of the ground-state rapidity function, $\varphi = \varphi_{1}(q)\in [-\pi,\pi]$. Its
inverse function $q_1 (\varphi)$ is the solution of the Bethe-ansatz equation, Eq. (\ref{BAqn})
with $\tilde{N}_1 (\varphi)=1$ for $\varphi\in [-B,B]$, $\tilde{N}_1 (\varphi)=0$ 
for $\vert\varphi\vert \in [B,\pi]$, and $\tilde{N}_n (\varphi)=0$ for $n>1$. Here
$\pm B = \varphi_{1} (\pm k_{F\downarrow})$ is such that $\lim_{m\rightarrow 0} B = \pi$
and $\lim_{m\rightarrow 1} B = 0$ for $\Delta > 1$. 

The corresponding $T=0$ critical magnetic fields $h_{c1}$ and $h_{c2}$ that define the interval $h\in [h_{c1},h_{c2}]$ 
associated with spin densities $m \in ]0,1]$ of the spin conducting quantum phase for which
the $T=0$ spin stiffness is plotted in Fig. \ref{figure3PR} read
\begin{equation}
h_{c1} = {2J \sqrt{1 - u_{\eta}^2}\over\pi\,g\mu_B \sinh \eta\,K (u_{\eta})}
\hspace{0.20cm}{\rm and}\hspace{0.20cm}
h_{c2} = {J (\Delta +1 )\over g\mu_B} \, ,
\label{criticalfields}
\end{equation}
respectively. Here $K (u_{\eta})$ is the complete elliptic integral and the $\eta$-dependence of the parameter 
$u_{\eta}$ is defined by the relation $\eta = \pi K' (u_{\eta}) /K (u_{\eta})$
where $K' (u_{\eta}) = K \left(\sqrt{1 - u_{\eta}^2}\right)$.
Their limiting behaviors are $K (u_{\eta}) = {\pi^2\over 2}{1\over\eta}$ and 
$K' (u_{\eta})  = {\pi\over 2}$ for $\eta\ll 1$ and
$K (u_{\eta}) = {\pi\over 2}\left(1 + 4\,e^{-\eta}\right)$ and
$K' (u_{\eta})  = {\eta\over 2}\left(1 + 4\,e^{-\eta}\right)$ for $\eta\gg 1$
where $u_{\eta} = 1 - 2\,e^{-{\pi^2\over\eta}}$ for $\eta\ll 1$ and
$u_{\eta} = 4\,e^{-\eta/2}$ for $\eta\gg 1$.

The $1$-band energy dispersion $\varepsilon_{1} (q)$ associated with the $1$-band
group velocity in Eq. (\ref{Dexpression}) and appearing in the spectrum, Eq. (\ref{deltaEk20}),
for zero field is for the whole field interval $h\in [0,h_{c2}]$ given by \cite{Carmelo_22,Carmelo_23}
\begin{eqnarray}
\varepsilon_{1} (q) & = & - {J\over\pi}\sinh \eta\,K (u_{\eta})\,\sqrt{1 - u_{\eta}^2\sin^2 q} + {1\over 2}\, g\mu_B\,h
\nonumber \\
&& \hspace{1.66cm}{\rm for}\hspace{0.20cm}h\in [0,h_{c1}]
\nonumber \\
\varepsilon_{1} (q) & = & {\bar{\varepsilon}_{1}} (\varphi_1 (q))\hspace{0.20cm}{\rm for}\hspace{0.20cm}h\in [h_{c1},h_{c2}]
\hspace{0.20cm}{\rm where}
\nonumber \\
&& {\bar{\varepsilon}_{1}} (\varphi) = {\bar{\varepsilon}_{1}^0} (\varphi) + g\mu_B\,h \hspace{0.20cm}
{\rm and}
\nonumber \\
\varepsilon_{1} (q) & = & J (1 - \cos q)+ J(1-m)\sin q 
\nonumber \\
& \times & \arctan\left(\coth\eta\tanh\left({\eta\over 2}\right)\tan\left({q\over 2}\right)\right)
\nonumber \\
&& {\rm for}\hspace{0.20cm}(1-m)\ll 1\hspace{0.20cm}{\rm and}\hspace{0.20cm}{(h_{c2}-h)\over (h_{c2}-h_{c1})}\ll 1 \, .
\nonumber \\
\label{equA4n}
\end{eqnarray}
The $1$-band rapidity-variable-dependent 
energy dispersion ${\bar{\varepsilon}_{1}^0} (\varphi)$ in this equation
is for $\Delta >1$ and magnetic fields $h\in [h_{c1},h_{c2}]$ defined by the equation
\begin{eqnarray}
&& {\bar{\varepsilon}_{1}^0} (\varphi) = \int_{0}^{\varphi }d\varphi^{\prime}2J\gamma_{1} (\varphi^{\prime}) + A_1^{0}
\hspace{0.20cm}{\rm where}
\nonumber \\
&& A_1^{0} = - J(1 + \cosh \eta) 
\nonumber \\
&& + {1\over\pi}\int_{-B}^{B}d\varphi^{\prime}\,2J\gamma_{1} (\varphi^{\prime})
\arctan\left(\coth \eta\tan\left({\varphi^{\prime}\over 2}\right)\right) \, .
\nonumber \\
\label{equA4n10}
\end{eqnarray}
Here the distribution $2J\gamma_{1} (\varphi)$ obeys the integral equation,
\begin{eqnarray}
2J\gamma_{1} (\varphi) & = & J\,{\sinh \eta\,\sinh (\eta)\sin (\varphi)\over (\cosh (\eta) - \cos (\varphi))^2} 
\nonumber \\
& + & \int_{-B}^{B}d\varphi^{\prime}\,G_{1\,1} (\varphi - \varphi^{\prime})\,2J\gamma_{1} (\varphi^{\prime})  \, .
\label{equA6}
\end{eqnarray}
Its kernel is given in Eq. (\ref{Gnn}) for $n=1$.
(See Ref. \onlinecite{Carmelo_20}, for corresponding $1$-band related quantities expressions for $\Delta =1$.) 

The $1$-band group velocity that appears in the expression of the $T=0$ spin stiffness, Eq. (\ref{Dexpression}),
is given by
\begin{equation}
v_1 (q) = {\partial\varepsilon_1 (q)\over\partial q} \, ,
\label{v1q}
\end{equation}
where the energy dispersion $\varepsilon_{1} (q)$ is defined in Eq. (\ref{equA4n}).

For $\Delta \geq 1$, $m=0$, and $0\leq h\leq h_{c1}$, where $h_{c1}=0$ for $\Delta =1$, it reads
\begin{eqnarray}
v_{1} (q) & = & J{\pi\over 2} \sin q \hspace{0.20cm}{\rm for}\hspace{0.20cm}\Delta = 1
\nonumber \\
v_{1} (q) & = & {J\over 2\pi}\sinh \eta\,K (u_{\eta}){u_{\eta}^2\sin 2q\over \sqrt{1 - u_{\eta}^2\sin^2 q}} 
\hspace{0.20cm}{\rm for}\hspace{0.20cm}\Delta > 1 \, .
\nonumber \\
\label{v1qm0}
\end{eqnarray}

In the opposite limit of $(1-m)\ll 1$ and for magnetic field $h$ values such that $(h_{c2}-h)/(h_{c2}-h_{c1})\ll 1$ it is 
for $\Delta \geq 1$ given by
\begin{eqnarray}
v_{1} (q) & = & J\sin q + J(1-m)\cos q 
\nonumber \\
& \times & \arctan\left(\coth \eta\tanh\left({\eta\over 2}\right)\tan\left({q\over 2}\right)\right)
+ {J\over 2}(1-m)
\nonumber \\
& \times & {\sinh (2\eta)\sinh\eta\,\sin q\over 
\cosh (2\eta)\cosh\eta - 1 + (\cosh (2\eta) - \cosh\eta)\cos q} \, .
\nonumber \\
\label{v1qm1}
\end{eqnarray}
At $\Delta =1$ and thus $\eta =0$, this gives $v_{1} (q) = J\sin q$.

The parameter $\xi$ also appearing in the $T=0$ spin-stiffness expression, Eq. (\ref{Dexpression}),
can be expressed as
\begin{equation}
\xi = {2\pi + 2\pi\Phi_{1,1} (k_{F\downarrow},k_{F\downarrow}) - 2\pi\Phi_{1,1}(k_{F\downarrow},-k_{F\downarrow})\over 2\pi} \, .
\label{x-aaPM}
\end{equation}
It has values in the intervals $\xi \in [1/2,1]$ for $\Delta >1$ and $\xi \in [1/\sqrt{2},1]$ at $\Delta = 1$. 
The corresponding limiting values are $\xi = 1/2$ for $\Delta > 1$ and $\xi = 1/\sqrt{2}$ for 
$\Delta =1$ in the $m\rightarrow 0$ limit and $\xi = 1$ for $\Delta \geq 1$ and $m\rightarrow 1$.

The quantity $2\pi\Phi_{1,1}(q,q')$ in Eq. (\ref{x-aaPM}) is the $1$-pair phase shift given by
\begin{equation}
2\pi\Phi_{1,1}(q,q') = 2\pi\,\bar{\Phi }_{1,1} \left(\varphi_1 (q),\varphi_1 (q')\right) \, .
\label{Phi-barPhi}
\end{equation}
The related rapidity-variable dependent phase shift $2\pi\bar{\Phi }_{1,1} \left(\varphi,\varphi'\right)$ is for 
$\Delta >1$ defined by the following integral equation:
\begin{eqnarray}
2\pi\bar{\Phi }_{1,1} \left(\varphi,\varphi'\right) & = & 
2\arctan\left(\coth \eta\tan\left({\varphi - \varphi'\over 2}\right)\right)
\nonumber  \\
& + & \int_{-B}^{B} d\varphi''\,G_{1\,1} (\varphi - \varphi'')\,2\pi\bar{\Phi }_{1,1} \left(\varphi'',\varphi'\right) \, ,
\nonumber \\
\label{Phis1n}
\end{eqnarray}
where the kernel is again given in Eq. (\ref{Gnn}) for $n=1$.
(See Ref. \onlinecite{Carmelo_20}, for corresponding phase-shift expression for $\Delta =1$.) 

\section{Largest spin elementary currents carried by the spin carriers for $\Delta >1$}
\label{B}

The main goal of this Appendix is the derivation for anisotropy $\Delta = \cosh\eta >1$ 
of a finite upper bound for the absolute value of the spin elementary current
$j_{\pm 1/2} = j_{\pm 1/2} (l_{\rm r}^{\eta},S_q)$, 
Eq. (\ref{elem-currents}), carried by one spin carrier of projection $\pm 1/2$.
That upper bound is given in Eq. (\ref{UB}).

The spin-current expectation value $\langle \hat{J}^z_{HWS} (l_{\rm r}^{\eta},S_q)\rangle$, Eq. (\ref{not-currents}), 
can be obtained from the HWS's energy eigenvalues of
the Hamiltonian, Eq. (\ref{HD1}), in the presence of a vector potential,
$\hat{H}=\hat{H} (\Phi/N)$, as follows:
\begin{equation}
\langle \hat{J}^z_{HWS} (l_{\rm r}^{\eta},S_q) \rangle  = \lim_{\Phi/N\rightarrow 0}
{d E (l_{\rm r}^{\eta},S_q,S_q,\Phi/N)\over d (\Phi/N)} \, .
\label{currentHWS}
\end{equation}
Its derivation for a given finite-$S_q$ energy eigenstate involves accounting for
the interplay of the Bethe-ansatz equations, Eq. (\ref{BAqn}) of Appendix \ref{A}, with the expression 
of the energy eigenvalues $E (l_{\rm r}^{\eta},S_q,S_q,\Phi/N)$, Eq. (\ref{Energy}) 
for $S^z=S_q$. In both such equations the $n$-band momentum values $q_{j}$ 
are to be replaced by $q_j - 2n{\Phi\over N}$. 

As shown in Appendix \ref{C}, the derivation of the spin-current expectation values
is much simplified both for HWSs, Eq. (\ref{currentHWS}), and non-HWSs,
Eq. (\ref{rel-currents-XXZ}), provided that the coupling of the spin carriers 
identified by the physical-spin representation
to the vector potential $\Phi/N$ is explicitly accounted for.

Since the spin elementary currents carried by the unpaired physical spins in
the multiplet configuration of both HWSs and the non-HWSs generated from them, Eq. (\ref{state}), are 
given by, $j_{\pm 1/2} = \pm \langle \hat{J}^z_{HWS}\rangle/(2S_q)$, Eq. (\ref{elem-currents}),
the finite-$S_q$ states considered in this Appendix are HWSs.

In each subspace with fixed numbers for the sets $\{N_n\}$ and $\{N_n^h\}$
of $n$-bands for which $N_n>0$, the different occupancies 
of the $N_n$ $n$-pairs and $N_n^h$ $n$-holes generate different such energy eigenstates.
As reported in Sec. \ref{SECIIA}, the corresponding $j=1,...,L_n$ discrete 
$n$-band momentum values $q_j$ where $L_n = N_n + N_n^h$ obey Pauli-like occupancies.
This implies that the $n$-band momentum distributions 
$N_n (q) = \tilde{N}_n (\varphi_n (q))$ [with $q_j$ replaced by $q$, in the thermodynamic limit]
can only have alternative $N_n (q)=1$ and $N_n (q)=0$ values, respectively.

The functional character of the Bethe-ansatz equations, Eq. (\ref{BAqn}) of Appendix \ref{A}, used in 
the studies of this paper is associated with the $n$-band momentum functions $q_n (\varphi)$ 
having  for $\varphi\in [-\pi,\pi]$ specific values for each choice of the set of 
$n=1,...,\infty$ $n$-band rapidity-variable distributions $\{\tilde{N}_n (\varphi)\}$. 
According to Eq. (\ref{tildeNN}) of Appendix \ref{A}, such values uniquely correspond to those of the
set of $n=1,...,\infty$ $n$-band momentum distributions $\{N_n (q)\}$ that describe each HWS and corresponding 
$q$-spin tower of non-HWSs, Eq. (\ref{state}). 

\subsection{General spin-current expectation value expressions}
\label{1}

As justified in Appendix \ref{C} for the more general case of both HWSs and non-HWSs, the general expression 
for the spin-current expectation value for HWSs in Eq. (\ref{currentHWS}) can be written as
\begin{eqnarray}
&& \langle \hat{J}^z_{HWS} (l_{\rm r}^{\eta},S_q) \rangle = \sum_{n=1}^{\infty}
c_n^{0}\int_{q_n^-}^{q_n^+}dq\,N_n^h (q) J^z_n (q,l_{\rm r}^{\eta},S_q)
\nonumber \\
&& = - {J N \sinh\eta\over\pi}\sum_{n=1}^{\infty}
\int_{-\pi}^{\pi}d\varphi\,c_n^{0}\tilde{N}_n^h (\varphi)
{n \sinh (n\eta)\sin \varphi\over (\cosh (n\eta) - \cos \varphi)^2} \, .
\nonumber \\
\label{jznn}
\end{eqnarray}
In the first expression of this equation,
\begin{equation}
J^z_n (q,l_{\rm r}^{\eta},S_q) = - {J N \sinh\eta\over\pi\,2\pi\sigma_n (\varphi_n (q))}{n \sinh (n\eta)\sin \varphi_n (q)\over
(\cosh (n\eta) - \cos \varphi_n (q))^2} \, ,
\label{Jznq}
\end{equation}
is the $n$-band spin-current spectrum, ${1\over 2\pi\sigma_n (\varphi_n (q))} = {d \varphi_n (q)\over dq}$
where $2\pi\sigma_n (\varphi)$ is the function, Eq. (\ref{sigmanderivative}) of Appendix \ref{A},
$N_n^h (q) = 1 - N_n (q)$ and $\tilde{N}_n^h (\varphi) = 1 - \tilde{N}_n (\varphi)$
are the $n$-hole momentum and $n$-hole rapidity-variable distributions, respectively, and
$c_n^{0} = 2S_q/N_n^h \leq 1$. 

The spin elementary currents $j_{\pm 1/2} = j_{\pm 1/2} (l_{\rm r}^{\eta},S_q)$,
Eq. (\ref{elem-currents}), carried by each of a number $M=2S_q$ of spin carriers in the 
multiplet configuration of all $S_q$-finite energy eigenstates then read
\begin{eqnarray}
&& j_{\pm 1/2} (l_{\rm r}^{\eta},S_q) = \pm {\langle \hat{J}^z_{HWS} (l_{\rm r}^{\eta},S_q) \rangle  \over 2S_q}
= \mp {J \sinh\eta\over\pi}
\nonumber \\
&& \times \sum_{n=1}^{\infty}{N\over N_n^h}\int_{q_n^-}^{q_n^+}dq\,{N_n^h (q)\over 2\pi\sigma_n (\varphi_n (q))} {n \sinh (n\eta)\sin \varphi_n (q)\over
(\cosh (n\eta) - \cos \varphi_n (q))^2} 
\nonumber \\
&& = \mp {J\sinh\eta\over\pi}\sum_{n=1}^{\infty}{N\over N_n^h}
\int_{-\pi}^{\pi}d\varphi\,\tilde{N}_n^h (\varphi)
{n \sinh (n\eta)\sin \varphi\over (\cosh (n\eta) - \cos \varphi)^2} \, ,
\nonumber \\
\label{jznnRAP}
\end{eqnarray}
where we used that $c_n^0 = 2S_q/N_n^h \leq 1$.

For compact $n$-hole rapidity-variable occupancies $\varphi\in [\varphi_{n,1},\varphi_{n,2}]$, this gives
\begin{eqnarray}
&& j_{\pm 1/2} (l_{\rm r}^{\eta},S_q) = \mp {J\sinh\eta\over\pi}\sum_{n=1}^{\infty}{N\over N_n^h}n \sinh (n\eta)
\nonumber \\
&& \times {\cos \varphi_{n,2} - \cos \varphi_{n,1}\over (\cosh (n\eta) - \cos \varphi_{n,2})(\cosh (n\eta) - \cos \varphi_{n,1})} \, .
\nonumber \\
\label{jznnCompact}
\end{eqnarray}

The use of the physical-spin representation in the general spin elementary current absolute value's expression, Eq. (\ref{jznnRAP}),
combined with manipulations of the Bethe-ansatz equations, Eq. (\ref{BAqn}) of Appendix \ref{A}, simplifies
the selection of some classes of $S_q >0$ energy eigenstates whose spin elementary currents have large 
absolute values $\vert j_{\pm 1/2}\vert$. Concerning
their $n$-bands occupancies, the following are found to be useful criteria/properties:

First, full $N_n = L_n$ $n$-bands with occupancies $\varphi\in [\varphi_{n,1},\varphi_{n,2}]$
such that $\varphi_{n,1}=-\pi$ and $\varphi_{n,2}=\pi$ in Eq. (\ref{jznnCompact}),
and empty $N_n^h = L_n$ $n$-bands such that $\varphi_{n,1}=\varphi_{n,2}=0$ in that equation, do not contribute
to the spin elementary current absolute values $\vert j_{\pm 1/2}\vert$.

Second, the dependence on the anisotropy $\Delta = \cosh\eta$ of the absolute values $\vert j_{\pm 1/2}\vert$ 
of the spin elementary currents given in Eq. (\ref{jznnRAP}) is such that in the 
thermodynamic limit all possible values of the $n$-band distributions $N_n^h (q)$ 
and $\tilde{N}_n^h (\varphi)=N_n^h(q_n(\varphi))$ in these general expressions cannot lead to divergences 
in $\vert j_{\pm 1/2}\vert$ when $\Delta >1$. This is consistent with the statement reported in Sec. \ref{SECIIIB}. 
Suitable analysis of the anisotropy-dependence of the 
spin elementary currents $j_{\pm 1/2}$ given in Eq. (\ref{jznnRAP}) shows indeed that this holds for
all state occupancies of these distributions except for $\Delta\rightarrow 1$.

Third, the absolute values $\vert j_{\pm 1/2}\vert$ are maximized by compact
occupancies, Eq. (\ref{jznnCompact}), of the $N_n^h$ $n$-holes of $n$-bands with finite $N_n>0$ occupancy.
In some cases, they are maximized by two separate intervals with compact occupancies, 
$\varphi\in [\varphi_{n,1},\varphi_{n,2}]$ and $\varphi\in [\varphi_{n,3},\varphi_{n,4}]$, 
respectively. However, the finite-$S_q$ states with larger absolute values $\vert j_{\pm 1/2}\vert$ have a single
interval with compact occupancy, Eq. (\ref{jznnCompact}).

Fourth, for each type of finite-$S_q$ states, the precise $n$-band location of the
compact rapidity-variable interval $\varphi\in [\varphi_{n,1},\varphi_{n,2}]$
and corresponding compact momentum interval $q \in [q_n (\varphi_{n,1}),q_n (\varphi_{n,2})]$
of the $n$-holes must be that which further maximizes the absolute values $\vert j_{\pm 1/2}\vert$.

In the following we consider several types of finite-$S_q$ states of the two classes (i) and (ii) of states, as defined in Sec. \ref{SECIIIB}.
To identify $S_q>0$ states whose spin carriers have large spin elementary current absolute values $\vert j_{\pm 1/2}\vert$,
Eq. (\ref{jznnRAP}), in the remaining of this Appendix we consider different types of $n$-band occupancies
and check their degree of contribution to such absolute values.

\subsection{The class (i) $q_1^*$-states}
\label{2}

The class (i) $q_1^*$-states considered in Sec. \ref{SECIIIC} and defined in the following are the states with 
largest absolute value for the spin elementary current $j_{\pm 1/2}$, Eq. (\ref{jznnRAP}), of a type of
states whose $1$-band is populated by two $1$-holes. Such states can be generated from 
zero-field ground states whose $1$-band has no holes. The two $1$-holes emerge at two 
$1$-band momentum values $q$ and $q'$ as a result of one $1$-pair breaking. Under it, 
the singlet $1$-pair gives rise to two unpaired physical spins.

The two $1$-holes describe the translational degrees of freedom of such two unpaired physical spins,
which are the spin carriers. When that process generates $S^z = \pm 1$ triplet states, 
it involves a spin flip of one of the emerging unpaired physical spins.
This is the case of the triplet pairs found in Sec. \ref{SECVB} to control for $\Delta >1$ the 
spin-diffusion constant at low temperatures, Eq. (\ref{Dlargebeta0}).

For completeness, we consider here four types of states: Three triplet states and one 
singlet state, whose $1$-band is populated by two $1$-holes.
The three triplet states have numbers $N_1 = N/2 -1$, $N_1^h=2$, $L_1 = N/2 + 1$,
$M = 2S_q = 2$, $2S^z = -2,0,2$, and $N_n =0$ for $n>1$.
The singlet state has numbers $N_1 = N/2 -2$, $N_1^h=2$, $L_1 = N/2$,
$M = 2S_q = 0$, $2S^z = 0$, $N_2 = 1$, $N_2^h=0$, and $N_n =0$ for $n>2$.

In the thermodynamic limit, the energy spectra of the $S^z =1$ triplet HWS state 
of more interest for our study and of the singlet state
are degenerate, their spin-current expectation values being given by
\begin{equation}
\langle \hat{J}^z (q,q')\rangle = - c_1^0\,2(v_1 (q) + v_1 (q')) \, ,
\label{J2states}
\end{equation}
where $v_1 (q)$ is the $1$-band group velocity given in Eq. (\ref{v1qm0}) of Appendix \ref{A} for $\Delta >1$
and $c_1^0=1$ and $c_1^0=0$ for the triplet and singlet states, respectively.
Indeed, the spin-current expectation value of the latter state vanishes. 
For the $S^z =-1$ and $S^z = 0$ triplet states, it reads 
$\langle \hat{J}^z (q,q')\rangle$ with $c_1^0 -1$ and $0$, respectively. 

Expression, Eq. (\ref{J2states}), is exactly the same for the corresponding triplet and singlet states 
at anisotropy $\Delta =1$. The procedures involving the combined use of Eq. (\ref{currentHWS})
and the Bethe-ansatz equations to obtain it are in this case similar to those used in 
Ref. \onlinecite{Zhang_07} for $\Delta =1$. 

In the case of HWS triplet states of interest for the problems addressed in
Secs. \ref{SECIIIC} and \ref{SECVB}, the expression of the absolute 
value $\vert j_{\pm 1/2} (q,q')\vert$ of the corresponding spin elementary current 
can be written as
\begin{eqnarray}
&& \vert j_{\pm 1/2} (q,q')\vert = {J\over 2\pi}\sinh \eta\,K (u_{\eta})\,u_{\eta}^2
\nonumber \\
&& \times \left({\sin 2q\over\sqrt{1 - u_{\eta}^2\sin^2 q}} + {\sin 2q'\over\sqrt{1 - u_{\eta}^2\sin^2 q'}}\right) \, .
\label{jznnqqqq}
\end{eqnarray}
Here $K (u_{\eta})$ and $u_{\eta}$ are in Eq. (\ref{J2states}) the same quantities as in 
Eqs. (\ref{criticalfields}), (\ref{equA4n}), and (\ref{v1qm0}) of Appendix \ref{A}. 
To obtain the expression, Eq. (\ref{jznnqqqq}), we used that given in Eq. (\ref{v1qm0}) of Appendix \ref{A} for $\Delta >1$
of the $1$-band group velocity $v_1 (q)$ appearing in the expression, Eq. (\ref{J2states}), for the related 
spin-current expectation values.

The spin-diffusion constant studied for very low temperature in Sec. \ref{SECVB} is controlled by
excited states populated by triplet pairs whose two unpaired physical spins carry 
spin elementary currents of this form with $q\approx \pm\pi/2$ and $q' = q \mp 2\pi/N$, Eq. (\ref{jznntoqeq}).
\begin{figure*}
\includegraphics[width=0.495\textwidth]{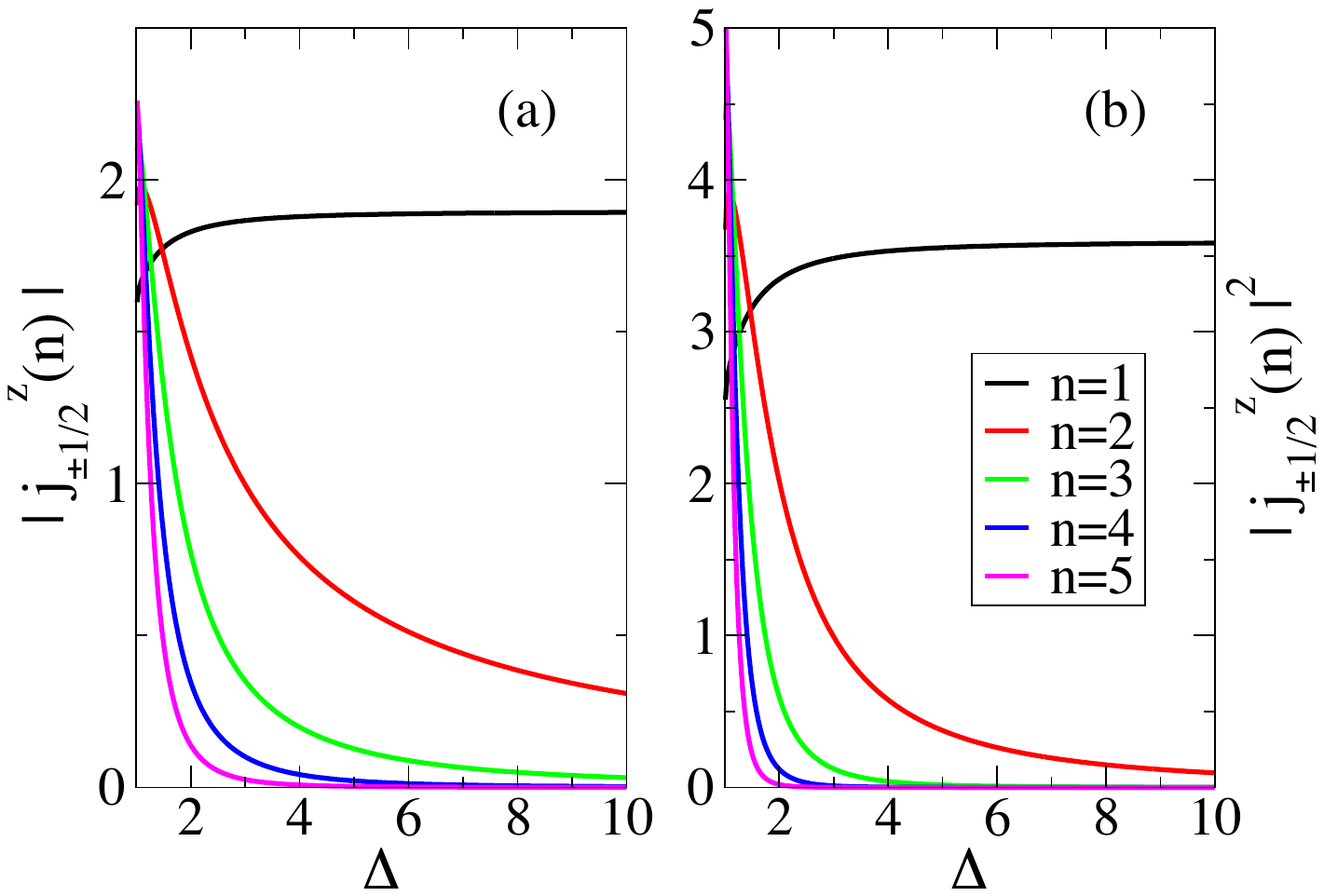}
\includegraphics[width=0.495\textwidth]{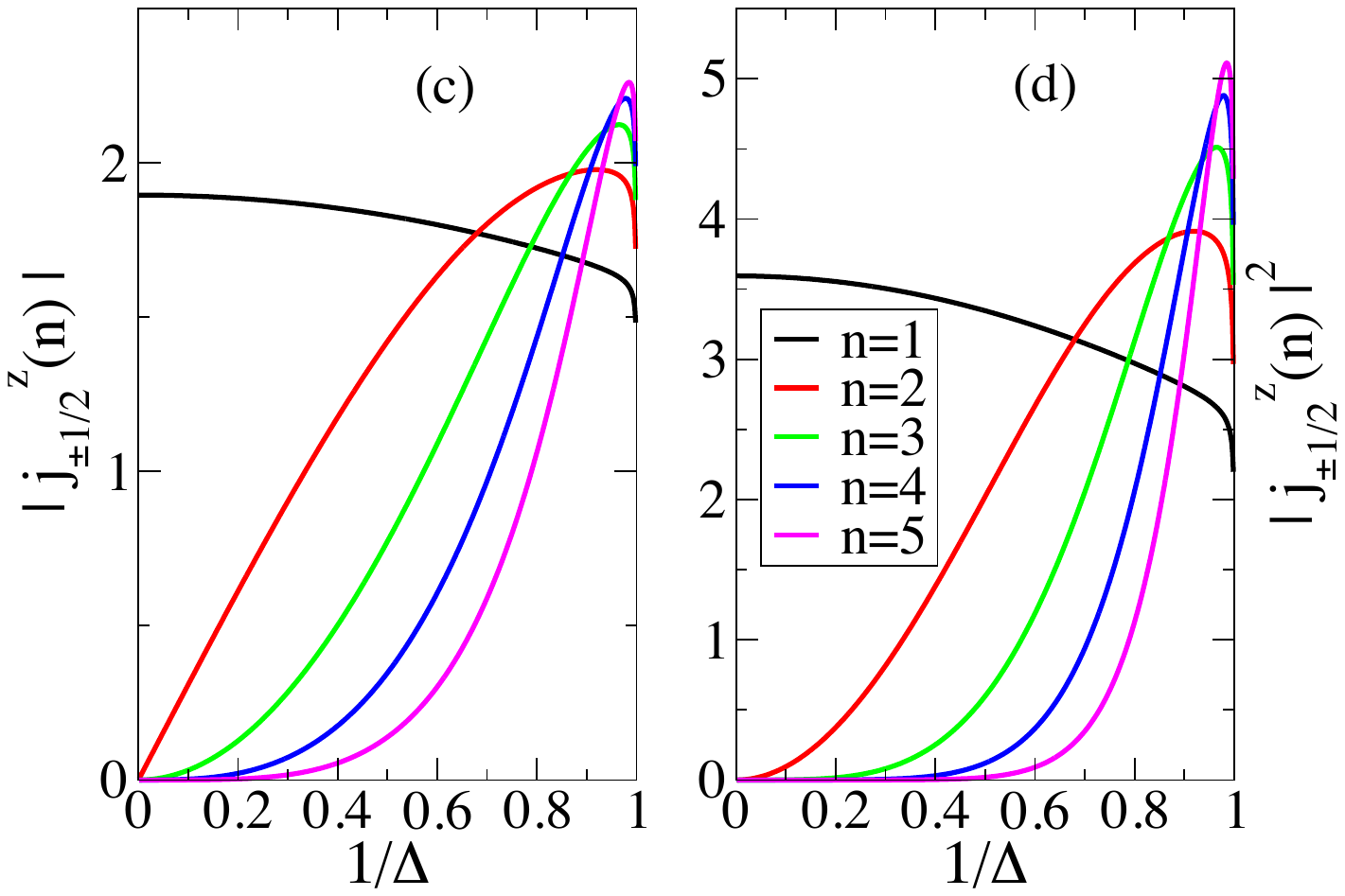}
\caption{The absolute values $\vert j_{\pm 1/2}(n)\vert$ of the spin elementary current carried by one spin 
carrier for $n$-states, Eq. (\ref{jznHBsimple}), with $n=1,2,3,4,5$ (a) and its square (b) as a function of anisotropy $\Delta$ and
(c) and (d) as a function of $1/\Delta$, respectively.}
\label{figure5PR}
\end{figure*}

On the other hand, the $q_1^*$-states considered here are another specific type of 
general $2S_q=2$ states for which $q=q_1^*$ and $q'=q_1^*-2\pi/N$. 
These two momentum values with $q_1^*$ given by
\begin{equation}
q_1^* = {\pi\over 4} + {1\over 2}\arcsin \left({2 (1 - \sqrt{1 - u_{\eta}^2})\over u_{\eta}^2} -1\right) \in \left[{\pi\over 4},{\pi\over 2}\right] \, ,
\label{q1*}
\end{equation}
maximize the absolute value in Eq. (\ref{jznnqqqq}). Consistent with the limiting behaviors 
of $u_{\eta}$ given in Appendix \ref{A}, we find that $q_1^*$ smoothly and 
continuously varies from $q_1^* = {\pi\over 2}$ for $\Delta\rightarrow 1$ to 
$q_1^* = {\pi\over 4}$ for $\Delta\rightarrow\infty$.

From the use of $q=q_1^*$ and $q'=q_1^* - 2\pi/N$ in the expression, Eq. (\ref{jznnqqqq}), with 
$q_1^*$ given in Eq. (\ref{q1*}), and accounting for the opposite sign of the spin elementary current
carried by the spin carriers of projection $+1/2$ and $-1/2$, respectively, 
we find that for $q_1^*$-states it is given by
\begin{equation}
j_{\pm 1/2}(q_1^*) = \mp {2J\over\pi}\sinh \eta\,K (u_{\eta})\left(1 - \sqrt{1 - u_{\eta}^2}\right) \, .
\label{jzq1*}
\end{equation}
Its absolute value and square are plotted in Fig. \ref{figure1PR} as a function of $\Delta$ (a), (b) and $1/\Delta$ (c), (d).
The finite spin-current expectation value $\langle\hat{J}^z_{HWS} (q_1^*)\rangle$ of $q_1^*$-states then reads
$\langle \hat{J}^z_{HWS} (q_1^*)\rangle = - 2\vert j_{\pm 1/2}(q_1^*)\vert$ for $2S_q=2$.

\subsection{Class (ii) states}
\label{3}

In the following we consider finite-$S_q$ states that are class (ii) states either for the whole
anisotropy interval $\Delta \in ]1,\infty]$ or for it except for $\Delta\gg 1$. 
We are particularly interested in class (ii) states with finite occupancy up to a given $n$-band 
whose integer $n$ we call $\bar{n}$, so that $N_n=0$ for $n>\bar{n}$. 

If in addition such states have compact $n$-hole rapidity-variable occupancies 
$\varphi\in [\varphi_{n,1},\varphi_{n,2}]$ for $n\leq \bar{n}$, as those in Eq. (\ref{jznnCompact}), the 
absolute value $\vert j_{\pm 1/2}\vert$ is for small $\eta \ll 1/\bar{n}$ given by
\begin{eqnarray}
\vert j_{\pm 1/2} (l_{\rm r}^{\eta},S_q)\vert & = & {\eta^2 J\over\pi}\sum_{n=1}^{\bar{n}}{N\over N_n^h}n^2
\nonumber \\
& \times & {\cos \varphi_{n,2} - \cos \varphi_{n,1}\over (1 - \cos \varphi_{n,2})(1 - \cos \varphi_{n,1})} \, .
\label{jznnCoeta0}
\end{eqnarray}

This applies to arbitrarily small $\eta$ even if $\bar{n}$ is very large provided that $\eta\ll1/\bar{n}$.
The $n=1,...,\bar{n}$ ratios $N/N_n^h$ in Eq. (\ref{jznnCoeta0}) obey the inequalities $N/N_n^h\leq N/2S_q$ 
and thus $N_n^h/N\geq 2S_q/N$. The equalities refer here to $N_{\bar{n}}^h=2S_q$. 
For such class (ii) states, the concentration $M/N=2S_q/N\leq N_n^h/N$ of spin carriers 
is finite, so that the absolute value $\vert j_{\pm 1/2}\vert$, Eq. (\ref{jznnCoeta0}), vanishes for
$\eta\rightarrow 0$. 

Concerning $n$-band occupancies of finite-$S_q$ states, we reported above four criteria/properties.
A fifth criterion is related to the absolute value $\vert j_{\pm 1/2}\vert$, Eq. (\ref{jznnCoeta0}),
vanishing for $\eta\rightarrow 0$ {\it not applying} when the compact $n$-hole rapidity-variable occupancies 
$\varphi\in [\varphi_{n,1},\varphi_{n,2}]$ are such that $\varphi_{n,1}=0$ and $\varphi_{n,2}=\pi$ or 
$\varphi_{n,1}=-\pi$ and $\varphi_{n,2}=0$. In that case, the absolute value $\vert j_{\pm 1/2} \vert$
of the spin elementary current, Eq. (\ref{jznnCompact}),
is finite for $\eta\rightarrow 0$ and thus $\Delta\rightarrow 1$, as confirmed in the following.

More generally, such a fifth criterion refers to $n$-bands for which $N_n^h/N$ and $N_n/N$ are finite for $N\rightarrow\infty$
and whose $n$-hole compact rapidity-variable distribution reads $\tilde{N}_{n}^h (\varphi) = 1$ (or 
$\tilde{N}_{n}^h (\varphi) = 0$) for $\varphi\in [-\pi,\varphi_h]$
and $\tilde{N}_{n}^h (\varphi) = 0$ [or $\tilde{N}_{n}^h (\varphi) = 1$] for $\varphi\in [\varphi_h,\pi]$.
This corresponds to $n$-hole rapidity-variable occupancies $\varphi\in [\varphi_{n,1},\varphi_{n,2}]$
such that $\varphi_{n,1}=-\pi$ and $\varphi_{n,2}=\varphi_h$ (or $\varphi_{n,1}=\varphi_h$ and $\varphi_{n,2}=\pi$)
in Eq. (\ref{jznnCompact}).

When $(\pi - \vert\varphi_h\vert)/\pi\ll 1$, the location of the $n$-hole compact 
rapidity-variable interval that maximizes the contribution to $\vert j_{\pm 1/2}\vert$ may
not have $-\pi$ as starting rapidity variable (or $\pi$ as ending rapidity variable) and
the following does not apply. If otherwise, the contribution from these $n$-bands to the absolute 
value $\vert j_{\pm 1/2}\vert$ is largest for small $(\Delta - 1)>0$ when $\varphi_h = 0$. 

In the case of class (ii) states, we have checked that such a largest contribution 
for small $(\Delta - 1)>0$ is indeed reached by a $\varphi_h = 0$ half-filled band in terms of the $n$-band 
rapidity-variable $\varphi \in [-\pi,\pi]$ and {\it not} of the
corresponding $n$-band momentum $q \in [q_n^-,q_n^+]$. 

Such two choices of half-filled occupancies
refer indeed to different states: We find that for states with half-filled $n$-band rapidity-variable occupancy 
one has for $\Delta >1$ that $N_n^h > N_n$
or $N_n^h < N_n$, the equality $N_n^h = N_n$ being reached only
in the $\Delta\rightarrow\infty$ limit. 
(For the specific states considered below one has that $N_n^h > N_n$.)

This can be confirmed by the use in the Bethe-ansatz equations, Eq. (\ref{BAqn}) of Appendix \ref{A}, of 
half-filled $n$-band rapidity-variable distributions. Since $q_n^{\Delta}\neq 0$ for $\Delta >1$ in Eq. (\ref{qqq}), 
it is often convenient to use shifted $n$-band momentum values that refer
to symmetrical intervals, $(q - q_n^{\Delta}) \in [-{\pi\over N}(L_n -1),{\pi\over N}(L_n -1)]$.

We find that the vanishing value $\varphi = 0$ of the above 
half-filled $n$-band rapidity-variable occupancies corresponds for $\Delta >1$ to 
$(q - q_n^{\Delta}) =q_n^0$. Here $q_n^0 = q_n^0 (\eta)$ is a finite $n$-band separation momentum $q_n^0= {\pi\over N}(N_n^h - N_n)$.
Indeed, $(q -  q_n^{\Delta}) = q_n^0$ separates intervals of non-occupied (or occupied) and
occupied (or non-occupied) shifted $n$-band momentum values $(q - q_n^{\Delta})$, respectively. The exception is for $\Delta\rightarrow\infty$, when 
$\varphi = 0$ refers to $q_n^0 = (q - q_n^{\Delta}) = 0$ and $N_n^h = N_n$ for such states. 

The numbers $N_n^h$ and $N_n$ and thus
the $n$-band separation momentum values $q_n^0 = q_n^0 (\eta)$ are found to depend 
on the anisotropy, the latter having maximum values for $\Delta\rightarrow 1$ and vanishing 
for $\Delta\rightarrow\infty$. 

On the one hand, Eq. (\ref{jznnCoeta0}) does not apply to class (i) states as for instance
the $q_1^*$-states whose absolute value $\vert j_{\pm 1/2}\vert$
can be finite in the $\eta\rightarrow 0$ limit yet {\it never} diverges in that limit.
\begin{figure*}
\includegraphics[width=0.495\textwidth]{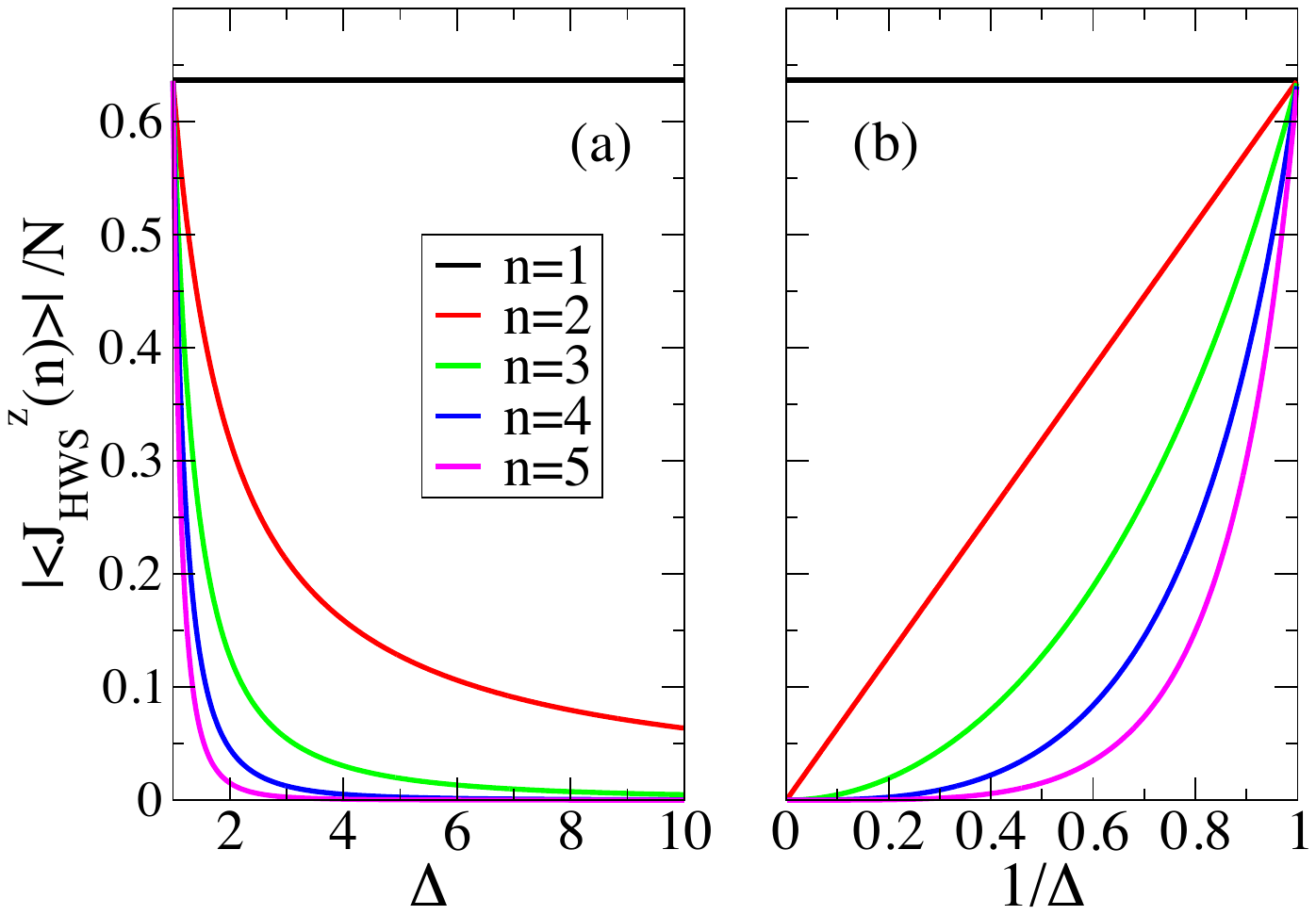}
\includegraphics[width=0.495\textwidth]{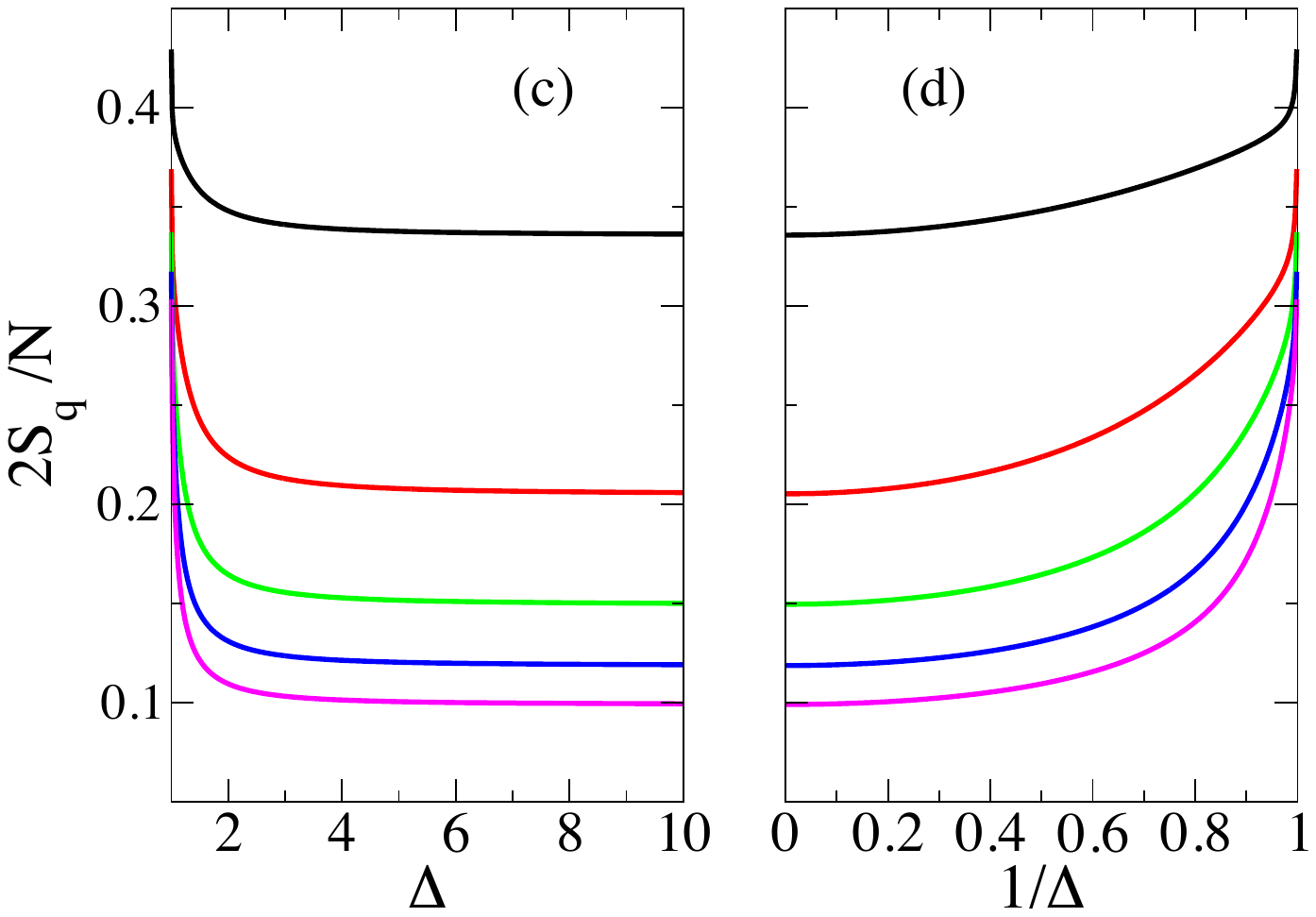}
\caption{The ratio $\vert\langle \hat{J}^z_{HWS}(n)\rangle\vert/N$ of the spin-current expectation value
over $N$ for $n$-states with $n=1,2,3,4,5$ as a function of $\Delta$ (a) and $1/\Delta$ (b) and the corresponding 
concentration of spin carriers $2S_q/N$ as a function of $\Delta$ (c) and $1/\Delta$ (d). Both such
quantities are given in Eq. (\ref{2Sqnst}).}
\label{figure6PR}
\end{figure*}

On the other hand, finite-$S_q$ states of class (ii) with at least some of their occupied $n$-bands having 
compact $n$-hole rapidity-variable occupancies $\varphi\in [\varphi_{n,1},\varphi_{n,2}]$
such that $\varphi\in [-\pi,0]$ or $\varphi\in [0,\pi]$ and thus $\varphi_{n,1}=-\pi$ and $\varphi_{n,2}=0$ or $\varphi_{n,1}=0$ 
and $\varphi_{n,2}=\pi$ in Eq. (\ref{jznnCompact}), are the only ones that in the $\eta\rightarrow 0$ limit
have finite and often very large absolute values $\vert j_{\pm 1/2}\vert$. 
These $n$-hole occupancies correspond to $n$-band rapidity-variable compact occupancies 
for $\varphi\in [0,\pi]$ or $\varphi\in [-\pi,0]$ respectively.

For simplicity, in the following we choose that the latter are $\tilde{N}_{n} (\varphi) = 0$ for $\varphi\in [-\pi,0]$ and 
$\tilde{N}_{n} (\varphi) = 1$ for $\varphi\in [0,\pi]$. This corresponds to $n$-band-momentum compact occupancies
$N_{n} (q) = 0$ for $(q-q_n^{\Delta}) \in [-{\pi\over N}(L_n -1),q_n^0]$ and 
$N_{n} (q) = 1$ for $(q-q_n^{\Delta}) \in [q_n^0,{\pi\over N}(L_n -1)]$. 

\subsection{$n$-states}
\label{4}

We have first considered the simplest of such finite-$S_q$ states of class (ii), which we call $n$-states. They have 
$n$-band rapidity variable occupancy $\tilde{N}_{n} (\varphi) = 0$ for $\varphi\in [-\pi,0]$ and $\tilde{N}_{n} (\varphi) = 1$ 
for $\varphi\in [0,\pi]$ in a single $n$-band and otherwise zero occupancies, $N_{n'}=0$ for $n'\neq n$.
For such states, Eq. (\ref{jznnRAP}) gives
\begin{eqnarray}
j_{\pm 1/2} (n) & = & \mp {J\sinh\eta\over\pi}{N\over N_n^h}
\int_{-\pi}^{0}d\varphi\,{n \sinh (n\eta)\sin \varphi\over (\cosh (n\eta) - \cos \varphi)^2} 
\nonumber \\
& = & \pm {2J\over\pi}{n\,\sinh\eta\over\sinh (n\eta)}{N\over N_n^h} \, .
\label{jznnRAPn}
\end{eqnarray}
Calculation of the ratio $N/N_n^h$ by use in Eq. (\ref{BAqn}) of Appendix \ref{A} 
of the distribution $\tilde{N}_{n} (\varphi)$ for these states and accounting
for the relation $N = 2S_q + 2n N_n$, gives the absolute values
\begin{equation}
\vert j_{\pm 1/2} (n)\vert = {(2n +1) 2J\over (\pi + 2n\,q_n^0)}{n\sinh\eta\over\sinh (n\eta)}  \, .
\nonumber \\
\label{jznHBsimple}
\end{equation}
The $n$-band separation momentum $q_n^0$ in this expression is given by $q_n^0 = {\pi\over N}(2S_q - N_n)$
where the values of both $2S_q$ and $N_n$ depend on $\Delta$. In the $\eta\rightarrow 0$ limit the absolute value,
Eq. (\ref{jznHBsimple}), is finite, in contrast to that given in Eq. (\ref{jznnCoeta0}), which vanishes for $\eta\rightarrow 0$.

In Fig. \ref{figure5PR}, the absolute value $\vert j_{\pm 1/2} (n)\vert$, Eq. (\ref{jznHBsimple}),
of the spin elementary current carried by one spin carrier
(a) and its square (b) are plotted for $n=1,2,3,4,5$ as a 
function of $\Delta$ and (c) and (d) of $1/\Delta$, respectively.
The largest absolute value is reached by $n=1$ $n$-states.

The concentration of spin carriers and the spin-current expectation value are 
for $n$-states given by
\begin{eqnarray}
{M\over N} = {2S_q\over N} & = & {(\pi + 2n\,q_n^0)\over \pi (2n +1)}
\hspace{0.20cm}{\rm and}
\nonumber \\
\vert\langle \hat{J}^z_{HWS}(n)\rangle\vert & = &
{2J\over\pi}{n\,\sinh\eta\over\sinh (n\eta)}N \, ,
\label{2Sqnst}
\end{eqnarray}
respectively. Consistent with $n$-states being class (ii) states, both 
the concentration $M/N=2S_q/N$ and the ratio $\vert\langle \hat{J}^z_{HWS}(n)\rangle\vert/N$ are finite
for $N\rightarrow\infty$. 

In Fig. \ref{figure6PR}, the ratio $\vert\langle \hat{J}^z_{HWS}(n)\rangle\vert/N$ (a) and (b) 
and the concentration of spin carriers $M/N=2S_q/N$ (c) and (d) are plotted  as a function of 
$\Delta$ (a), (c) and $1/\Delta$ (b), (d), respectively. Analysis of the curves plotted in the figure 
shows that both $\vert\langle \hat{J}^z_{HWS}(n)\rangle\vert/N$ and $M/N=2S_q/N$ are largest for $n=1$.

\subsection{$\bar{n}$-states and $\bar{n}_*$-states}
\label{5}

We have found that some finite-$S_q$ states with half-filled rapidity-variable occupancy 
in {\it several} $n$-bands and vanishing occupancy $N_{n'}=0$ in the remaining $n'$-bands
have larger absolute values $\vert j_{\pm 1/2}\vert$ than the $n$-states.

We considered several orders for the set of $n$-bands with such a rapidity-variable occupancy.
We then found that among the states with several $n$-bands with half-filled rapidity-variable occupancy,
those that have largest absolute values $\vert j_{\pm 1/2}\vert$ belong to a type of states that 
we call $\bar{n}$-states. 

Such $\bar{n}$-states have finite $n$-band rapidity-variable occupancy 
$\tilde{N}_{n} (\varphi) = 0$ for $\varphi\in [-\pi,0]$ and $\tilde{N}_{n} (\varphi) = 1$ for $\varphi\in [0,\pi]$ 
in a set of successive $n=1,2,3,...,\bar{n}$ bands starting at $n=1$ and ending at $n=\bar{n}$
and zero $n'$-band occupancy, $N_{n'}=0$, for $n'>\bar{n}$. 
(Note that $n$-states for $n=1$ and $\bar{n}$-states for $\bar{n}=1$ are exactly the same states.)

$\bar{n}$-states are class (ii) states whose concentration $M/N=2S_q/N$ 
of spin carriers is finite for $N\rightarrow\infty$. Only for $\bar{n}\rightarrow\infty$
and $\Delta\rightarrow\infty$ where $\bar{n}/\bar{n}_*\leq 1$ they become class (i) states such that $M/N=2S_q/N = 0$ 
for $N\rightarrow\infty$.

The $n$-band momentum occupancies of the ${\bar{n}}$-states that correspond
to their above $n$-band rapidity-variable occupancies are for $n=1,...,\bar{n}$ given by
$N_{n} (q) = 0$ for $(q-q_n^{\Delta}) \in [-{\pi\over N}(L_n -1),q_n^0[$ and 
$N_{n} (q) = 1$ for $(q-q_n^{\Delta}) \in [q_n^0,{\pi\over N}(L_n -1)]$ where
$q_n^0 = q_n^0 (\eta) = {\pi\over N}(N_n^h - N_n)$. As justified below, here $q_n^0\rightarrow 0$
for $\eta\rightarrow\infty$ and $q_1^0\rightarrow\pi/4$ and $q_n^0\rightarrow\pi/2$ 
for $n>1$ in the opposite $\eta\rightarrow 0$ limit.

The dependence on the anisotropy of the $\bar{n}$-states's $n$-band numbers $N_n = N_n (\eta)$ and $N_n^h = N_n^h (\eta)$ 
and corresponding $n$-band separation momentum $q_n^0 = q_n^0 (\eta)$ where $n=1,...,\bar{n}$ is fully determined by 
the Bethe-ansatz equations, Eq. (\ref{BAqn}) of Appendix \ref{A}. This refers to using in these
equations the above $n$-band rapidity-variable distributions $\tilde{N}_{n} (\varphi)$ for $n=1,...,\bar{n}$ specific to
$\bar{n}$-states. Solution of the equations provides the corresponding needed dependence on 
$\eta$ and $\Delta $ of the numbers $N_n^h$ and $N_n$ in $q_n^0 = {\pi\over N}(N_n^h - N_n)$.

Such a procedure leads for $\bar{n}$-states to the following expressions
for the concentration $M/N=2S_q/N$ of spin carriers and corresponding ratios $N_n^h/N$ and $N_n/N$:
\begin{eqnarray}
{2S_q\over N} & = & {M \over N} = {N_{\bar{n}}^h\over N} = {e^{-\bar{n}\ln (2+\sqrt{3})}\over c_{\bar{n}}}
\Bigl(1 + \sum_{n=1}^{\bar{n}}2n {Q_{n}^0\over\pi}\Bigr)
\nonumber \\
{N_n^h\over N} & = & c_{\bar{n} -n}\,e^{(\bar{n} -n)\ln (2+\sqrt{3})}\,{2S_q\over N} - {Q_{n}^0-q_n^0\over\pi}
\hspace{0.20cm}{\rm and}
\nonumber \\
{N_n\over N} & = & {N_n^h\over N} - {q_{n}^0\over\pi} \hspace{0.20cm}{\rm for}\hspace{0.20cm}
n = 1,...,\bar{n} \, .
\label{2Sq}
\end{eqnarray}

The $n$-dependent coefficients $c_n$ that appear here for $n$ given by $\bar{n}$ and $\bar{n} -n$ and the
quantities $Q_n^0$ are defined in the following. The former coefficients are related to quantities $x_n$ 
that obey the following recursive relation whose use provides their general expression for $n = 1,...,\infty$,
\begin{eqnarray}
x_n & = & 1 + \sum_{n'=1}^{n-1}2 (n-n') x_{n'} = \prod_{n'=1}^{n-1} A_{n-n'} 
\nonumber \\
& = &  \prod_{n'=1}^{n-1} A_{n'} = c_{n - 1}\,e^{(n - 1)\ln (2+\sqrt{3})} \, ,
\label{xn}
\end{eqnarray}
where $x_1 = 1$. 

The coefficients $c_n$ read $c_0 = 1$ for $n=0$ and for $n>0$ involve the ratios
$A_n = x_{n+1}/x_n$ as follows:
\begin{eqnarray}
&& c_ 0 = 1 \hspace{0.20cm}{\rm and}\hspace{0.20cm}c_n = {x_{n + 1}\over (2+\sqrt{3})^n} =
\prod_{n'=1}^n \left({A_{n'}\over 2+\sqrt{3}}\right)
\hspace{0.20cm}{\rm where}
\nonumber \\
&& A_{n} = {x_{n+1}\over x_n} 
\hspace{0.20cm}{\rm such}\hspace{0.20cm}{\rm that}\hspace{0.20cm}
A_{n+1} - A_n = {2\over x_{n+1}\,x_n} \, .
\label{Anxnnj}
\end{eqnarray}

Here, we used that $e^{-n\ln (2+\sqrt{3})} = (2+\sqrt{3})^{-n}$.
The ratios $A_n = x_{n+1}/x_n$ slightly increase upon increasing $n$, with limiting 
values $A_1=3$ and $\lim_{n\rightarrow\infty}A_{n} = 2+\sqrt{3}$. For $n>0$ the values of the coefficients $c_n$ also slightly increase 
upon increasing $n$. They are given in Table \ref{table1} for $n=0,1,...,11$. For $n=\infty$ they read $c_{\infty} = 0.78867513459481...$.

The set of coefficients $c_n$ defined in Eq. (\ref{Anxnnj}) obey for $n =1,...,\bar{n}$ the following exact sum-rule,
\begin{equation}
\sum_{n=1}^{\bar{n}}2n\,c_{\bar{n}-n}\,e^{-n\ln (2+\sqrt{3})} = c_{\bar{n}}  - e^{-\bar{n}\ln (2+\sqrt{3})} \, .
\label{sumrulecn}
\end{equation}
It is obtained by combining $N = 2S_q + \sum_{n=1}^{\bar{n}}2n\,N_n$ with the expressions for $2S_q/N$
and  $N_n/N$ given in Eq. (\ref{2Sq}). 

The expression of the quantities $Q_n^0$ in Eq. (\ref{2Sq}) involves sums of $n$-band separation momentum 
values $q_n^0$ for $n=1,...,\bar{n}$. They also obey a recursive relation,
\begin{equation}
Q_n^0 = q_n^0 + \sum_{n'=n+1}^{\bar{n}}2 (n'-n) Q_{n'}^0
\hspace{0.20cm}{\rm for}\hspace{0.20cm}
n = 1,...,\bar{n} \, ,
\label{f0n}
\end{equation}
under the boundary condition $Q_{\bar{n}}^0 = q_{\bar{n}}^0$. 

The dependence for $n=1,...,\bar{n}$ of the numbers $N_n^h$ and $N_n$ 
on the anisotropy defined by Eq. (\ref{2Sq}) implies that
$\bar{n}$-states are different energy eigenstates for each $\Delta$ value.
For such states the use of Eq. (\ref{jznnRAP}) leads to the following expression
for the spin elementary currents carried by the spin carriers:
\begin{eqnarray}
j_{\pm 1/2} (\bar{n}) = & \mp & {J\sinh\eta\over\pi}\sum_{n=1}^{\bar{n}}{N\over N_n^h}
\int_{-\pi}^{0}d\varphi\,{n \sinh (n\eta)\sin \varphi\over (\cosh (n\eta) - \cos \varphi)^2} 
\nonumber \\
= & \pm & {2J\over\pi}\sum_{n=1}^{\bar{n}}{n\,\sinh\eta\over\sinh (n\eta)}{N\over N_n^h} \, .
\nonumber \\
\label{jznnRAPbarn}
\end{eqnarray}
The corresponding HWS spin-current expectation value is given by
$\langle \hat{J}^z_{HWS}(\bar{n})\rangle = 2S_q\,j_{+1/2} (\bar{n})$.
\begin{table}
\begin{center}
\begin{tabular}{|c|c|c|c|c|c|c|c|} 
\hline
$c_0$ & $c_1$ & $c_2$ & $c_3$ & $c_4$ & $c_5$ \\
\hline
$1$ & ${3\over 2+\sqrt{3}}$ & ${11\over (2+\sqrt{3})^2}$ & ${41\over  (2+\sqrt{3})^3}$ & ${153\over  (2+\sqrt{3})^4}$ & ${571\over  (2+\sqrt{3})^5}$ \\
\hline
$1$ & $0.803848$ & $0.789764$ & $0.788753$ & $0.788681$ & $0.788676$ \\
\hline
\hline
$c_6$ & $c_7$ & $c_8$ & $c_9$ & $c_{10}$ & $c_{11}$ \\
\hline
${2131\over (2+\sqrt{3})^6}$ & ${7953\over  (2+\sqrt{3})^7}$ & ${29681\over  (2+\sqrt{3})^8}$ & ${110771\over  (2+\sqrt{3})^9}$ & ${413403\over  (2+\sqrt{3})^{10}}$ & ${1542841\over  (2+\sqrt{3})^{11}}$ \\
\hline
$0.788675$ & $0.788675$ & $0.788675$ & $0.788675$ & $0.788675$ & $0.788675$ \\
\hline
\end{tabular}
\caption{The coefficients $c_n$, Eq. (\ref{Anxnnj}), for $n=0,1,...,11$. The first and second rows give their
exact and approximate values, respectively. For $n=\infty$ they read $c_{\infty} = 0.78867513459481...$.}
\label{table1}
\end{center}
\end{table}

Valuable information on general $\bar{n}$-states and $\bar{n}_*$-states is contained in the separation momentum values 
$q_1^0,q_{2}^0,...,q_{\bar{n}}^0$. The use of Eqs. (\ref{qnDelta1}) and (\ref{qnDeltainfty}) of Appendix \ref{A} leads for 
these states to the following limiting behaviors of the function $q_n (\varphi)$, Eq. (\ref{BAqn}) of Appendix \ref{A}, 
\begin{eqnarray}
q_n (\varphi) & = & \varphi {L_n\over N} + {\pi\over 2}\left(1 - {L_n\over N}\right)
\hspace{0.20cm}{\rm for}\hspace{0.20cm}
\Delta \rightarrow\infty\hspace{0.20cm}{\rm and}
\nonumber \\
& = & 2\pi\Theta (\varphi) - \sum_{n'=1}^{\bar{n}}c_{n,n'} \int_0^{\pi}d\varphi'
\Theta (\varphi-\varphi')\,2\pi\sigma_{n'} (\varphi')
\nonumber \\
&& \hspace{0.20cm}{\rm for}\hspace{0.20cm} \Delta \rightarrow 1 \, ,
\label{qnlimits}
\end{eqnarray}
where $\Theta (\varphi)$ and $c_{n,n'}$ are given in Eq. (\ref{qnDelta1}) of that Appendix. 

For $\Delta  \rightarrow \infty$ we thus have that $q_{n}^{\Delta} = {\pi\over 2}\left(1 - {L_n\over N}\right)$ 
where $L_n/N = {2c_{\bar{n} -n}\over c_{\bar{n}}}(2+\sqrt{3})^{-n}$ and {\it all} $n = 1,...,{\bar{n}}$
separation momentum values vanish, $q_{n}^0 = q_{n} (0) - q_{n}^{\Delta} = 0$. 

In the opposite $\Delta  \rightarrow 1$ limit, the use of Eqs. (\ref{LnNnh}), (\ref{MM}), (\ref{sumNn}) of Appendix \ref{A}, 
and (\ref{qnlimits}) gives $q_{n}^{\Delta} \rightarrow 0$, $q_1^{\pm} \rightarrow \pm {3\pi\over 4}$, 
$q_n^{\pm} \rightarrow \pm {\pi\over 2}$, $q_{1}^0\rightarrow \pi/4$, and $q_{n}^0\rightarrow \pi/2$ where $n = 2,...,\bar{n}$,
so that the $n$-band separation momentum values $q_{2}^0,...,q_{\bar{n}}^0$
have the same value. We then find that the relation $q_n^0= 2q_1^0$ for $n=2,...,\bar{n}$
is for finite $\Delta >1$ compatible with the $2S_q/N$'s expression, Eq. (\ref{2Sq}).

If the separation momentum values $q_1^0,q_{2}^0,...,q_{\bar{n}}^0$ had the same 
$\Delta$ dependent value for $\Delta \in ]1,\infty]$, the related quantities $Q_n^0$, Eq. (\ref{f0n}), would 
read $Q_n^0 = \tilde{Q}_n^0 = c_{\bar{n} -n}\,(2+\sqrt{3})^{\bar{n} -n}\,q_{{\bar{n}}}^0$. 

Accounting for the corrections due to the $\bar{n}>1$ boundary condition $q_n^0= 2q_1^0$ for $n=2,...,\bar{n}$,
which refers both to the $\Delta\rightarrow 1$ and $\Delta\rightarrow\infty$ limits, imposes that the quantities $Q_n^0$ are such that
\begin{eqnarray}
1 & + & \sum_{n=1}^{\bar{n}} 2n {Q_n^0\over \pi} = 1 + \sum_{n=1}^{\bar{n}} 2n {\tilde{Q}_n^0\over \pi}
- (1-\delta_{\bar{n},1}){q_{\bar{n}}^0\over\pi}
\nonumber \\
& = & \Bigl(1 - {\gamma_{\bar{n}}\,q_{\bar{n}}^0\over\pi}\Bigr)
+ {q_{{\bar{n}}}^0\over\pi}{N \over 2S_q^{\infty}}  \, .
\label{sumQ}
\end{eqnarray}
Here $2S_q^{\infty}/N = \lim_{\Delta \rightarrow\infty}2S_q/N$ and $\gamma_{\bar{n}}$ read
\begin{eqnarray}
{2S_q^{\infty}\over N} & = & {1\over c_{\bar{n}}(2+\sqrt{3})^{\bar{n}}} \in \left[0,{1\over 3}\right]
\hspace{0.20cm}{\rm and}\hspace{0.20cm}
\gamma_{\bar{n}} = 2 - \delta_{\bar{n},1}
\nonumber \\
{\rm where} && \bar{n} = {\ln\left({N\over c_{\bar{n}}\,2S_q^{\infty}}\right)
\over \ln (2+\sqrt{3})}\hspace{0.20cm}{\rm for}\hspace{0.20cm}\Delta > 1 \, ,
\label{2Sinfgaban}
\end{eqnarray}
and $\bar{n}$ is the integer number closest to the number given by the latter expression.
For $\bar{n}$-states the $\eta$-spin value limit $S_q^{\infty} = \lim_{\Delta \rightarrow\infty}S_q$ only 
depends on that number.

The relation $q_n^0= 2q_1^0$ for $n=2,...,\bar{n}$, which is exact in the $\Delta\rightarrow 1$ and $\Delta\rightarrow\infty$ limits, 
is exact or a good approximation for $\Delta \in ]1,\infty]$. 
Importantly, its use leads to exact results in
the case of $\bar{n}_*$-states for which  $\bar{n} = \bar{n}_*\rightarrow\infty$, Eq. (\ref{n*}), as in that case
the dominant processes stem from $\bar{n}$-large bands that obey such a relation. 

We have then used the relation $q_n^0= 2q_1^0$ where $n=2,...,\bar{n}$ for $\Delta \in ]1,\infty]$ and
performed the summations $\sum_{n=1}^{\bar{n}}$ in Eq. (\ref{sumQ}) by use of the 
exact sum-rule, Eq. (\ref{sumrulecn}). Combining the corresponding obtained result with the expressions in Eq. (\ref{2Sq}) leads to the 
following expressions for the concentration of spin carriers $2S_q/N$ and ratio $N/N_n^h$:
\begin{eqnarray}
{2S_q (\eta)\over N} & = & 
{q_{\bar{n}}^0 (\eta)\over\pi} + \Bigl(1 - {\gamma_{\bar{n}}\,q_{\bar{n}}^0(\eta)\over\pi}\Bigr){2S_q^{\infty}\over N}
\hspace{0.20cm}{\rm and}
\nonumber \\
{N\over N_n^h (\eta)} & = & {1\over {q_{{\bar{n}}}^0(\eta)\over\pi} + 
\Bigl(1 - {\gamma_{\bar{n}}\,q_{\bar{n}}^0(\eta)\over\pi}\Bigr){c_{\bar{n} -n}\over c_{\bar{n}}(2+\sqrt{3})^n}} \, ,
\label{setEq}
\end{eqnarray}
respectively. 

To obtain this expression of the ratio $N/N_n^h (\eta)$ for $n \neq \bar{n}$ and $\bar{n}>1$, the above corrections 
associated with the relation $q_n^0= 2q_1^0$ for $n=2,...,\bar{n}$  
impose that $Q_n^0$ is replaced by $\tilde{Q}_n^0$. Indeed, the coefficient $\gamma_{\bar{n}} = 2$ in that ratio expression already 
accounts for such corrections.

The $2S_q/N$'s expression in Eq. (\ref{setEq}) is equivalent to the following related expressions:
\begin{equation}
{q_{\bar{n}}^0(\eta)\over\pi} = {{2S_q (\eta) - 2S_q^{\infty}\over N}\over 
1 - \gamma_{\bar{n}}\,{2S_q^{\infty}\over N}}
\hspace{0.20cm}{\rm and}\hspace{0.20cm}
{2S_q^{\infty}\over N} = {{2S_q (\eta) \over N} - {q_{\bar{n}}^0(\eta)\over\pi}\over
1 - {\gamma_{\bar{n}}\,q_{\bar{n}}^0(\eta)\over\pi}} \, .
\label{q2Sinf}
\end{equation}
The first expression shows that the anisotropy dependence of the $\bar{n}$-band separation
momentum $q_{\bar{n}}^0$ is fully controlled by that of the spin concentration $2S_q/N$.
The second expression reveals that in spite of $q_{\bar{n}}^0$ and $2S_q/N$
depending on the anisotropy, the quantity on its right-hand side does not.

The use in Eq. (\ref{jznnRAPbarn}) of the expression of $N/N_n^h$ provided in Eq. (\ref{setEq}) gives
the following general expression for the spin-elementary current absolute value $\vert j_{\pm 1/2} (\bar{n})\vert$:
\begin{eqnarray}
\vert j_{\pm 1/2}(\bar{n})\vert & = & \sum_{n=1}^{\bar{n}}
{2nJ \sinh\eta\over \sinh (n\eta)
\Bigl(q_{{\bar{n}}}^0(\eta) + {c_{\bar{n} -n}\over c_{\bar{n}}}{(\pi - \gamma_{\bar{n}}\,q_{\bar{n}}^0(\eta))\over (2+\sqrt{3})^{n}}\Bigr)}
\nonumber \\
& = & {2J\over\pi}{c_{\bar{n}}(2+\sqrt{3})\over c_{\bar{n}-1}} 
\hspace{0.20cm}{\rm for}\hspace{0.20cm}\Delta\rightarrow\infty
\nonumber \\
& = & {4J\over\pi}(
\bar{n} - 1 + \delta_{\bar{n},1})
\hspace{0.20cm}{\rm for}\hspace{0.20cm}\Delta\rightarrow 1 \, .
\label{jzlarger}
\end{eqnarray}

Using $\lim_{\eta\rightarrow 0}N_n^h/N = 1/2$ and $\lim_{\eta\rightarrow 0}{n\,\sinh\eta\over\sinh (n\eta)} = 1$ in 
Eq. (\ref{jznnRAPbarn}) seems to give $\vert j_{\pm 1/2}(\bar{n})\vert = \bar{n}{4J\over\pi}$ in 
the $\Delta\rightarrow 1$ limit. Why as given in Eq. (\ref{jzlarger}) for $\bar{n}>1$ the result rather is 
$\vert j_{\pm 1/2}(\bar{n})\vert = (\bar{n} - 1) {4J\over\pi}$  
is an issue that deserves clarification. 

To obtain it we used that
${q_n^0\over\pi} = {2q_1^0\over\pi} = {1\over 2} - {2\pi\over N}{\cal{N}}$ 
for small $(\Delta-1)\approx \eta^2/2$ where ${\cal{N}}/N\ll 1$ and
${\cal{N}}/N\rightarrow 0$ for $\eta\rightarrow 0$.
For $\bar{n}>1$ and $n\geq 1$ the number $N_n = N_n^h - {q_n^0\over\pi}N$
of $n$-pairs then reads
\begin{eqnarray}
N_n & = & 2S_q - \Bigl({N\over 2} - 2\pi{\cal{N}}\Bigr)\Bigl(1 - {\delta_{n,1}\over 2}\Bigr) + {\cal{N}}_n
\hspace{0.20cm} {\rm where}
\nonumber \\
{\cal{N}}_n & = & 4\pi (c_{\bar{n}-n}-1){{\cal{N}}\over N} 2S_q^{\infty} 
\hspace{0.20cm}{\rm for}\hspace{0.20cm}\bar{n} > 1 \, .
\label{Nnlarger1}
\end{eqnarray}

For very small, yet finite, $(\Delta-1)\approx \eta^2/2$, we have that $2S_q = N/2 - 2\pi{\cal{N}}$. 
Since $c_0=1$ (see Table \ref{table1}), Eq. (\ref{Nnlarger1}) gives $N_{\bar{n}} = 0$ and
$N_n = {\cal{N}}_n + ({N\over 4} - \pi{\cal{N}})\delta_{n,1}$ for 
$n = 1,...,\bar{n}-1$ when $\bar{n} >1$. In addition, $N_1 = {N\over 4} - {2\pi {\cal{N}}\over 3}$ for $\bar{n}=1$. 

Empty $n$-bands for which $N_n =0$ do not contribute to spin transport: In the corresponding
$n$-squeezed effective lattices considered in Appendix \ref{C}, spin transport results from processes 
where a number $N_n^h$ of $n$-holes that describe the translational degrees of freedom of the spin carriers 
interchange position with a finite number $N_n$ of $n$-pairs. 

It follows that there is for $\bar{n} >1$ a qualitative difference when 
for arbitrarily small yet finite values of $(\Delta-1)\approx \eta^2/2$
the $n$-pairs numbers $N_n$ remain finite for $n = 1,...,\bar{n}-1$ and vanish at $n = \bar{n}$, respectively.

Indeed, when the $\Delta \rightarrow 1$ limit is approached the $\bar{n} > 1$ $\bar{n}$-band for which $N_{\bar{n}} = 0$
does not contribute to spin transport. Therefore, only the $n$-bands with $n = 1,...,\bar{n}-1$ up to
$n = \bar{n}-1$ contribute to it. The use of $\lim_{\eta\rightarrow 0}N_n^h/N = 1/2$ 
and $\lim_{\eta\rightarrow 0}{n\,\sinh\eta\over\sinh (n\eta)} = 1$ in 
Eq. (\ref{jznnRAPbarn}) only for $n$-bands such that $n = 1,...,\bar{n}-1$
for which $N_n>0$ then leads to $\vert j_{\pm 1/2}(\bar{n})\vert = {4(\bar{n} - 1 + \delta_{\bar{n},1})J\over\pi}$ 
in the $\eta\rightarrow 0$ and $\Delta\rightarrow 1$ limit, as given in Eq. (\ref{jzlarger}).

As shown in Fig. \ref{figure1PR}, the absolute value $\vert j_{\pm 1/2} (\bar{n})\vert$, Eq. (\ref{jzlarger}),
increases upon increasing $\bar{n}$. Our interest is thus in the $\bar{n}_*$-states where $\bar{n}_*$ is
the largest physically allowed value of $\bar{n}$.
The expression of the number $\bar{n}$ in Eq. (\ref{2Sinfgaban}) shows that the physical allowed maximum value
of $\bar{n}$ refers to the minimum finite value of $q$-spin $2S_q^{\infty}/N = \lim_{\Delta \rightarrow\infty}2S_q/N$.
For $N$ and $2S_q$ even, that minimum value is $S_q^{\infty} = 1$. This justifies why $\bar{n}_*$
such that $\bar{n}_* \rightarrow\infty$ for $N \rightarrow\infty$ is the integer number closest to that given
by the expression, Eq. (\ref{n*}). 

The expression given in Eq. (\ref{jzlarger}) for the absolute value $\vert j_{\pm 1/2} (\bar{n})\vert$
is exact for $\bar{n} = \bar{n}_*\rightarrow\infty$, Eq. (\ref{n*}). For the corresponding $\bar{n}_*$-states, the spin carrier
concentration reads $2S_q (\eta)/N = q_{\bar{n}_*}^0 (\eta)/\pi$, so that $\vert j_{\pm 1/2} (\bar{n}_*)\vert$ can be written as 
provided in Eq. (\ref{limitsjz12n*}). 

$\vert j_{\pm 1/2} (\bar{n}_*)\vert$ is the largest absolute value of spin elementary currents of all ${\bar{n}}$-states for 
$\Delta \in ]1,\infty]$. It continuously increases upon decreasing $\Delta$ in that interval.
Therefore, before reaching infinity in the $\Delta  \rightarrow 1$ limit, it acquires finite arbitrarily large values at very small values
of $(\Delta - 1)>0$ that are larger than those of the spin elementary currents carried by the carriers of any other finite-$S_q$ state for $\Delta >1$.
Their spin elementary currents $j_{\pm 1/2}$ have the general form given in Eq. (\ref{jznnRAP}).

\subsection{Final considerations}
\label{6}

Merging all information provided in this Appendix, we arrive to the following final considerations:\\

(a) Finite-$S_q$ states of class (i), such as for instance the $q_1^*$-states, may have finite 
absolute value $\vert j_{\pm 1/2}\vert$ for small $(\Delta -1)$. However, it never diverges for
$\eta\rightarrow 0$.\\

(b) In the case of finite-$S_q$ states of class (ii), only states with a large number 
$N_b = \sum_{n=1}^{\infty}\theta (N_n) \leq \bar{n}_*$ of $n$-bands with
rapidity-variable half-filled occupancy can reach absolute values $\vert j_{\pm 1/2}\vert$ 
that diverge for $\eta\rightarrow 0$. Here $\bar{n}_*\rightarrow\infty$, Eq. (\ref{n*}),
and $\theta (x)$ is the Heaviside step function.\\

(c) Out of the latter states, those whose absolute value $\vert j_{\pm 1/2}\vert$ is larger
for arbitrarily small finite $\eta$ and a given fixed $N_b$ are the $\bar{n}$-states with
$\bar{n} = N_b \rightarrow\infty$ in the thermodynamic limit and $\bar{n}/\bar{n}_*\leq 1$.\\

(d) Finally, out of such $\bar{n}$-states, the largest absolute value $\vert j_{\pm 1/2}\vert$ 
for arbitrarily small finite $\eta$ is reached by the $\bar{n}_*$-states where
$\bar{n}_*\rightarrow\infty$ for $N\rightarrow\infty$, Eq. (\ref{n*}).\\

{\it No} finite-$S_q$ states have absolute values $\vert j_{\pm 1/2}\vert$ that diverge at finite values of $(\Delta-1)$.
As justified in Sec. \ref{SECIIIC}, the absolute value $\vert j_{\pm 1/2}(\bar{n}_*)\vert$, Eq. (\ref{limitsjz12n*}), continuously 
increases upon decreasing $\Delta$ and reads $\vert j_{\pm 1/2}(\bar{n}_*)\vert\rightarrow\infty$ for $\Delta\rightarrow 1$.

We can thus choose the width of an anisotropy interval, $\Delta \in ]1,\Delta_*]$, to be arbitrarily small yet finite and thus 
$\vert j_{\pm 1/2}(\bar{n}_*^{*})\vert = \vert j_{\pm 1/2}(\bar{n}_*)\vert_{\Delta = \Delta_*}$ to be 
arbitrarily large yet no infinity. Our above criteria then show that for $\Delta \in ]1,\Delta_*]$ 
the $\bar{n}_*$-states have the largest absolute value $\vert j_{\pm 1/2}\vert$ of all finite-$S_q$ states. 

In summary, our systematic analysis of the spin elementary currents of 
selected classes of finite-$S_q$ states with large absolute values $\vert j_{\pm 1/2}\vert$ 
has revealed that:\\

(i) For the anisotropy interval $\Delta \in ]1,\Delta_*]$ as defined above the $\bar{n}_*$-states are in 
the thermodynamic limit those with largest spin elementary current absolute value
$\vert j_{\pm 1/2} (\bar{n}_*)\vert$, Eq. (\ref{jzlarger}) for $\bar{n}_*\rightarrow\infty$,
of all finite-$S_q$ states.\\

(ii) Since {\it no} finite-$S_q$ states have absolute values $\vert j_{\pm 1/2}\vert$ that 
diverge at finite values of $(\Delta-1)$, the absolute values $\vert j_{\pm 1/2}\vert$ of all
finite-$S_q$ states obey in the thermodynamic limit the inequality,
\begin{equation}
\vert j_{\pm 1/2}\vert_{\Delta \in [\Delta_*,\infty]} \leq \vert j_{\pm 1/2}(\bar{n}_*^*)\vert \, ,
\label{ineAll}
\end{equation}
where $\vert j_{\pm 1/2}(\bar{n}_*^*)\vert = \vert j_{\pm 1/2}(\bar{n}_*)\vert_{\Delta = \Delta_*}$.

\section{Squeezed space and spin-current expectation values}
\label{C}

The concept of a squeezed effective lattice is well known in 1D correlated systems 
\cite{Ogata_90,Penc_97,Kruis_04}. In the case of the spin-$1/2$ $XXZ$ chain with anisotropy 
$\Delta >1$, it applies to subspaces for which the values of the set of $N_n>0$ numbers $\{N_n\}$ of $n$-pairs is fixed. 
Such $n$-pairs then move in the $n$-squeezed effective lattices with as many sites, $j=1,...,L_n$, 
as discrete $n$-band momentum values $q_j$, $L_n = N_n + N_n^h$, Eq. (\ref{LnNnh}).

Upon moving in the $n$-squeezed effective lattice, the $n$-pairs interchange position with the unpaired 
physical spins in a number $M_{+1/2} + M_{-1/2}=M$ of sites 
and a subclass of paired physical spins in a number ${\cal{M}}_{n,+ 1/2}+{\cal{M}}_{n,-1/2}={\cal{M}}_{n}$ of sites.
Here ${\cal{M}}_{n,\pm 1/2}<{\cal{M}}_{\pm 1/2}$, so that ${\cal{M}}_{n}<{\cal{M}}= \sum_{n=1}^{\infty}2n\,N_n$,
Eq. (\ref{MM}).

Indeed, upon moving in their $n$-squeezed effective lattice the $n$-pairs only ``see'' 
and interchange position with paired physical spins contained in $n'$-pairs such that $n'>n$. And out of the 
number $2n'$ of unpaired physical spins of each such $n'$-pairs, they only ``see'' and interchange position with 
a smaller number $2(n'-n)$ of unpaired physical spins.
[That $n$-pairs contain a number $2n$ of paired physical spins can be shown to be related to they only ``seeing'' a number $2(n'-n)$ 
of paired physical spins out of the $2n'$ sites of each such $n'$-pairs for which $n'>n$.]

Hence, out of the ${\cal{M}}_{\pm 1/2} = \sum_{n=1}^{\infty}n\,N_n$ paired physical spins of projection $\pm 1/2$
that populate an energy eigenstate, Eq. (\ref{MM}), the $n$-pairs only ``see'' and interchange position with a number 
${\cal{M}}_{n,\pm 1/2}<{\cal{M}}_{\pm 1/2}$ of such paired physical spins given by
\begin{eqnarray}
&& {\cal{M}}_{n,\pm 1/2} = \sum_{n'=n+1}^{\infty}(n'-n)N_{n'}\hspace{0.20cm}{\rm so}\hspace{0.20cm}{\rm that}
\nonumber \\
&& {\cal{M}}_{n} = {\cal{M}}_{n,+1/2} + {\cal{M}}_{n,-1/2} = \sum_{n'=n+1}^{\infty}2(n'-n)N_{n'}
\nonumber \\
&& N_n^h = M + {\cal{M}}_{n} \, .
\label{Nnh}
\end{eqnarray}
Indeed, $N_n^h = 2S_q + \sum_{n'=n+1}^{\infty}2(n'-n)N_{n'}$, Eq. (\ref{LnNnh}).

Consistent with the $n$-band discrete momentum values having separation $q_{j+1}-q_j = {2\pi\over N}$,
the $n$-squeezed effective lattices have the same length ($L=N$ in our units) as the original lattice.
In the thermodynamic limit, such lattices spacing corresponds to the average distance between 
its $L_n = N_n + N_n^h$ sites, $a_n = {N\over L_n}\,a > a$. Here, $a$ denotes the original lattice spacing, 
which in the units of this paper is one.

We can either consider that upon moving in their $n$-squeezed effective lattice the $n$-pairs 
interchange position with a number $M + {\cal{M}}_{n}$ of physical spins
or that upon moving in that lattice the latter physical spins interchange position with a number $N_n$ of $n$-pairs. 

The latter description is of interest for spin transport. Indeed, the translational degrees of freedom of
the $M = 2S_q$ unpaired physical spins, which are the spin carriers, are described by the
unoccupied sites of the $n$-squeezed effective lattice whose number is $N_n^h = M + {\cal{M}}_{n}$, Eq. (\ref{Nnh}).

For simplicity, let us first consider energy eigenstates for which the $1$-squeezed effective lattice
has a number $L_1 = M + N_1$ of sites, $N_1^h = M$ of which are empty.
Upon moving in the $1$-squeezed effective lattice, the $M = M_{+1/2} + M_{-1/2}$ unpaired physical spins 
interchange position with the $N_1$ $1$-pairs. The identification of the $M_{+1/2}$ and $M_{-1/2}$  unpaired physical spins 
with projections $+1/2$ and $-1/2$, respectively, is clear in each corresponding $1$-squeezed effective lattice
sites occupancy configuration.

The quantum numbers represented by $l_{\rm r}^{\eta}$ in the
spin-current expectation value $\langle \hat{J}^z (l_{\rm r}^{\eta},S_q,S^z) \rangle$,
Eq. (\ref{rel-currents}), include a number $N_1^h= M_{+1/2} + M_{-1/2}$ of
$1$-band momentum values $q_j$. Those are associated with the number $N_1^h = M$ of $1$-holes that describe
the translational degrees of freedom of the $M = M_{+1/2} + M_{-1/2}$ unpaired physical spins.

The expression of that expectation value in terms of the $1$-squeezed effective lattice
sites occupancy configurations is involved. Consistently, there is quantum uncertainty about which of the number $N_1^h = M$
of $1$-band momentum values $q_j$ of the $1$-holes describe the translational degrees of freedom of the $M_{+1/2}$
physical spins with projection $+1/2$ and the $M_{-1/2}$ physical spins with projection
$-1/2$, respectively. 

As justified below, the quantum-system information on the opposite-sign coupling
to spin of the unpaired physical spins of projection $+1/2$ and $-1/2$, respectively,
is then stored in the coupling factor ${S^z\over S} = {M_{+1/2} - M_{-1/2}\over M_{+1/2} + M_{-1/2}}$  
on the right-hand side of Eq. (\ref{rel-currents-XXZ}).

As discussed in Appendix \ref{B}, in the case of general energy eigenstates only the $n$-bands and 
corresponding $n$-squeezed effective lattices with finite $N_n$ occupancy contribute to spin transport. 
The corresponding general expression of the spin-current expectation values, 
Eq. (\ref{jznn}) of Appendix \ref{B}, can be exactly written as
\begin{eqnarray}
&& \langle \hat{J}^z (l_{\rm r}^{\eta},S_q,S^z) \rangle = 
\nonumber \\
&& \sum_{n=1}^{\infty}c_n^{n_z}\int_{q_n^-}^{q_n^+}dq\,N_n^h (q)\,J^z_n (q,l_{\rm r}^{\eta},S_q)
\label{Jzgeneral}
\end{eqnarray}
where as justified below the coupling factors $c_n^{n_z}$ are given by
\begin{eqnarray}
&& c_n^{n_z} = {M_{+1/2} - M_{-1/2} + {\cal{M}}_{n,+1/2} - {\cal{M}}_{n,-1/2}\over
M_{+1/2} + M_{-1/2} + {\cal{M}}_{n,+1/2} + {\cal{M}}_{n,-1/2}}
\nonumber \\
&& \hspace{0.56cm} = {M_{+1/2} - M_{-1/2}\over M + {\cal{M}}_{n}} = {2S^z\over N_n^h}  \, .
\label{cnnz}
\end{eqnarray}

The $n$-band spin-current spectrum $J^z_n (q,l_{\rm r}^{\eta},S_q)$ in this expression is given in Eq. (\ref{Jznq}) of 
Appendix \ref{B}. Although ${\cal{M}}_{n,+1/2} - {\cal{M}}_{n,-1/2} = 0$, we have included that term
in the first expression of $c_n^{n_z}$ given in Eq. (\ref{cnnz}): It provides useful information
on how the unpaired and paired physical spins contribute and do not contribute, respectively, to the
spin-current expectation values.

Upon moving in each of the $n$-squeezed effective lattices for which $N_n>0$,
a number $M + {\cal{M}}_{n} = N_n^h$ of physical spins interchange position with 
a number $N_n$ of $n$-pairs. The identification of the $M_{+1/2}$ and $M_{-1/2}$ 
unpaired physical spins of projection $+1/2$ and $-1/2$, respectively,
that carry spin current and of the ${\cal{M}}_{n}={\cal{M}}_{n,+1/2}+{\cal{M}}_{n,-1/2}$ paired physical 
spins that do not carry it is again clear in each corresponding $n$-squeezed effective lattice
sites occupancy configuration.

The $N_n^h = M + {\cal{M}}_{n}$ $n$-band momentum values $q_j$
of a number $N_n^h$ of $n$-holes describe the translational degrees of freedom of 
$M + {\cal{M}}_{n} = N_n^h$ physical spins. There is again quantum uncertainty about which
of the $n$-band momentum values $q_j$ of the $N_n^h = M + {\cal{M}}_{n}$ $n$-holes describe 
the translational degrees of freedom of the $M_{+1/2}$ and $M_{-1/2}$ unpaired physical spins whose coupling
to spin has opposite sign and those of the ${\cal{M}}_{n,+1/2} + {\cal{M}}_{n,-1/2}$ 
paired physical spins that do not couple to spin.

The spin-current expectation values, Eq. (\ref{Jzgeneral}), are expressed in terms of
the $n$-band momentum values $q_j$ of the $N_n^h$ $n$-holes rather than of $n$-squeezed effective lattice sites
occupancies. The quantum-system information on the couplings
to spin of the unpaired physical spins of projections $\pm 1/2$ is then
again stored in the coupling factor $c_n^{n_z}$, Eq. (\ref{cnnz}).

The use of the physical-spin representation, which refers to all finite-$S_q$ energy eigenstates, has
allowed the identification of the spin carriers. This combined with a symmetry reported below
much simplifies the evaluation of the general expression in functional form for the spin-current 
expectation values of such states given in Eq. (\ref{Jzgeneral}). 

The usual procedure used to evaluate such expectation values involves the interplay
of their expression for HWSs, Eq. (\ref{currentHWS}) of Appendix \ref{B}, with the coupled Bethe-ansatz 
equations, Eq. (\ref{BAqn}) of Appendix \ref{A}. And this is usually done for each specific energy 
eigenstate and becomes a complex problem for those with finite occupancy in several $n$-bands.

Within the physical-spin representation, the translational degrees of freedom of the number
$M= M_{+1/2} + M_{-1/2} = 2S_q$ of unpaired physical spins are described in each $n$-band of a $S_q >0$ energy eigenstate
for which $N_n>0$ by a number $N^h_{n} = 2S_q +\sum_{n'=n+1}^{\infty}2(n'-n)N_{n'}$ of $n$-holes.
To account for the coupling to spin of the $M= M_{+1/2} + M_{-1/2} = 2S_q$ 
unpaired physical spins, we must thus use a representation in terms of $n$-holes distributions $N_{n}^h (q_j)$
rather than of $n$-pairs distributions $N_{n} (q_j)$.

The identification and knowledge of the spin carriers permits the separation of the problem to 
obtain the general expression of spin-current expectation values, Eq. (\ref{Jzgeneral}),
into two complementary processes: The $\Phi/N\rightarrow 0$ limit of the derivative $d E/d (\Phi/N)$ 
of the energy eigenvalues, Eq. (\ref{currentHWS}) of Appendix \ref{B}, provides the $\Delta$-dependent current spectra 
$J^z_n (q,l_{\rm r}^{\eta},S_q)$, Eq. (\ref{Jznq}) of Appendix \ref{B}, appearing in Eq. (\ref{Jzgeneral});
The coupling factors $c_n^{n_z}$ of the spin carriers to spin, Eq. (\ref{cnnz}), also appearing in that equation
are accounted for by the $\Phi/L\rightarrow 0$ limit of the derivative $d P/d (\Phi/N)$ of the momentum eigenvalues. 

This second process occurs through the transformation of the $n$-band momentum summations 
$\sum_{j=1}^{L_n}N_n^h (q_j) = {N\over 2\pi}\int_{q_n^-}^{q_n^+}dq\,N_n^h (q)$
in the spin-current expectation values expression. That transformation is directly related to
the dependence on the vector potential $\Phi/N$ of the general momentum eigenvalues.
As justified and reported in Sec. \ref{SECIIB}, in the case of general non-HWSs such eigenvalues read,
\begin{equation}
P = \pi\sum_{j=1}^{L_n}N_{n} - \sum_{n=1}^{\infty}\sum_{j=1}^{L_n}N_{n}^h (q_j)\,q_j + {\Phi\over N}\,(M_{+1/2} - M_{-1/2}) \, .
\label{PPhinonHWS}
\end{equation}

An important symmetry that much simplifies the problem is that, in contrast to the Bethe-ansatz equations, 
such momentum eigenvalues are for $\Delta >1$ independent of the anisotropy $\Delta$. 
Only the spin carriers, which are the unpaired physical spins in the multiplet configuration
of all $S_q>0$ energy eigenstates, couple to the vector potential $\Phi/N$ and
thus contribute the spin-current expectation values, Eq. (\ref{Jzgeneral}). 

Within periodic boundary conditions, such a selective coupling occurs through effective fluxes 
$\Phi_{\rm eff}^{n} $. In the presence of the vector potential, they pierce each of the rings associated 
with the $n$-squeezed effective lattices of length $L=N$ for which $N_n >0$. 

Such effective fluxes are determined by the derivative $d P/d (\Phi/N)$ of the momentum eigenvalues, 
Eq. (\ref{PPhinonHWS}), through the following sum rules,
\begin{eqnarray}
&& {d\Phi_{\rm eff}^{n}\over d (\Phi/L)}\sum_{j=1}^{L_n}N_n^h (q_j) =
\nonumber \\
&& {d\Phi_{\rm eff}^{n}\over d (\Phi/L)}{N\over 2\pi}\int_{q_n^-}^{q_n^+}dq\,N_n^h (q) = {d P\over d (\Phi/N)} \, .
\label{cnnzring}
\end{eqnarray}
Here the derivative $d P/d (\Phi/N)= 2S^z = M_{+1/2} - M_{-1/2}$ involves the numbers of spin carriers of
opposite projection $\pm 1/2$ multiplied by their coupling to spin of opposite sign. 

The physical solutions of Eq. (\ref{cnnzring}) are under the boundary conditions $\Phi_{\rm eff}^{n}=0$ 
for $\Phi/L\rightarrow 0$ given by
\begin{equation}
\Phi_{\rm eff}^{n} = {\Phi\over L} {2S^z\over N_n^h} =  {\Phi\over L} c_n^{n_z} 
\hspace{0.20cm}{\rm and}\hspace{0.20cm}{d\Phi_{\rm eff}^{n}\over d (\Phi/L)}
= {2S^z\over N_n^h} = c_n^{n_z} \, .
\label{Phieffsn}
\end{equation}
The derivatives $d\Phi_{\rm eff}^{n}/d (\Phi/L)$ are the coupling factors $c_n^{n_z}$, Eq. (\ref{cnnz}), in the expression 
of the spin-current expectation values, Eq. (\ref{Jzgeneral}). 

The form of Eq. (\ref{cnnzring}) confirms that the selective mechanism under which only the unpaired physical
spins that couple to the vector potential are those that contribute to 
the spin-current expectation values is indeed controlled by the derivative $d P/d (\Phi/N)$. 

Under that mechanism, the summations ${N\over 2\pi}\int_{q_n^-}^{q_n^+}dq\,N_n^h (q) = N_n^h$
where $N_n^h = M + {\cal{M}}_{n}$ involves both the $M$ unpaired physical spins
and the ${\cal{M}}_{n}$ paired physical spins ``seen'' by the $n$-pairs, are replaced
by $c_n^{n_z}{N\over 2\pi}\int_{q_n^-}^{q_n^+}dq\,N_n^h (q) = M_{+1/2} - M_{-1/2}$,
i.e. the number of spin carriers of projection $\pm 1/2$ multiplied by their
couplings of opposite sign to spin.

The coupling factors $c_n^{n_z}$, Eq. (\ref{cnnz}), in the spin-current expectation values in 
functional form, Eq. (\ref{Jzgeneral}), ensure, for instance, that such expectation values
vanish when $M = M_{+1/2} + M_{-1/2} = 2S_q =0$ or $M_{+1/2} = M_{-1/2}$.

In summary, the general spin-current expectation value expression, Eq. (\ref{Jzgeneral}), can 
be obtained by first replacing $N_{n}(q_{j})$ by $1 - N_{n}^h (q_{j})$ in 
the energy eigenvalue expression, Eq. (\ref{Energy}). Accounting for the above two
processes separation, that expectation value can be written as
\begin{eqnarray}
\langle \hat{J}^z (l_{\rm r}^{\eta},S_q,S^z) \rangle & = & - \sum_{n=1}^{\infty}\sum_{j=1}^{L_n}
(1 - {d P\over d (\Phi/N)} {1\over N_{n}^h}N_{n}^h (q_{j}))
\nonumber \\
& \times & {d\over d (\Phi/N)}{J\sinh\eta\sinh (n\,\eta)\,\over \cosh (n\,\eta) - \cos\varphi_{n} (q_{j})} \, .
\label{DerivativesEnergy}
\end{eqnarray}
Performing the derivative $d/d (\Phi/N)$ of the energy spectrum factor
and replacing $\sum_{j=1}^{L_n}$ in the term without the factor $N_{n}^h (q_{j})$ 
by $n$-band rapidity-variable integrals, we find that term vanishes, i.e.,
\begin{equation}
- {N\over\pi}\sum_{n=1}^{\infty}\int_{-\pi}^{\pi }d\varphi\,
{J n \sinh\eta\sinh (n\eta)\sin \varphi\over (\cosh (n\eta) - \cos \varphi)^2} = 0 \, .
\label{J0}
\end{equation}

Finally, replacing $\sum_{j=1}^{L_n}$ by $n$-band momentum integrals in the remaning term
that contains the distribution $N_{n}^h (q_{j})$, we 
obtain the general expression, Eq. (\ref{Jzgeneral}).

\end{document}